\documentclass[sn-nature]{sn-jnl}

\usepackage{graphicx}%
\usepackage{multirow}%
\usepackage{amsmath,amssymb,amsfonts}%
\usepackage{amsthm}%
\usepackage{mathrsfs}%
\usepackage[title]{appendix}%
\usepackage{xcolor}%
\usepackage{textcomp}%
\usepackage{manyfoot}%
\usepackage{booktabs}%
\usepackage[ruled,lined,linesnumbered]{algorithm2e}
\usepackage{algpseudocode}% http://ctan.org/pkg/algorithmicx
\usepackage{listings}%
\usepackage{subcaption}
\usepackage{longtable}
\let\oldnl\nl% Store \nl in \oldnl
\newcommand{\nonl}{\renewcommand{\nl}{\let\nl\oldnl}}% Remove line number for one line

\newcommand{\beginsupplement}{%
        \setcounter{table}{0}
        \renewcommand{\thetable}{S-\arabic{table}}%
        \setcounter{figure}{0}
        \renewcommand{\thefigure}{S-\arabic{figure}}
        \setcounter{section}{0}
        \renewcommand{\thesection}{S-\arabic{section}}%
        \setcounter{page}{1}
        \renewcommand{\thepage}{\arabic{page}}
        \setcounter{equation}{1}
        \renewcommand{\theequation}{S-\arabic{equation}}
     }

\usepackage[resetlabels]{multibib}
\newcites{SM}{References}

\begin{document}

\title[Article Title]{Using Scalable Computer Vision to Automate High-throughput Semiconductor Characterization}

\author*[1]{\fnm{Alexander E.} \sur{Siemenn}}\email{asiemenn@mit.edu}
\equalcont{These authors contributed equally to this work.}

\author*[1]{\fnm{Eunice} \sur{Aissi}}\email{eunicea@mit.edu}
\equalcont{These authors contributed equally to this work.}

\author[1]{\fnm{Fang} \sur{Sheng}}\email{shengf22@mit.edu}

\author[1,2]{\fnm{Armi} \sur{Tiihonen}}\email{armi.tiihonen@gmail.com}

\author[1,3]{\fnm{Hamide} \sur{Kavak}}\email{hkavak@mit.edu}

\author[1]{\fnm{Basita} \sur{Das}}\email{dasb@mit.edu}

\author[1]{\fnm{Tonio} \sur{Buonassisi}}\email{buonassisi@mit.edu}

\affil[1]{\orgdiv{Department of Mechanical Engineering}, \orgname{Massachusetts Institute of Technology}, \orgaddress{\street{77 Massachusetts Avenue}, \city{Cambridge}, \postcode{02139}, \state{Massachusetts}, \country{USA}}}

\affil[2]{\orgdiv{Department of Applied Physics}, \orgname{Aalto University}, \orgaddress{\street{Otakaari 24}, \city{Espoo}, \postcode{02150}, \country{Finland}}}

\affil[3]{\orgdiv{Department of Physics}, \orgname{Cukurova University}, \orgaddress{\city{Adana}, \postcode{01330}, \country{Turkiye}}}

%%%%%%%%% ABSTRACT
\abstract{
High-throughput materials synthesis methods have risen in popularity due to their potential to accelerate the design and discovery of novel functional materials, such as solution-processed semiconductors. After synthesis, key material properties must be measured and characterized to validate discovery and provide feedback to optimization cycles. However, with the boom in development of high-throughput synthesis tools that champion production rates up to $10^4$ samples per hour with flexible form factors, most sample characterization methods are either slow (conventional rates of $10^1$ samples per hour, approximately 1000x slower) or rigid (\textit{e.g.}, designed for standard-size microplates), resulting in a bottleneck that impedes the materials-design process. To overcome this challenge, we propose a set of automated material property characterization (autocharacterization) tools that leverage the adaptive, parallelizable, and scalable nature of computer vision to accelerate the throughput of characterization by 85x compared to the non-automated workflow. We demonstrate a generalizable composition mapping tool for high-throughput synthesized binary material systems as well as two scalable autocharacterization algorithms that (1) autonomously compute the band gap of 200 unique compositions in 6 minutes and (2) autonomously compute the degree of degradation in 200 unique compositions in 20 minutes, generating ultra-high compositional resolution trends of band gap and stability. We demonstrate that the developed band gap and degradation detection autocharacterization methods achieve 98.5\% accuracy and 96.9\% accuracy, respectively, on the FA$_{1-x}$MA$_{x}$PbI$_3$, $0\leq x \leq 1$ perovskite semiconductor system.

}


%%\pacs[JEL Classification]{D8, H51}

%%\pacs[MSC Classification]{35A01, 65L10, 65L12, 65L20, 65L70}

\maketitle

%%%%%%%%% BODY TEXT
\section*{Introduction} % Rename to Main for submission

To discover commercially relevant semiconductor materials, \textit{e.g.}, for solar applications \cite{Mazumdar2021, Siegler2022, Duan2023}, vast compositional search spaces must be rapidly synthesized and characterized, \textit{e.g.}, for band gap \cite{Hu2019, Prasanna2017, Baloch2022} and stability \cite{Sun2021, keesey_tiihonen_siemenn_colburn_sun_hartono_serdy_zeile_he_gurtner_et}. Recently, several new methods have been developed for high-throughput (HT) combinatorial synthesis across a wide range of material domains including perovskites, nanomaterials, porous media, aerosols, and lithium-ion batteries \cite{Wang2023, Langner2020, Ludwig2019, MacLeod2022, Moradi2022, Sun2019, Yao2020, Clayson2020, Zeng2023, Liu2017}. Although these HT manufacturing methods have shown great progress in driving the rapid screening of large material search spaces in an automated fashion, much of the materials characterization process is still hindered due to its manual nature \cite{Sun2019, Makula2018} or rigid microplate-based form factors \cite{Wang2023, Langner2020, Du2021, Surmiak2020, Reinhardt2020}. This results in a significant bottleneck in which the rate of synthesis can achieve throughputs over 800x faster than those of characterization (\textit{e.g.}, Supplementary Figure \ref{sfig:times}). The importance of developing rapid and accurate methods of characterization for HT materials discovery and optimization derives from the intractable nature of exhaustively testing every material within a functional material's search space using these conventional tools \cite{Sun2019}.

\begin{figure}[h!p]
\begin{center}
\includegraphics[width=0.8\columnwidth]{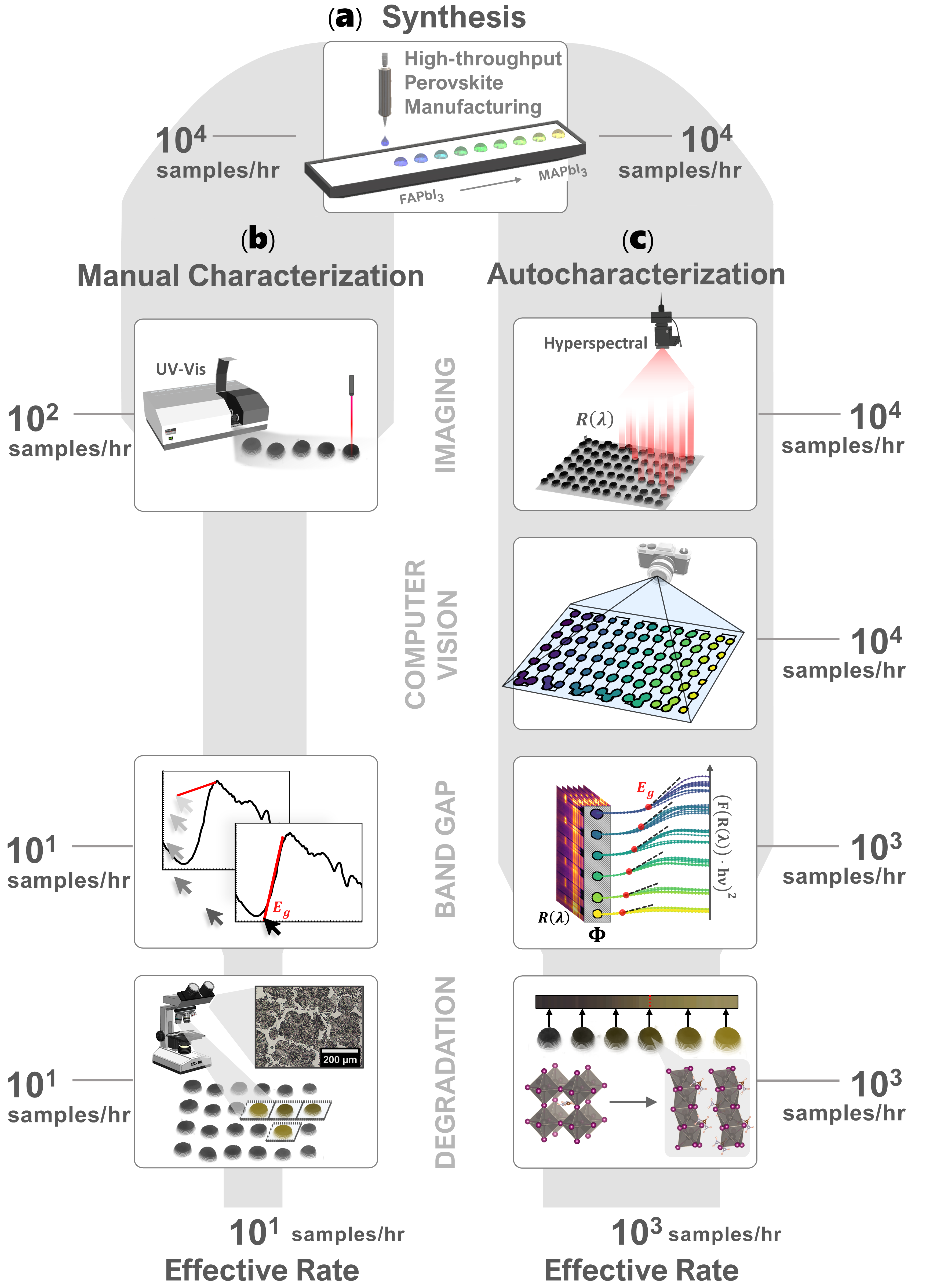}
\end{center}
   \caption{\textbf{a-c}, Overview of the synthesis and characterization of perovskite semiconductors. \textbf{a}, High-throughput combinatorial synthesis of FA$_{1-x}$MA$_{x}$PbI$_3$ perovskites attain throughputs of $10^4$ samples/hr. \textbf{b}, Manual characterization of the high-throughput-manufactured materials using UV-Vis spectroscopy and manual determination of band gap and degradation bottlenecks the pipeline down to a throughput of $10^1$ samples/hr. \textbf{c}, Autocharacterization, developed in this paper, of the high-throughput-manufactured material's band gaps and degradation attain throughputs of $10^3$ samples/hr using scalable and parallelizable computer vision measurement. Band gap is determined by automatically segmenting and fitting the material reflectance spectra while the degradation pathway is detected by the material yellowing, in the case depicted above due to a phase change from $\alpha$-FAPbI$_3$ to $\delta$-FAPbI$_3$. The widths of the gray backgrounds visualize the process throughputs.}
   \label{fig:workflow}
\end{figure}

% The workflow for autocharacterization of perovskite semiconductors using computer vision. From left to right: A fluid handling system mixes and deposits each unique semiconductor material onto a plate. The vision system captures images in RGB and hyperspectral channels (wavelengths $\lambda \in [380\textrm{nm}, 1020\textrm{nm}]$), then segments the pixels of each material deposit ($X, Y, \lambda$) from the background. The (i) compositional information, (ii) direct band gap, and (iii) stability of each unique sample are automatically characterized using the optical data segmented by the vision system.

The metal halide perovskite material search space is both highly dimensional and vast, hence, as a result, it is intractable to map using conventional synthesis or characterization methods. The common metal halide perovskite subspace explored in literature consists of an eight-component system: $(\textrm{FA}_x \textrm{MA}_y \textrm{Cs}_{1-x-y})(\textrm{Pb}_z \textrm{Sn}_{1-z}) (\textrm{Br}_a \textrm{Cl}_b \textrm{I}_{1-a-b})_3$ \cite{Wang2023, Ahmadi2021, Sun2021, Wang2017, Sun2019, Liu2023}. For one to exhaustively search this subspace at 1\% compositional steps would require synthesizing and characterizing $7\times10^{12}$ unique material samples (Supplementary Figure \ref{sfig:search}). Thus, only small regions of this search space can be explored experimentally with current methods, given the large discrepancy between search space size and characterization throughput. For example, Escobedo \textit{et al.} \cite{escobedo2019} develop automated programs to compute the band gap of semiconductor materials, however, the throughput of the software is insufficient for HT combinatorial synthesis applications as it can only compute the band gap of one sample at a time using pre-collected optical data. Surmiak \textit{et al.} \cite{Surmiak2020} and Reinhardt \textit{et al.} \cite{Reinhardt2020} expand on these methods by employing HT optical characterization of perovskites for up to 24 unique samples per batch, however, their measurements occur in serial and are rigidly hard-coded to measure only specified locations, in turn, lacking scalability and generalizability. Similarly, Langner \textit{et al.} \cite{Langner2020} and Wang \textit{et al.} \cite{Wang2023} develop significantly higher throughput drop casting tools capable of synthesizing and characterizing up to 6048 organic photovoltaic films per day, while Du \textit{et al.} \cite{Du2021} develop an automated robotic synthesis and characterization platform for full organic photovoltaic devices. However, these prior methods all utilize rigid characterization which hard-codes measurement locations, in turn, locking the method of characterization to a specific platform architecture. Unlike the aforementioned literature, Wu \textit{et al.} \cite{Wu2023_advmat} develop a HT and scalable tool, automated from end-to-end, to characterize the optical properties of organic molecules. However, their developed tool is designed to only characterize organic molecules actively suspended in solution \cite{Wu2023_advmat}, thus, leaving the task of rapidly and automatically characterizing semiconductors deposited with flexible form factors onto a substrate as an open research gap.

% Many of the existing HT approaches work for a fixed system and sample architecture, however, many R\&D laboratories frequently modify their process and invest a significant effort into re-building setups. Thus, developing pipelines that are not locked to specific systems or sample architectures accelerates the research cycles.

Debottlenecking the materials screening pipeline is not only a matter of accelerating the characterization time per sample \cite{Siemenn2022, Zhu2020} but also a matter of scaling the characterization procedure to many samples in parallel \cite{Tung2010, Jain_2021_ICCV}. Computer vision methods have the capacity to scale a measurement to arbitrarily many samples, each with differing form factor geometries, without significantly slowing characterization time \cite{garnot2021panoptic, Li_2022_CVPR, park_ding_2019}. Material science applications of computer vision are gaining traction in current literature, specifically in the use case of rapid morphological analysis of large microscopy datasets \cite{park_ding_2019}. In turn, several analytical computer vision tools have been developed to access and characterize this morphological information, often focused on identifying microstructures. For example, Park \textit{et al.} \cite{doi:10.1080/0740817X.2011.587867} create a semi-automated image segmentation and analysis algorithm to classify the morphology of nanoparticles from image data. Likewise, Chowdhury \textit{et al.} \cite{CHOWDHURY2016176} utilize computer vision and machine learning to detect dendritic microstructures from a database of solder alloy micrographs. There is also extensive work surrounding materials recognition in non-micrographic images using computer vision \cite{chen_wolff_1998, leung_malik_2001, Siemenn2022, Ma2023}. In order to support the growing need for the analysis of micrographs and other experimental images, advances like those made by Li \textit{et al.} \cite{Li_2022_CVPR}, Wang \textit{et al.} \cite{Wang_2022_CVPR}, Neshatavar \textit{et al.} \cite{Neshatavar_2022_CVPR}, Tung \textit{et al.} \cite{Tung2010}, and Jain \textit{et al.} \cite{Jain_2021_ICCV} on object segmentation, denoising, and scalability allow for further use cases of computer vision in scientific research. The throughput dichotomy between characterization and HT synthesis motivates the integration of computer vision into the semiconductor characterization pipeline to parallelize measurements and, in turn, match or exceed the rate of synthesis while achieving accuracies comparable to those attained by domain expert evaluation.

In this paper, we address the unresolved challenge of characterizing deposited materials quickly and automatically in parallel by developing a suite of computer vision-based automated characterization (autocharacterization) tools that enable the quantification of three key properties within minutes: composition \cite{Tian2020, Jeon2015}, optical band gap \cite{Hu2019, Prasanna2017, Baloch2022}, and environmental degradation \cite{Wang2023, Sun2021, keesey_tiihonen_siemenn_colburn_sun_hartono_serdy_zeile_he_gurtner_et}. We propose the following in this contribution: (1) sample detection by a scalable computer vision tool that segments arbitrarily many, spatially non-uniform material samples, (2) a tool to map the elemental composition of HT-manufactured material arrays, (3) a scalable autocharacterization algorithm for the computation of direct band gaps from hyperspectral reflectance data, and (4) a scalable autocharacterization algorithm for quantifying the environmental stability of perovskite samples from optical degradation data. The performances of the developed autocharacterization methods are demonstrated on 200 unique HT-manufactured FA$_{1-x}$MA$_{x}$PbI$_3$ perovskite semiconductor samples, generating ultra-high compositional resolution trends of band gap and stability, and are benchmarked against X-ray diffraction \cite{Massuyeau2022, Mundt2020}, X-ray photoelectron spectroscopy \cite{Zhidkov2021, Ahmad2017}, and domain expert evaluation.

\section*{Results}

\subsection*{Computer Vision Parallelization}

\begin{figure}[h!]
\begin{center}
\includegraphics[width=0.6\columnwidth]{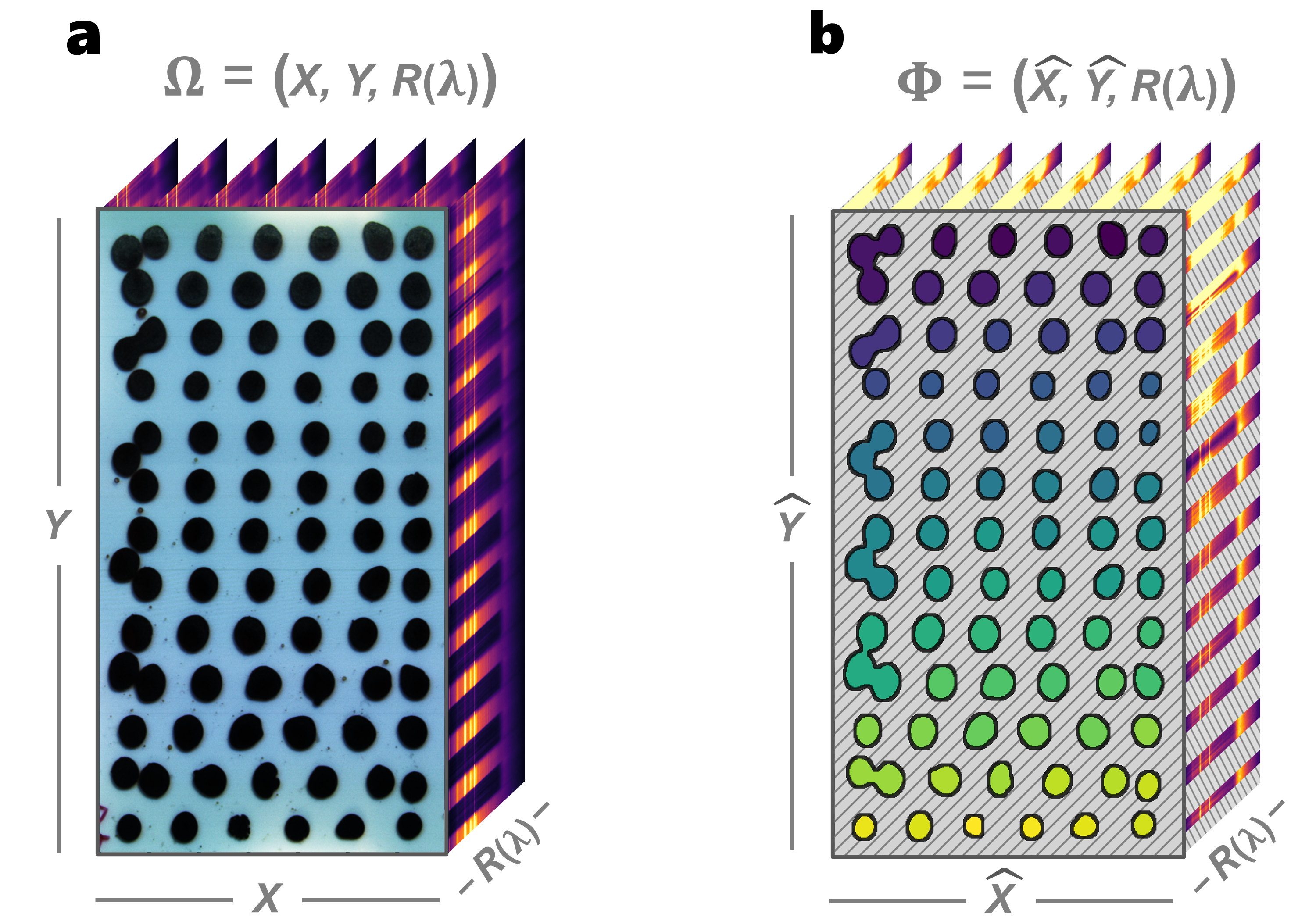}
\end{center}
\caption{\textbf{a}, Raw hyperspectral datacube, $\Omega$, captured using a hyperspectral imager (Resonon, Pika L) of HT-deposited FA$_{1-x}$MA$_{x}$PbI$_3$ perovskites. $(X, Y)$ represents the pixel coordinates, and $R(\lambda)$ represents the reflectance spectra for each pixel. Each sample is deposited onto the glass substrate with a unique composition $0\leq x \leq 1$ and flexible form factor geometry. \textbf{b}, Computer-vision segmented datacube, $\Phi$, that pairs each unique sample's pixels, $(\widehat{X},\widehat{Y})_n \in N$, to its reflectance spectra, $R(\lambda)$. The gray hatched region indicates the discarded background pixels.}
\label{fig:segment}
\end{figure}

With the integration of computer vision into the materials characterization workflow, data across many samples can be captured and analyzed in parallel as a fast and scalable process \cite{Tung2010, Jain_2021_ICCV, garnot2021panoptic, Li_2022_CVPR, park_ding_2019}. Computer vision can be applied to both standard RGB and hyperspectral image data types. Figure \ref{fig:segment} illustrates the segmentation of a hyperspectral datacube, taken for one batch of HT-manufactured perovskite samples \cite{Siemenn2022}. In this study, we synthesize three batches of samples, amounting to a total of $N=201$ unique semiconductors along the FA$_{1-x}$MA$_{x}$PbI$_3$ compositional series with $0 \leq x \leq 1$. Using Algorithm \ref{alg:segmentation}, we generate unique pairings for each discrete semiconductor sample, $(\widehat{X},\widehat{Y})_n \in N$, and its corresponding reflectance spectra, $R(\lambda)$, \textit{via} parallel image segmentation and mapping, shown in Figure \ref{fig:segment}b.

The process of parallel segmentation uses a sequence of edge-detection filters \cite{opencv_library} to first identify each island of material and then uniquely index each island based on its position within a graph connectivity network \cite{batagelj2003efficient, Meyer1994}. The pixel coordinates of each segmented sample are then spatially mapped to their corresponding reflectance spectra, in turn, generating the segmented datacube $\Phi=(\widehat{X},\widehat{Y}, R(\lambda))$. Parallel segmentation and mapping of these reflectance spectra accelerate a once point-by-point measurement process \cite{Sun2019, Makula2018}, to a rapid and scalable process that can rate-match the throughput of HT synthesis (Supplementary Figure \ref{sfig:times}). Furthermore, this segmentation process is shown to scale to more than 80 unique samples in parallel (Supplementary Figure \ref{sfig:comp-extract}). Algorithm \ref{alg:segmentation} illustrates the sample size-agnostic nature of the method, highlighting its further scalable potential. This segmented reflectance data serves as the starting point for automating the mapping of composition as well as the automating the computation of optical band gap and degradation for all $N=201$ unique semiconductors.

\subsection*{Composition Mapping}

\begin{figure*}[h!]
\begin{center}
\includegraphics[width=1.\columnwidth]{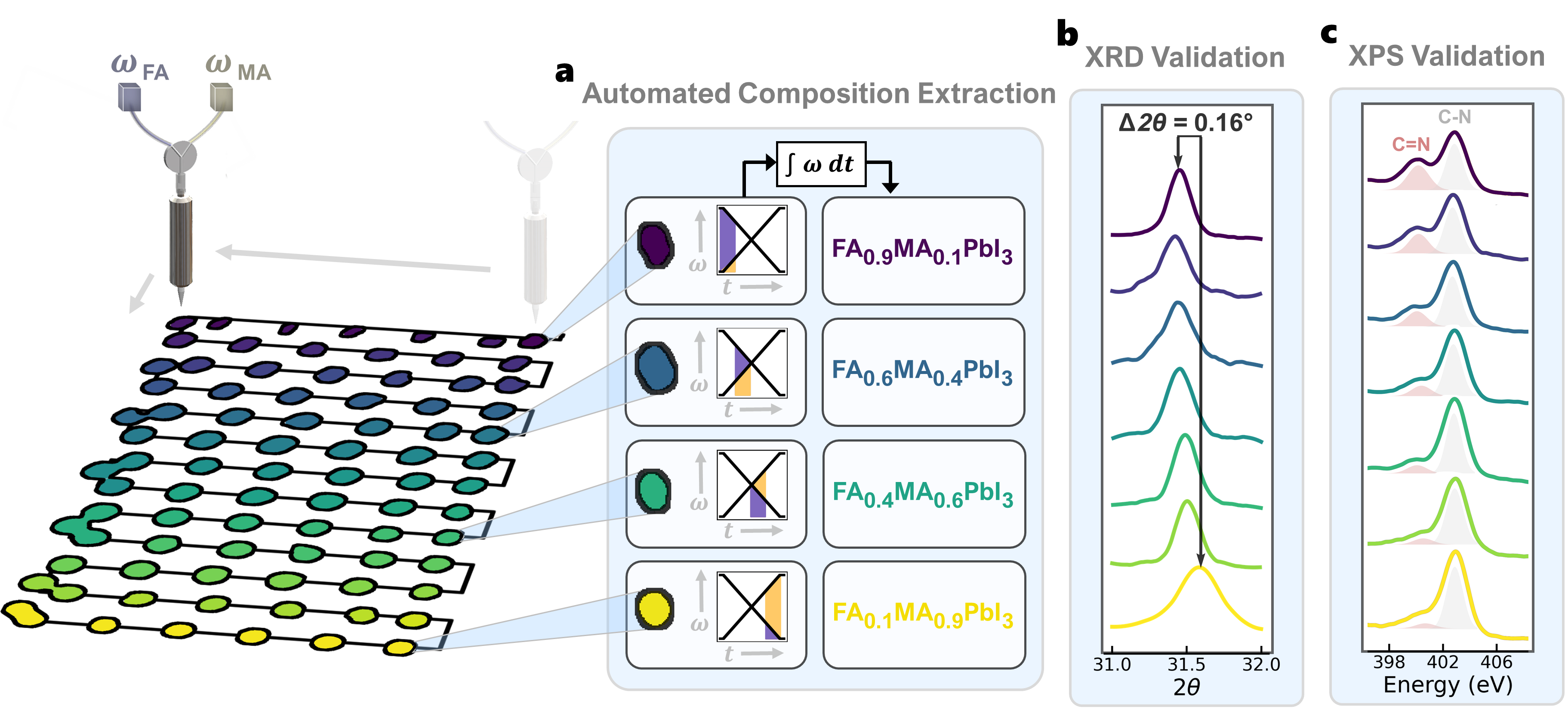}
\end{center}
   \caption{\textbf{a}, High-throughput combinatorial synthesis of one batch of perovskite samples illustrated with its corresponding computer vision-segmented composition map. The print head rasters in a serpentine pattern (black connecting lines) to print a gradient of FA$_{1-x}$MA$_x$PbI$_3$ deposits onto a glass substrate, where the purple labels indicate FA-rich deposits, and the yellow labels indicate MA-rich deposits. Integrating the pump speeds over time determines the proportion of MA, $x$, in the composition. \textbf{b}, XRD peak traces at the (012) crystallographic plane measured at uniformly spaced compositions in the batch print. The peak shifts towards a higher $2\theta$ angle gradually as the proportion of MA increases in the composition. \textbf{c}, XPS traces of the C=N bond peak (red area under the curve) and C-N bond peak (gray area under the curve) measured at uniformly spaced compositions in the batch print. The C=N peak intensity decreases as the proportion of MA increases. Purple-labeled traces are FA-rich while yellow-labeled traces are MA-rich.}
\label{fig:integral}
\end{figure*}

Semiconductor properties, such as band gap and stability, are largely governed by the chemical composition of the material \cite{Sun2021, keesey_tiihonen_siemenn_colburn_sun_hartono_serdy_zeile_he_gurtner_et, Tian2020}. Thus, determining the composition accurately enough and mapping the composition to a correct sample is essential for scalable HT setups that deposit variable sample geometries. In this work, we use HT deposition of solution-processed precursors to synthesize perovskite semiconductor samples. The variable rotational velocities of two pumps, $\omega_\textrm{FA}$ and $\omega_\textrm{MA}$, determine the output sample composition by combining the liquid-based FAPbI$_3$ and MAPbI$_3$ precursors. Figure \ref{fig:integral}a illustrates this printing process as a function of both space and time. The FA-rich materials are printed first and then proportions of MA are added gradually as the print head rasters in a serpentine pattern. Thus, to determine the composition of each material deposit, $\omega_\textrm{FA}$ and $\omega_\textrm{MA}$ are spatially and temporally mapped onto the computer vision-segmented samples, $(\widehat{X},\widehat{Y})_n \in N$, and then integrated over time to determine each sample's computed FA$_{1-x}$MA$_x$PbI$_3$ composition:
\begin{equation}
    \label{eq:composition}
    x(t) \approx \int_{t_a}^{t_b}\frac{\omega_\textrm{MA}(t)}{\left(\omega_\textrm{MA}(t) + \omega_\textrm{FA}(t)\right)}dt,
\end{equation}
where $x$ is the proportion of MA, $t_a$ and $t_b$ are the starting and ending timesteps for the deposition of a single sample, and $\omega_\textrm{FA}(t)$ and $\omega_\textrm{MA}(t)$ are the pump velocities at a given timestep for the FAPbI$_3$ and MAPbI$_3$ precursors, respectively.

Figure \ref{fig:integral}b-c illustrates the composition validation results using X-ray diffraction (XRD) and X-ray photoelectron spectroscopy (XPS). XRD is used to validate the crystalline structure  \cite{Massuyeau2022, Mundt2020} while XPS is used to validate the elemental composition of the manufactured perovskite deposits \cite{Zhidkov2021, Ahmad2017}. By assessing the gradated shifts in both XRD and XPS peaks, the composition mapping can be validated. For crystal structure validation, the crystal lattice size of MAPbI$_3$ is smaller than that of FAPbI$_3$ \cite{Jeon2015}, thus, we expect the XRD peaks of the MA-rich deposits to shift toward higher angles \cite{Elsayed2023}. In Figure \ref{fig:integral}b, the (012) crystallographic plane at $2\theta \approx 31.5 ^\circ$ \cite{Murugadoss2021, Zhang2019} is shown, for uniformly-spaced samples across the batch print, to gradually increase in angle from the FA-rich to the MA-rich compositions of FA$_{1-x}$MA$_x$PbI$_3$, amounting to a total shift of approximately $0.16^\circ$ (quantitative shifts shown in Supplementary Figure \ref{sfig:quant-xrd-xps}a). For elemental validation, the A-site MA and FA cations are distinguished by the presence of a carbon-nitrogen double bond (C=N), where FA contains a C=N bond while MA contains only a C-N single bond \cite{Elsayed2023}. In the high-resolution XPS scans, shown in Figure \ref{fig:integral}c, the C=N bond peak appears at approximately 400eV \cite{Maqsood2020}. This C=N bond peak is shown, for uniformly-spaced samples across the batch print, to gradually decrease in intensity from the FA-rich to the MA-rich compositions of FA$_{1-x}$MA$_x$PbI$_3$ (quantitative shifts shown in Supplementary Figure \ref{sfig:quant-xrd-xps}b). These results validate both the structural and elemental composition gradients synthesized and mapped using computer vision. With the developed composition map, we can now automatically compute the band gap and detect degradation across the 201 uniquely synthesized compositions.

\subsection*{Automated Band Gap Extraction}

\begin{figure*}[h!]
\begin{center}
\includegraphics[width=1.\columnwidth]{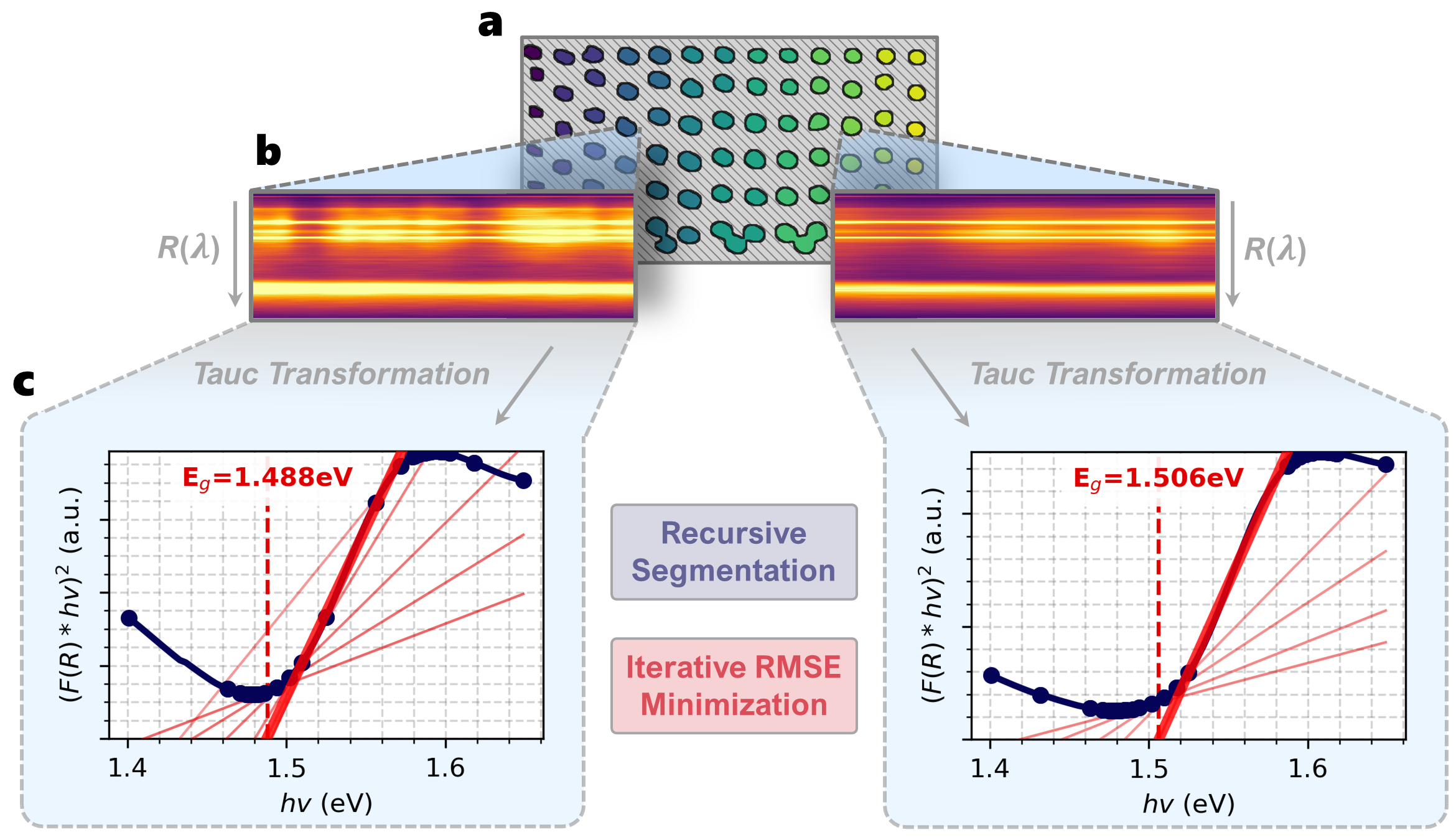}
\end{center}
   \caption{\textbf{a-c}, Automatic band gap computation shown for two unique, computer vision-segmented perovskite deposits. \textbf{b}, The reflectance intensities, $R(\lambda)$, are acquired for each sample from \textbf{a}, the vision-segmented hypercube, $\Phi$. \textbf{c}, The Tauc curves are computed from the median reflectance spectra for each deposit, recursively segmented into line segments, and then iteratively fit with linear regression lines. The best-fit regression line that minimizes the RMSE between the detected Tauc peaks is illustrated by the thick red line, which determines the band gap, $E_g$, from the x-intercept.}
\label{fig:bandgap}
\end{figure*}

Band gap is essential in defining the light-harvesting potential of a semiconductor material \cite{Hu2019, Jeon2015}. However, conventionally computing the band gap of a material is a laborious process, requiring a user to manually curve-fit the Tauc-transformed UV-Vis spectra \cite{Tauc1966, Makula2018}. In this paper, we automate and accelerate the computation of band gap across 201 unique semiconductor samples by leveraging hyperspectral imaging and computer vision segmentation, such that the characterization process is parallelized across batches of HT-manufactured semiconductor samples.

Figure \ref{fig:bandgap} illustrates the autocharacterization process of band gap by first, (a) extracting the reflectance spectra from each computer vision-segmented sample within the hyperspectral datacube, then, (b) transforming each reflectance spectra to its Tauc curve, and finally, (c) recursively segmenting the Tauc curves into linear segments ($R^2\geq0.990$) to find the regression line of best fit between peaks, which determines band gap. The Tauc curves are obtained from a hyperspectral reflectance datacube using the following transformation \cite{Makula2018}:
\begin{equation}
    F(R(\lambda)\cdot h\nu)^{1/ \gamma}=B(h\nu-E_g),
    \label{eq:bg}
\end{equation}
where $ F(R(\lambda))$ is the Kubelka-Munk function \cite{Kubelka1931} applied to reflectance spectra, $R(\lambda)$, for each wavelength, $\lambda$, with $h$ as the Planck constant, $\nu$ as the photon frequency, $B$ as a constant, and $\gamma=\frac{1}{2}$ for direct band gap and $\gamma=2$ for indirect band gap.

\begin{figure}[h!]
\begin{center}
\includegraphics[width=0.65\columnwidth]{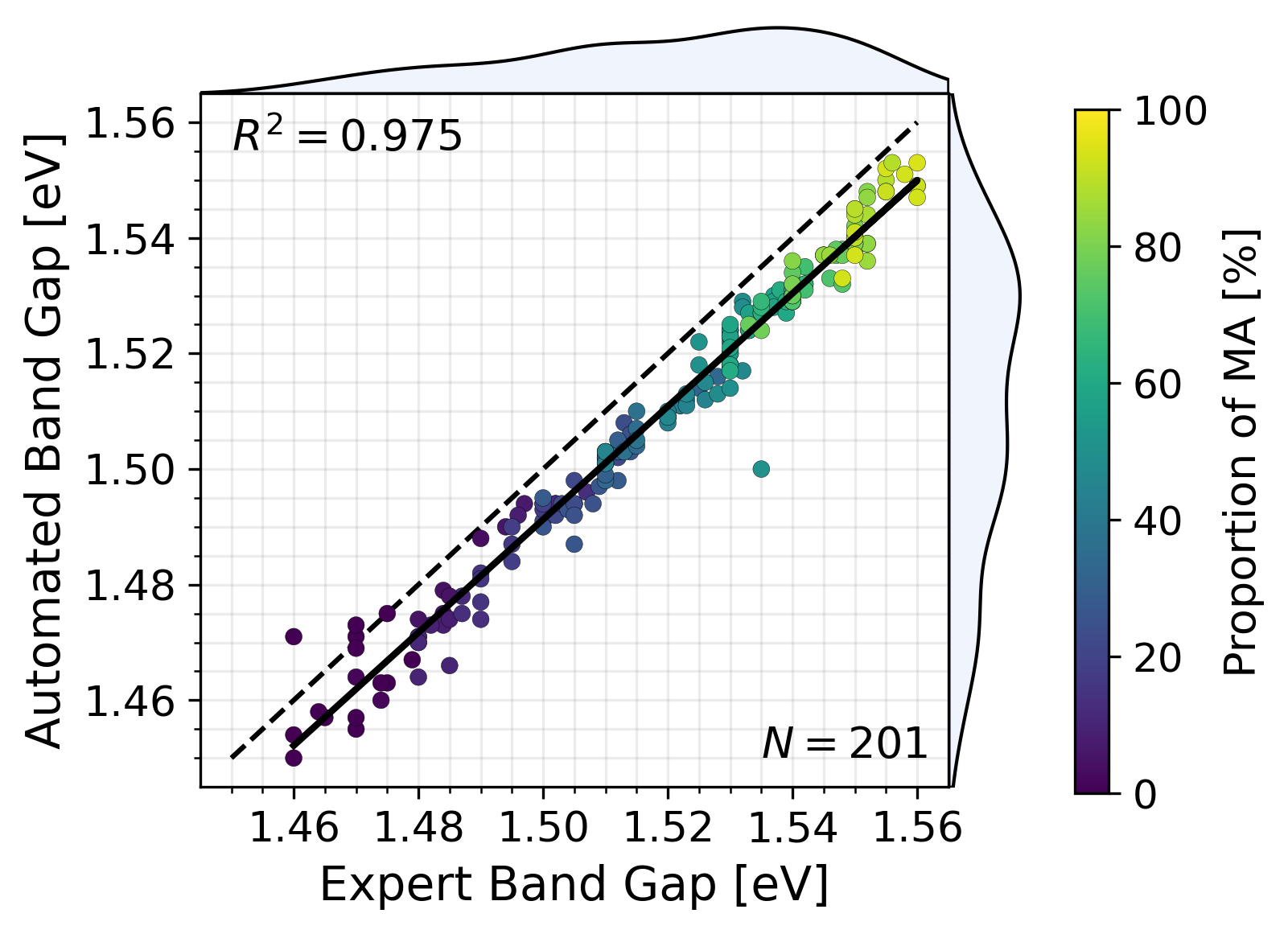}
\end{center}
   \caption{Performance of the autocharacterization of band gap relative to the domain expert-compute band gap for $N=201$ unique perovskite samples across 3 independent trials. The solid black line is the regression fit to the band gap data and the dashed black line is the $y=x$ line. Histogram distributions of both autocharacterization and domain expert band gaps are shown on the right and top of the plot area, respectively. The color of the scatter points corresponds to the proportion of MA, $x$, in the composition FA$_{1-x}$MA$_x$PbI$_3$.}
\label{fig:bg-result}
\end{figure}

To demonstrate the performance of the band gap autocharacterization developed in this paper, the algorithm's output band gaps are compared with those calculated by a domain expert using the manual fitting process described in Makula \textit{et al.} \cite{Makula2018}. The autocharacterization output was withheld from the domain expert. Figure \ref{fig:bg-result} illustrates the performance of the autocharacterization-calculated band gaps relative to the expert-calculated band gaps for $N=201$ FA$_{1-x}$MA$_x$PbI$_3$ compositions across three independent batches of samples. The autocharacterization output achieves a strong linear fit of $R^2=0.975$ with the expert-calculated results, however, a systematic underprediction of the autocharacterization algorithm is noted. Relative to the expert-computed band gap, the autocharacterization method achieves 98.5\% accuracy within a 0.02eV range on the FA$_{1-x}$MA$_x$PbI$_3$ system (Supplementary Figure \ref{sfig:bandgap-acc}). In addition to the autocharacterization achieving high similitude with the domain expert, it also provides significant speedups in band gap determination. The domain expert takes approximately 510 minutes to compute the band gap of 200 unique samples while the autocharacterization takes 6 minutes to compute the band gap of 200 samples, resulting in the developed autocharacterization tool achieving 85x faster throughput \textit{via} the parallel-processing power of computer vision.

Using the fast and accurate band gap autocharacterization tool developed in this paper, we tractably generate an ultra-high resolution band gap trend for the FA$_{1-x}$MA$_x$PbI$_3$, $0\leq x \leq 1$  series, shown in Figure \ref{fig:bg-result}, where 120 of the 201 compositions are unique, a resolution that has not yet been reported in literature. Prior literature reports band gap compositional resolutions from $0\leq x \leq 1$ for 9 compositions \cite{Weber2016}, 7 compositions \cite{gil2020efficient, lu2020compositional}, 6 compositions \cite{slimi2016perovskite}, and 5 compositions \cite{Sun2021} using conventional characterization methods. Thus, with autocharacterization, we achieve over a 13x increase in the compositional resolution of FA$_{1-x}$MA$_x$PbI$_3$ band gap, to our knowledge.

\subsection*{Automated Degradation Detection}

\begin{figure*}[h]
\begin{center}
\includegraphics[width=1.\columnwidth]{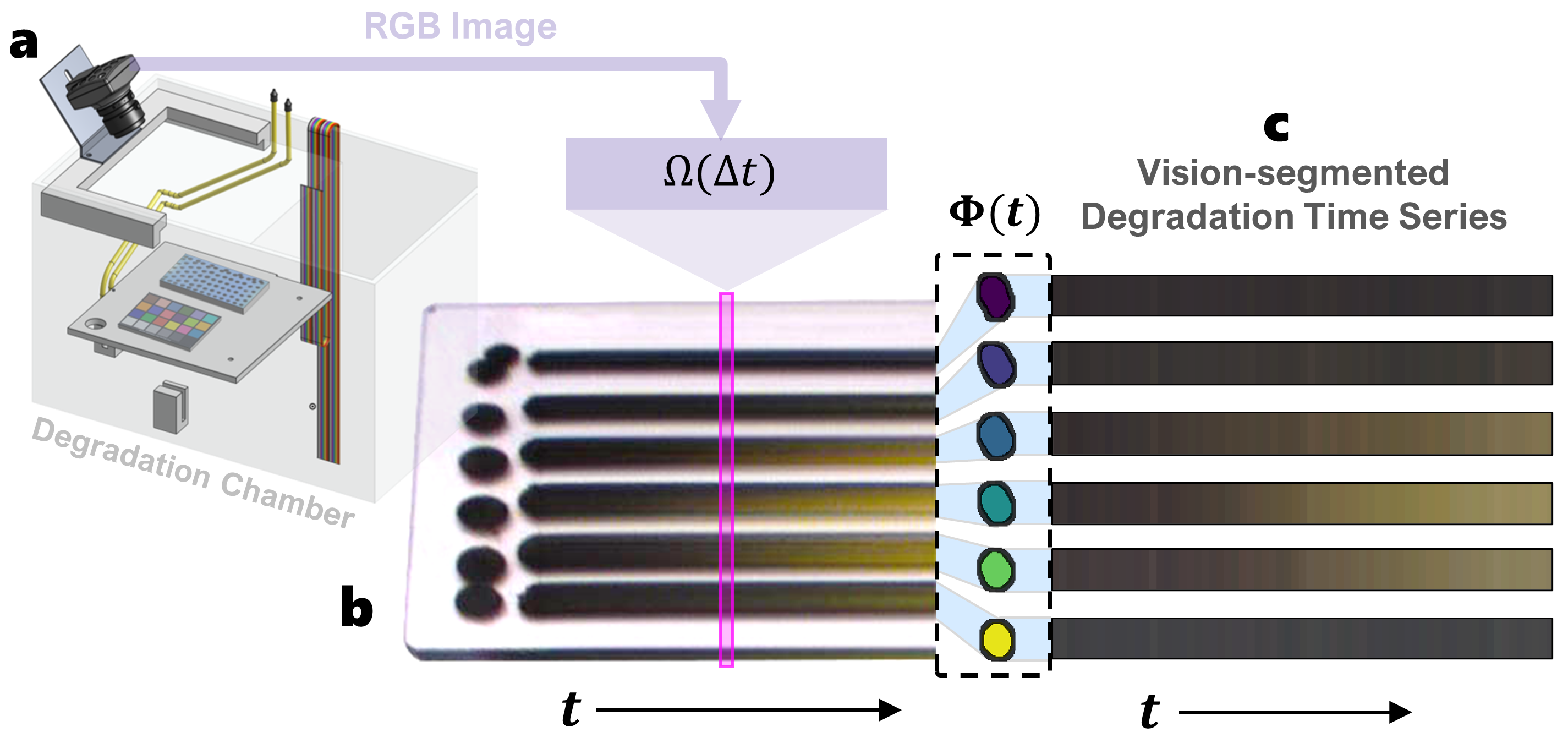}
\end{center}
   \caption{\textbf{a-c} Automatic degradation testing and measurement of computer vision-segmented perovskite deposits. \textbf{a}, The samples are placed in the degradation chamber with specified environmental conditions for a total of two hours. \textbf{b}, RGB images of the samples are taken every 30 seconds for two hours to resolve the time-dependent color change in material. \textbf{c}, Computer vision is used to segment each deposited sample over time, $\Phi(t)$, to compute the degradation intensity metric, $I_c$.}
\label{fig:degradation}
\end{figure*}

Sufficient stability of perovskite semiconductors is required for the material to be utilized in solar cell applications \cite{Wang2016, Mazumdar2021, Siegler2022, Duan2023}. As a lead halide perovskite degrades, it changes color from black to yellow, a result of a phase change and/or decomposition of the structure \cite{Sun2021,Nan2021, Wu2023}. We leverage this RGB-detectable degradation mechanism \cite{keesey_tiihonen_siemenn_colburn_sun_hartono_serdy_zeile_he_gurtner_et} and use parallelized computer vision segmentation to automate the detection of degradation within perovskites, as shown in Figure \ref{fig:degradation}c. Three independent degradation experiments are conducted across the $N=201$ samples by placing each batch of samples within a degradation chamber, shown in Figure \ref{fig:degradation}a, for 2 hours at an illumination of 0.5 suns, temperature of $34.5^\circ \mathrm{C} \pm 0.5^\circ \mathrm{C}$, and relative humidity of
$40\% \pm 1\%$ (Supplementary Figure \ref{sfig:deg}). We compute the degradation intensity, $I_c$, of each HT-manufactured perovskite composition by integrating the change in color, $R$, for each sample over time, $t$ \cite{Sun2021}:
\begin{equation}
    \label{eq:5}
    I_c(\widehat{X},\widehat{Y})= \sum_{R = \{r,g,b\}} \int_{0}^{T} |R(t; \widehat{X}, \widehat{Y}) - R(0;\widehat{X},\widehat{Y})| dt,
\end{equation}
where $T$ is the duration of the degradation and the three reflectance color channels are red, $r$, green, $g$, and blue, $b$, for each sample, $(\widehat{X},\widehat{Y})_n \in N$. High $I_c$ indicates high color change, corresponding to high degradation; $I_c$ close to zero indicates low color change and low degradation.

\begin{figure*}[h]
\begin{center}
\includegraphics[width=.83\columnwidth]{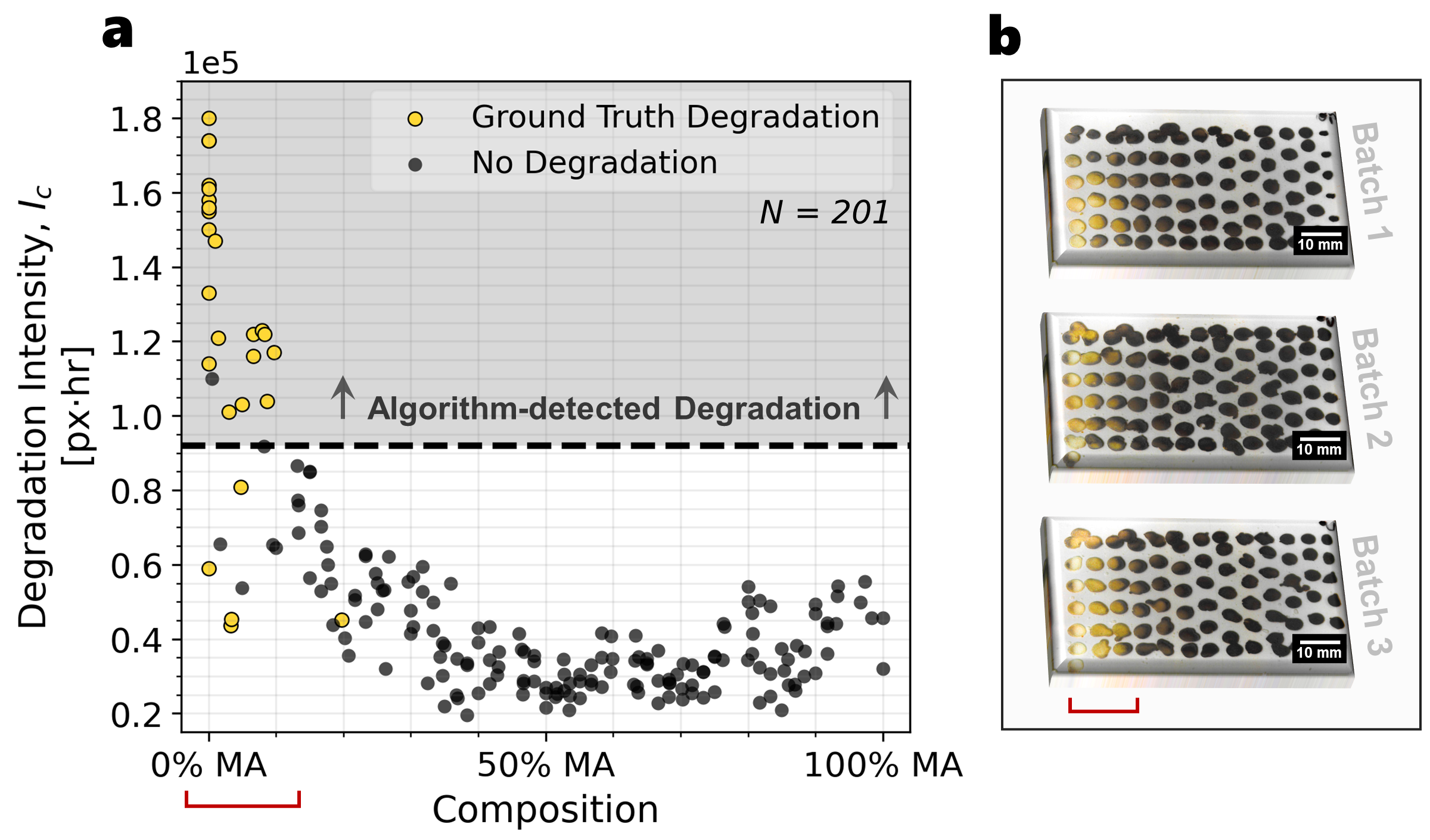}
\end{center}
   \caption{\textbf{a}, Performance of the autocharacterization of degradation intensity, $I_c$, relative to the ground truth degradation determined by a domain expert (yellow scatter points) on $N=201$ unique perovskite samples across 3 independent trials. The black dashed line indicates the split between high and low $I_c$ values, corresponding to high and low degrees of degradation, respectively. \textbf{b}, Images of the three batches of FA$_{1-x}$MA$_x$PbI$_3$ gradient samples after the 2-hour controlled degradation. The leftmost samples are FA-rich and the rightmost samples are MA-rich. The yellowed FA-rich compounds have undergone a phase transition from $\alpha$-FAPbI$_3$ to $\delta$-FAPbI$_3$ and are considered as ``ground truth" degradation samples if they exhibit a deviation of $>0.02$eV in band gap from pre- to post-degradation, evaluated by a domain expert.}
\label{fig:deg-result}
\end{figure*}

The performance of the degradation autocharacterization is demonstrated by comparing the output $I_c$ to the ground truth degradation, obtained from the pre- and post-band gap deviation \cite{Stoumpos2017, Sun2021} (Supplementary Figure \ref{sfig:deg-acc}a). Figure \ref{fig:deg-result}a illustrates the output of the autocharacterization where high computed $I_c$ values strongly correspond to the occurrence of the ground truth degradation in the samples (yellow scatter points). The determination of ground truth degradation is conducted by a human domain expert, further described in Supplementary Figure \ref{sfig:deg-acc}a. This classification performance of the autocharacterization algorithm achieves a precision-recall AUC (area under the curve) of $0.853$ (Supplementary Figure \ref{sfig:deg-acc}c), and a maximal accuracy of 96.9\%, relative to the ground truth (Supplementary Figure \ref{sfig:deg-acc}d). The yellowing pattern of the FA-rich samples is shown in Figure \ref{fig:deg-result}b as a result of the phase change from favorable cubic phase $\alpha$-FAPbI$_3$ to the non-perovskite hexagonal phase $\delta$-FAPbI$_3$ \cite{Nan2021} (Supplementary Figure \ref{sfig:xrd}). Furthermore, running a full degradation detection computation using autocharacterization takes only 20 minutes per 200 samples, given 48000 total degradation images over the 2-hour degradation experiment. This is a significant speedup from the standard microscopy or XRD methods of determining degradation, which can take hours or days to identify the degradation of an equivalent number of samples.

Using the fast and accurate stability autocharacterization tool developed in this paper, we tractably generate an ultra-high resolution stability trend for the FA$_{1-x}$MA$_x$PbI$_3$  series, shown in Figure \ref{fig:deg-result}a where, similar to band gap, this trend has not been reported at such a high resolution yet in literature. Prior literature reports stability compositional resolutions from $0\leq x \leq 1$ for 11 compositions \cite{Charles2017}, 9 compositions \cite{Weber2016}, and 7 compositions \cite{Pisanu2017} using conventional characterization methods. Moreover, Charles \textit{et al.} \cite{Charles2017} reports the stability at $x\approx0.1$ compositional increments from $0\leq x \leq 1$ using 6 timesteps, amounting to a total of 66 temporal data points. Comparatively, this study reports the stability at $x\approx0.008$ unique compositional increments from $0\leq x \leq 1$ using 240 timesteps, amounting to 28800 unique temporal data points (with 48000 total temporal data points). Thus, with autocharacterization, we achieve over a 10x increase in the compositional resolution and a 40x increase in the temporal resolution for a total of a 436x increase in the number of unique data points reported for the FA$_{1-x}$MA$_x$PbI$_3$ stability series, to our knowledge. Furthermore, with this high-resolution stability trend, we note the same regions of high-degradation appear in Figure \ref{fig:deg-result}a as do in the literature for the $\alpha$-FAPbI$_3 \rightarrow \delta$-FAPbI$_3$ degradation pathway at $0.0\leq x \leq 0.15$, with the optimal low-degradation region occurring at $x\approx0.40$ \cite{Binek2015,  Charles2017}. Through the generation of ultra-high resolution trends, we may achieve a better understanding of complex semiconductor composition-property relationships to enable higher-performance design of materials in the future.

\section*{Conclusion}
% Computer vision is used to segment batches of up to 80 unique high-throughput-deposited perovskite semiconductor samples at once to map them to their measured optical spectra.

Accelerating the characterization of key material properties relevant to semiconductor engineering, such as band gap and stability, is a necessary step to enable the high-throughput discovery and optimization of perovskite materials. Conventional methods of characterization bottleneck the materials screening pipeline when high-throughput synthesis is utilized, inhibiting optimally efficient high-throughput experimentation. For example, computing the band gap of 200 unique halide perovskite samples takes a domain expert over 8 hours to complete. In this work, we demonstrate the fast and accurate characterization of band gap and detection of degradation within the FA$_{1-x}$MA$_x$PbI$_3$, $0\leq x \leq 1$ perovskite system using parallelized, scalable computer vision segmentation. From the segmented data, the band gaps and degrees of degradation are automatically computed using the developed autocharacterization algorithms, and ultra-high compositional resolution trends are generated. The band gaps of 200 unique perovskite samples are determined in 6 minutes at 98.5\% accuracy within a 0.02eV range using the band gap autocharacterization tool. The degrees of degradation of 200 unique perovskite samples are determined within 20 minutes at 96.9\% accuracy using the degradation autocharacterization tool. Overall, the developed autocharacterization methods achieve 85x faster throughputs than conventional domain expert evaluation, in turn, contributing to the debottlenecking of high-throughput and autonomous materials discovery and optimization workflows. Therefore, through the wider application of autocharacterization methods, the ability to scan through larger material search spaces is unlocked, in turn, enabling the tractable design of higher-performance functional materials

\section*{Methods}

\subsection*{Materials}
3"$\times$2"$\times$1mm glass slides (C\&A Scientific) are cleaned using deionized water (DI, $<1.0\mu$S/cm, VWR), Hellmanex III (VWR), and isopropyl alcohol (IPA, $\geq99.5$\%, VWR) to be used as substrates. Lead iodide powder (PbI$_2$, 99.999\% trace metal basis, Sigma-Aldrich), formamidinium iodide powder (FAI, $>$99.9\%, Greatcell Solar Materials), methylammonium iodide (MAI, $>$99.9\%, Greatcell Solar Materials), dimethylformamide (DMF, $\geq$99.8\%, Sigma-Aldrich), and dimethylsulfoxide (DMSO, $\geq$99.9\%, Sigma-Aldrich) are used to prepare the perovskites.

\subsection*{Computer Vision Segmentation of Hyperspectral Datacubes}

A hyperspectral datacube of size $X\times Y \times R(\lambda) \rightarrow$ 900px$\times$800px$\times$300, where 300 is the number of wavelengths, $\lambda$, is captured as a raw image, $\Omega=(X,Y,R(\lambda))$, from the hyperspectral camera (Resonon, Pika L). This datacube is passed through several filters to find the edges and segment each material deposit sample ($\widehat{X}$, $\widehat{Y}$)$_n \in N$ and then index the features appropriately, such that each pixel is mapped to its reflectance spectra, $R(\lambda)$. Once segmented, each deposited material contains an area of approximately $1000$px worth of spatial spectral data. Inputting images or features of different sizes may require tuning the kernel sizes, $\kappa$, of the filters.

Algorithm \ref{alg:segmentation} describes this segmentation process of $(X,Y,R(\lambda)) \rightarrow (\widehat{X},\widehat{Y},R(\lambda))$. First, the input image is cropped and converted to greyscale, then it is passed through several layers of thresholding and smoothing. By thresholding, eroding, blurring, and then thresholding again, we capture the edges of each feature while removing edge effects. The background is indexed as zeros, hence, all features split by zeros are assigned a unique index using island-finding graph network methods from the \texttt{OpenCV} Python library \cite{opencv_library}: \texttt{LabelFeatures}$(\cdot)$ and \texttt{Watershed}$(\cdot)$ \cite{batagelj2003efficient, Meyer1994}. Once segmented, the features are smoothed, and any improperly segmented aberrations are pruned with the user-selected variables $\Theta_\textrm{min}$, $\Theta_\textrm{max}$, where features of size $s<\Theta_\textrm{min}$ or $s>\Theta_\textrm{max}$ are removed. Finally, a boolean mask is created for all pixels encoded with non-zero values to output $\Phi$, where each uniquely indexed material deposit is directly mappable to the $R(\lambda)$ measured for that deposit.

\IncMargin{2em}
\begin{algorithm}[h!]

\DontPrintSemicolon
   \caption{Scalable Computer Vision Segmentation}
   \label{alg:segmentation}
  \SetKwData{Left}{left}
  \SetKwData{Up}{up}
  \SetKwFunction{FindCompress}{FindCompress}
  \SetKwInOut{Input}{Input}
  \SetKwInOut{Output}{Output}
  \SetKwRepeat{Repeat}{repeat}{end}

\small
\Indm\Indm
    \BlankLine
  \Input{$(X,Y,R(\lambda))$: All image pixels and reflectance\\
    $\Theta_\textrm{min}$: Minimum segmented feature size\\
    $\Theta_\textrm{max}$: Maximum segmented feature size}
  \BlankLine
  \Output{$(\widehat{X},\widehat{Y},R(\lambda))$: Segmented pixels and reflectance}
\Indp\Indp
  \BlankLine
    \nonl \textup{Let} $\kappa$ \textup{be kernel size}\\
    \nl img \: $\leftarrow (X,Y,R(\lambda))$\\
    \nl img \: $\leftarrow$ \texttt{Crop}(\textrm{img})\\
    \nl img \: $\leftarrow$ \texttt{Grayscale}(\textrm{img})\\
    \nl img \: $\leftarrow$ 255 - \textrm{img}\\
    \nl mask $\leftarrow$ \texttt{Binarize}(\textrm{img})\\
    \nl mask $\leftarrow$ \texttt{MorphGradient}(\textrm{mask}, $\kappa=12$)\\
    \nl mask $\leftarrow$ \texttt{Erode}(\textrm{mask}, $\kappa=3$)\\
    \nl mask $\leftarrow$ \texttt{MedianBlur}(\textrm{mask}, $\kappa=7$)\\
    \nl mask $\leftarrow$ \texttt{DistTransform}(\textrm{mask}, $\kappa=3$)\\
    \nl mask $\leftarrow$ \texttt{LabelFeatures}(\textrm{mask})\\
    \nl $(\widehat{X},\widehat{Y})$ \: $\leftarrow$ \texttt{Watershed}(\textrm{img, mask})\\
    \nl $(\widehat{X},\widehat{Y})$ \: $\leftarrow$ \texttt{Dilate}($(\widehat{X},\widehat{Y})$, $\kappa=5$)\\
    \nl $(\widehat{X},\widehat{Y})$ \: $\leftarrow$ \texttt{MedianBlur}($(\widehat{X},\widehat{Y})$, $\kappa=7$)\\
    \nl \For{$\phi$ \textup{in} \texttt{Feature}($(\widehat{X},\widehat{Y})$)}{
        \If{\texttt{Size}$(\phi) < \Theta_\mathrm{min}$ \textbf{\textup{or}} \texttt{Size}$(\phi) > \Theta_\mathrm{max}$}{
        $(\widehat{X},\widehat{Y})$\texttt{.Prune}$(\phi)$
        }}
    bool $\leftarrow$ \texttt{Boolean}($(\widehat{X},\widehat{Y})$)\\
    \nl $(\widehat{X},\widehat{Y},R(\lambda)) \leftarrow \left(X,Y,R(\lambda)\right)$\texttt{.Mask}$ \left(\textrm{bool}\right)$\\
    \Indm\Indm
    \nonl \textbf{Return} $(\widehat{X},\widehat{Y},R(\lambda))$
\end{algorithm}

\subsection*{Perovskite Elemental Composition Mapping onto Computer Vision Segmented Data}

The composition for each segmented deposit, $(\widehat{X},\widehat{Y})_n \in N$ is determined by mapping the encoded pump speeds of the high-throughput combinatorial printer to the time step, $\Delta t = [t_a,t_b]$, at which each deposit is printed. This is done by mapping the segmented image $\Phi$ to the printer head raster path, acquired from the G-code controlling the printer motion. Hence, each deposited material now has its pixel coordinates $(\widehat{X},\widehat{Y})$ mapped to a timestamp along the raster path. Then, this positional timestamp is mapped to its corresponding pump speed timestamp, $t(\widehat{X},\widehat{Y}) \rightarrow t(\omega_\textrm{FA}, \omega_\textrm{MA})$. Since these variable pump speeds $\omega_\textrm{FA}$ and $\omega_\textrm{MA}$ are both monotonic along the time series, they are deterministic of the proportion of FA and MA within the FA$_{1-x}$MA$_{x}$PbI$_3$ compositional structure by integrating pump speed over the $\Delta t$ for each material deposit \textit{via} Equation \ref{eq:composition}.

\subsection*{Automating Band Gap Computation using Hyperspectral Reflectance Data}

The direct band gap for all $N=201$ FA$_{1-x}$MA$_{x}$PbI$_3$ perovskite compositions is computed from the vision-segmented reflectance data, $\Phi = (\widehat{X},\widehat{Y},R(\lambda))$, since all the FA$_{1-x}$MA$_{x}$PbI$_3$ compositions are direct band gap materials at atmospheric pressure \cite{Wang2017b, Targhi2018}. For every segmented sample, $(\widehat{X},\widehat{Y})_n \in N$, the spatial median $R(\lambda)$ spectra is used for computing the band gap. $R(\lambda)$ spans across wavelengths, $\lambda$, where $\{\lambda \in \mathbb{Z}:380\textrm{nm} \leq \lambda \leq 1020\textrm{nm}\}$ for hyperspectral imaging and $\lambda = \{r,g,b\}$ for the red, green, and blue color channels of RGB imaging.  First, the median $R(\lambda)$ spectra are transformed into a Tauc curve using Equation \ref{eq:bg}, with $\gamma=\frac{1}{2}$ for direct band gap. Then, transformed Tauc curves are recursively segmented in half until each segment achieves a fit of $R^2\geq0.990$, indicating that each segment is near-linear:
\begin{equation}
    R^2=1-\frac{\sum_{i}^N(y_i-\widehat{y}_i)^2}{\sum_{i}(y_i-\overline{y})^2},
\end{equation}
where $\widehat{y}$ is the predicted value and $\overline{y}$ is the average value of the set. Once the recursion is complete, each pair of adjacent line segments is iteratively fit to a linear regression line, generating a set of candidate fit lines to use for computing band gap. To determine the best candidate fitted line, RMSE is used rather than using the inclination angles of Tauc curves \cite{escobedo2019} to improve generalizability across different materials, \textit{e.g.} FAPbI$_3$ and MAPbI$_3$:
\begin{equation}
    \textrm{RMSE}=\sqrt{\frac{\sum_i^N(y_i-\widehat{y}_i)^2}{N}}.
\end{equation}
We implement an iterative root-mean-square error (RMSE) minimization routine that automatically identifies the Tauc curve peaks to fit between. Then, the RMSE is computed between each regression line and the Tauc curve within the lower bound of the regression $x$-intercept and the upper bound of the Tauc peak location minus one-half of the peak width. Enforcing the RMSE computation to occur within these bounds was shown to increase fitting accuracy with the Tauc slope. The band gap, $E_g$, is then extracted from $x$-intercept point of the regression line that achieves the minimum RMSE. 

\subsection*{Detecting Perovskite Degradation from RGB Time Series Data}

In order to use color as a reproducible and repeatable quantitative proxy for degradation, color calibration needs to be applied because the illumination conditions in the aging test chamber may create distortions to the true sample color. At the beginning of the degradation study, an image of a reference color chart (X-Rite Colour Checker Passport; 28 reference color patches), $I_R$, is taken under the same illumination conditions as the perovskite semiconductor samples. Images at each time step, $\Omega (\Delta t)$, are transformed into CIELAB colorspace and subsequently into a stable reference color space, CIE 1931 color space with a 2-degree standard observer and standard illuminant D50, by applying a 3D-thin plate spline distortion matrix $D$ \cite{Sun2021,s120607063} defined by $I_R$ and known colors of the reference color chart:

\begin{equation}
    \label{eq:4}
    D= \begin{bmatrix}
    V\\
    O(4,3)
    \end{bmatrix}{\begin{bmatrix}
    K & P \\
    P^T & O(4,4)
    \end{bmatrix}}^{-1}
\end{equation}
% :)

 Here, $O(n,m)$ is an $n \times m$ zero matrix, $V$ is a matrix of the color checker reference colors in the stable reference color space, $P$ is a matrix of the color checker RGB colors obtained from $I_R$, and $K$ is a distortion matrix between the color checker colors in the reference space and in $I_R$. Using the color-calibrated images and droplet pixel locations given by $\Phi$, a final array, $R(t; \widehat{X}, \widehat{Y})$ of the average color at time $t$ for perovskite semiconductor of composition FA$_{1-x}$MA$_x$PbI$_3$ is created. The color of each droplet is measured to determine the stability metric $I_c$ \cite{Sun2021}, calculated using Equation \ref{eq:5}.

\subsection*{Experimental Section}

\subsubsection*{Substrate Preparation}

Glass slide substrates are prepared for printing the perovskite samples using a three-step cleaning process: (1) ultrasonication for 5 minutes in DI water with 2\%vol. Hellmanex III solution, (2) ultrasonication for 5 minutes in DI water only, and (3) ultrasonication for 5 minutes in IPA. Once cleaned, the substrates are transferred to an inert nitrogen environment glovebox with moisture levels $<10$ppm.

\subsubsection*{Perovskite Preparation}

FAPbI$_3$ (formamidinium lead iodide) and MAPbI$_3$ (methylammonium lead iodide) are prepared as 0.6M liquid-based precursors for high-throughput printing. For printing, 2mL of each precursor is prepared in an inert nitrogen environment glovebox with moisture levels $<10$ppm. First, 3.2mL DMF is mixed with 0.8mL of DMSO to make 4mL of $4:1$ DMF:DMSO solution. Then, $1.106$g of PbI$_2$ powder is dissolved into the 4mL of $4:1$ DMF:DMSO to make a PbI$_2$ stock. Next, the 4mL PbI$_2$ stock is split in half, pipetting 2mL of stock per vial. Lastly, $0.206$g of FAI powder is dissolved into one of the 2mL PbI$_2$ stock vials and $0.191$g of MAI powder is dissolved into the other making 0.6M FAPbI$_3$ and 0.6M MAPbI$_3$, respectively.

\subsubsection*{High-throughput Perovskite Synthesis}

The liquid-based FAPbI$_3$ and MAPbI$_3$ precursor solutions are used in the high-throughput combinatorial printer to synthesize $N=201$ unique FA$_{1-x}$MA$_x$PbI$_3$ composition samples. The high-throughput printer is custom-made, and parts of its construction are documented in Siemenn \textit{et al.} \cite{Siemenn2022}. To begin printing, first, all printer plumbing lines are flushed twice with the $4:1$ ratio DMF:DMSO solution. Then, the FAPbI$_3$ and MAPbI$_3$ precursors are extracted into syringes using a microcontroller to communicate with the pumps. These syringes contain prismatic motion plungers that use positive displacement to fill and eject solution. Next, all plumbing lines are primed with the precursor solution. After priming, the precursors are purged at equal rates from both syringes for 50 seconds to remove air bubbles. Finally, motor encoders pump the precursors out of the syringes at pre-programmed rates, illustrated in Supplementary Figure \ref{sfig:motor-traces}, and enter a mixing chamber prior to deposition. A pinch valve breaks up the fluid flow within a 1/32"ID $\times$ 3/32"OD silicone tube, actuating at 11Hz frequency and 5\% duty cycle to deposit each sample as a discrete droplet onto the cleaned glass slide. The print head translates at a speed of 38mm/s over the 3"$\times$2" glass slide in a serpentine pattern, depositing approximately 70-80 unique composition samples per batch in $16.5$ seconds. Approximately 0.15mL of total precursor volume is consumed per print, which includes the volume required for purging and priming the plumbing lines. After the droplets have been deposited onto the substrate, the substrate is transferred to a hotplate to anneal for 15 minutes at $150^\circ$C.

\section*{Data Availability}
The raw experimental data that support the findings and plots of this study are available publicly. The band gap data (\href{https://osf.io/qe9ax}{https://osf.io/qe9ax}) and the degradation data (\href{https://osf.io/y2xds}{https://osf.io/y2xds}) are available from OSF.

\section*{Code Availability}
All code used to develop the autocharacterization and computer vision algorithms are available publicly with complete working examples. The band gap autocharacterization code (\href{https://github.com/PV-Lab/Autocharacterization-Bandgap}{https://github.com/PV-Lab/Autocharacterization-Bandgap}) and the degradation detection autocharacterization code (\href{https://github.com/PV-Lab/Autocharacterization-Stability}{https://github.com/PV-Lab/Autocharacterization-Stability}) are available from GitHub.

\pagebreak[1]
\clearpage
\newpage

\section*{Acknowledgements}

We thank First Solar for support and fruitful discussions. The automation portion of this work was undertaken thanks in part to funding provided to the University of Toronto's Acceleration Consortium. This material is partially based upon work supported by the U.S. Department of Energy’s Office of Energy Efficiency and Renewable Energy (EERE) under the Advanced Manufacturing Office (AMO) Award Number DE-EE0009096. B.D. acknowledges the support of the Simons Foundation. A.T. was supported by the Research Council of Finland Flagship programme: Finnish Center for Artificial Intelligence FCAI. H.K. received financial support from the Scientific and Technological Research Council of Turkey (TÜBİTAK) BİDEB-2219 (Project No: 1059B192100703) during this study. This work made use of the MRSEC Shared Experimental Facilities at MIT, supported by the National Science Foundation under award number DMR-1419807.

\section*{Author Contributions}

A.E.S. and E.A. conceptualized the work. A.E.S., E.A., A.T., B.D., and T.B. designed the methodology. A.E.S., E.A., and A.T. wrote the software. F.S. and H.K. prepared the experimental materials. A.E.S., E.A., F.S., and B.D. conducted experiments. A.E.S. and E.A. performed the analysis. H.K. performed expert human benchmarking. A.E.S. and E.A. wrote the manuscript. All authors reviewed and edited the manuscript. A.T., B.D., and T.B. provided guidance.

{\small
\bibliography{references}

\begin{thebibliography}{10}
\expandafter\ifx\csname url\endcsname\relax
  \def\url#1{\burl{#1}}\fi
\expandafter\ifx\csname urlprefix\endcsname\relax\def\urlprefix{URL }\fi
\providecommand{\bibinfo}[2]{#2}
\providecommand{\eprint}[2][]{\url{#2}}
\providecommand{\doi}[1]{\url{https://doi.org/#1}}
\bibcommenthead

\bibitem{Wang2017b}
\bibinfo{author}{Wang, T.} \emph{et~al.}
\newblock \bibinfo{title}{Indirect to direct bandgap transition in
  methylammonium lead halide perovskite}.
\newblock \emph{\bibinfo{journal}{Energy Environ. Sci.}}
  \textbf{\bibinfo{volume}{10}}, \bibinfo{pages}{509--515}
  (\bibinfo{year}{2017}).

\bibitem{Targhi2018}
\bibinfo{author}{Targhi, F.~F.}, \bibinfo{author}{Jalili, Y.~S.} \&
  \bibinfo{author}{Kanjouri, F.}
\newblock \bibinfo{title}{Mapbi3 and fapbi3 perovskites as solar cells: Case
  study on structural, electrical and optical properties}.
\newblock \emph{\bibinfo{journal}{Results Phys.}}
  \textbf{\bibinfo{volume}{10}}, \bibinfo{pages}{616--627}
  (\bibinfo{year}{2018}).

\bibitem{Kubelka1931}
\bibinfo{author}{Kubelka, P.} \& \bibinfo{author}{Munk, F.}
\newblock \bibinfo{title}{A contribution to the optics of pigments}.
\newblock \emph{\bibinfo{journal}{Z. Technol. Phys.}}
  \textbf{\bibinfo{volume}{12}}, \bibinfo{pages}{593--599}
  (\bibinfo{year}{1931}).

\bibitem{Makula2018}
\bibinfo{author}{Makuła, P.}, \bibinfo{author}{Pacia, M.} \&
  \bibinfo{author}{Macyk, W.}
\newblock \bibinfo{title}{How to correctly determine the band gap energy of
  modified semiconductor photocatalysts based on uv-vis spectra}.
\newblock \emph{\bibinfo{journal}{J. Phys. Chem. Lett.}}
  \textbf{\bibinfo{volume}{9}}, \bibinfo{pages}{6814--6817}
  (\bibinfo{year}{2018}).

\bibitem{Tauc1966}
\bibinfo{author}{Tauc, J.}, \bibinfo{author}{Grigorovici, R.} \&
  \bibinfo{author}{Vancu, A.}
\newblock \bibinfo{title}{Optical properties and electronic structure of
  amorphous germanium}.
\newblock \emph{\bibinfo{journal}{Phys. Status Solidi B}}
  \textbf{\bibinfo{volume}{15}}, \bibinfo{pages}{627--637}
  (\bibinfo{year}{1966}).

\bibitem{keesey_tiihonen_siemenn_colburn_sun_hartono_serdy_zeile_he_gurtner_et}
\bibinfo{author}{Keesey, R.} \emph{et~al.}
\newblock \bibinfo{title}{An open-source environmental chamber for
  materials-stability testing using an optical proxy}.
\newblock \emph{\bibinfo{journal}{Digit. Discov.}}  (\bibinfo{year}{2023}).

\bibitem{Sun2021}
\bibinfo{author}{Sun, S.} \emph{et~al.}
\newblock \bibinfo{title}{A data fusion approach to optimize compositional
  stability of halide perovskites}.
\newblock \emph{\bibinfo{journal}{Matter}} \textbf{\bibinfo{volume}{4}},
  \bibinfo{pages}{1305–1322} (\bibinfo{year}{2021}).

\bibitem{Stoumpos2017}
\bibinfo{author}{Stoumpos, C.~C.}, \bibinfo{author}{Mao, L.},
  \bibinfo{author}{Malliakas, C.~D.} \& \bibinfo{author}{Kanatzidis, M.~G.}
\newblock \bibinfo{title}{Structure-band gap relationships in hexagonal
  polytypes and low-dimensional structures of hybrid tin iodide perovskites}.
\newblock \emph{\bibinfo{journal}{Inorg. Chem.}} \textbf{\bibinfo{volume}{56}},
  \bibinfo{pages}{56--73} (\bibinfo{year}{2017}).

\bibitem{Nan2021}
\bibinfo{author}{Nan, Z.~A.} \emph{et~al.}
\newblock \bibinfo{title}{Revealing phase evolution mechanism for stabilizing
  formamidinium-based lead halide perovskites by a key intermediate phase}.
\newblock \emph{\bibinfo{journal}{Chem}} \textbf{\bibinfo{volume}{7}},
  \bibinfo{pages}{2513--2526} (\bibinfo{year}{2021}).

\bibitem{Wu2023}
\bibinfo{author}{Wu, J.}, \bibinfo{author}{Chen, J.} \& \bibinfo{author}{Wang,
  H.}
\newblock \bibinfo{title}{Phase transition kinetics of mapbi3 for
  tetragonal-to-orthorhombic evolution}.
\newblock \emph{\bibinfo{journal}{JACS Au}} \textbf{\bibinfo{volume}{3}},
  \bibinfo{pages}{1205--1212} (\bibinfo{year}{2023}).

\bibitem{Elsayed2023}
\bibinfo{author}{Elsayed, M.~R.}, \bibinfo{author}{Elseman, A.~M.},
  \bibinfo{author}{Abdelmageed, A.~A.}, \bibinfo{author}{Hashem, H.~M.} \&
  \bibinfo{author}{Hassen, A.}
\newblock \bibinfo{title}{Synthesis and numerical simulation of
  formamidinium-based perovskite solar cells: a predictable device performance
  at nis-egypt}.
\newblock \emph{\bibinfo{journal}{Sci. Reports}} \textbf{\bibinfo{volume}{13}},
  \bibinfo{pages}{1--16} (\bibinfo{year}{2023}).

\bibitem{Wang2023}
\bibinfo{author}{Wang, T.} \emph{et~al.}
\newblock \bibinfo{title}{Sustainable materials acceleration platform reveals
  stable and efficient wide-bandgap metal halide perovskite alloys}.
\newblock \emph{\bibinfo{journal}{Matter}} \textbf{\bibinfo{volume}{6}},
  \bibinfo{pages}{2963--2986} (\bibinfo{year}{2023}).

\bibitem{Ahmadi2021}
\bibinfo{author}{Ahmadi, M.}, \bibinfo{author}{Ziatdinov, M.},
  \bibinfo{author}{Zhou, Y.}, \bibinfo{author}{Lass, E.~A.} \&
  \bibinfo{author}{Kalinin, S.~V.}
\newblock \bibinfo{title}{Machine learning for high-throughput experimental
  exploration of metal halide perovskites}.
\newblock \emph{\bibinfo{journal}{Joule}} \textbf{\bibinfo{volume}{5}},
  \bibinfo{pages}{2797--2822} (\bibinfo{year}{2021}).

\bibitem{Wang2017}
\bibinfo{author}{Wang, Z.} \emph{et~al.}
\newblock \bibinfo{title}{Efficient ambient-air-stable solar cells with 2d–3d
  heterostructured butylammonium-caesium- formamidinium lead halide
  perovskites}.
\newblock \emph{\bibinfo{journal}{Nat. Energy}} \textbf{\bibinfo{volume}{2}},
  \bibinfo{pages}{1--10} (\bibinfo{year}{2017}).

\bibitem{Sun2019}
\bibinfo{author}{Sun, S.} \emph{et~al.}
\newblock \bibinfo{title}{Accelerated development of perovskite-inspired
  materials via high-throughput synthesis and machine-learning diagnosis}.
\newblock \emph{\bibinfo{journal}{Joule}} \textbf{\bibinfo{volume}{3}},
  \bibinfo{pages}{1437--1451} (\bibinfo{year}{2019}).

\bibitem{Liu2023}
\bibinfo{author}{Liu, A.} \emph{et~al.}
\newblock \bibinfo{title}{High-performance metal halide perovskite
  transistors}.
\newblock \emph{\bibinfo{journal}{Nat. Electron.}} \bibinfo{pages}{1--13}
  (\bibinfo{year}{2023}).

\end{thebibliography}


\begin{thebibliography}{10}
\expandafter\ifx\csname url\endcsname\relax
  \def\url#1{\burl{#1}}\fi
\expandafter\ifx\csname urlprefix\endcsname\relax\def\urlprefix{URL }\fi
\providecommand{\bibinfo}[2]{#2}
\providecommand{\eprint}[2][]{\url{#2}}
\providecommand{\doi}[1]{\url{https://doi.org/#1}}
\bibcommenthead

\bibitem{Mazumdar2021}
\bibinfo{author}{Mazumdar, S.}, \bibinfo{author}{Zhao, Y.} \&
  \bibinfo{author}{Zhang, X.}
\newblock \bibinfo{title}{Stability of perovskite solar cells: Degradation
  mechanisms and remedies}.
\newblock \emph{\bibinfo{journal}{Front. Electron.}}
  \textbf{\bibinfo{volume}{2}}, \bibinfo{pages}{712785} (\bibinfo{year}{2021}).

\bibitem{Siegler2022}
\bibinfo{author}{Siegler, T.~D.} \emph{et~al.}
\newblock \bibinfo{title}{The path to perovskite commercialization: A
  perspective from the united states solar energy technologies office}.
\newblock \emph{\bibinfo{journal}{ACS Energy Lett.}}
  \textbf{\bibinfo{volume}{7}}, \bibinfo{pages}{1728--1734}.

\bibitem{Duan2023}
\bibinfo{author}{Duan, L.} \emph{et~al.}
\newblock \bibinfo{title}{Stability challenges for the commercialization of
  perovskite–silicon tandem solar cells}.
\newblock \emph{\bibinfo{journal}{Nat. Rev. Mater.}}
  \textbf{\bibinfo{volume}{8}}, \bibinfo{pages}{261--281}
  (\bibinfo{year}{2023}).

\bibitem{Hu2019}
\bibinfo{author}{Hu, Z.} \emph{et~al.}
\newblock \bibinfo{title}{A review on energy band-gap engineering for
  perovskite photovoltaics}.
\newblock \emph{\bibinfo{journal}{Sol. RRL}} \textbf{\bibinfo{volume}{3}},
  \bibinfo{pages}{1900304} (\bibinfo{year}{2019}).

\bibitem{Prasanna2017}
\bibinfo{author}{Prasanna, R.} \emph{et~al.}
\newblock \bibinfo{title}{Band gap tuning via lattice contraction and
  octahedral tilting in perovskite materials for photovoltaics}.
\newblock \emph{\bibinfo{journal}{J. Am. Chem. Soc.}}
  \textbf{\bibinfo{volume}{139}}, \bibinfo{pages}{11117--11124}
  (\bibinfo{year}{2017}).

\bibitem{Baloch2022}
\bibinfo{author}{Baloch, A.~A.}, \bibinfo{author}{Albadwawi, O.},
  \bibinfo{author}{AlShehhi, B.} \& \bibinfo{author}{Alberts, V.}
\newblock \bibinfo{title}{Impact of mixed perovskite composition based silicon
  tandem pv devices on efficiency limits and global performance}.
\newblock \emph{\bibinfo{journal}{Energy Reports}}
  \textbf{\bibinfo{volume}{8}}, \bibinfo{pages}{504--510}
  (\bibinfo{year}{2022}).

\bibitem{Sun2021}
\bibinfo{author}{Sun, S.} \emph{et~al.}
\newblock \bibinfo{title}{A data fusion approach to optimize compositional
  stability of halide perovskites}.
\newblock \emph{\bibinfo{journal}{Matter}} \textbf{\bibinfo{volume}{4}},
  \bibinfo{pages}{1305–1322} (\bibinfo{year}{2021}).

\bibitem{keesey_tiihonen_siemenn_colburn_sun_hartono_serdy_zeile_he_gurtner_et}
\bibinfo{author}{Keesey, R.} \emph{et~al.}
\newblock \bibinfo{title}{An open-source environmental chamber for
  materials-stability testing using an optical proxy}.
\newblock \emph{\bibinfo{journal}{Digit. Discov.}}  (\bibinfo{year}{2023}).

\bibitem{Wang2023}
\bibinfo{author}{Wang, T.} \emph{et~al.}
\newblock \bibinfo{title}{Sustainable materials acceleration platform reveals
  stable and efficient wide-bandgap metal halide perovskite alloys}.
\newblock \emph{\bibinfo{journal}{Matter}} \textbf{\bibinfo{volume}{6}},
  \bibinfo{pages}{2963--2986} (\bibinfo{year}{2023}).

\bibitem{Langner2020}
\bibinfo{author}{Langner, S.} \emph{et~al.}
\newblock \bibinfo{title}{Beyond ternary opv: High‐throughput experimentation
  and self‐driving laboratories optimize multicomponent systems}.
\newblock \emph{\bibinfo{journal}{Adv. Mater.}} \textbf{\bibinfo{volume}{32}},
  \bibinfo{pages}{1907801} (\bibinfo{year}{2020}).

\bibitem{Ludwig2019}
\bibinfo{author}{Ludwig, A.}
\newblock \bibinfo{title}{Discovery of new materials using combinatorial
  synthesis and high-throughput characterization of thin-film materials
  libraries combined with computational methods}.
\newblock \emph{\bibinfo{journal}{npj Comput. Mater.}}
  \textbf{\bibinfo{volume}{5}}, \bibinfo{pages}{1--7} (\bibinfo{year}{2019}).

\bibitem{MacLeod2022}
\bibinfo{author}{MacLeod, B.~P.} \emph{et~al.}
\newblock \bibinfo{title}{A self-driving laboratory advances the pareto front
  for material properties}.
\newblock \emph{\bibinfo{journal}{Nat. Commun.}} \textbf{\bibinfo{volume}{13}},
  \bibinfo{pages}{1--10} (\bibinfo{year}{2022}).

\bibitem{Moradi2022}
\bibinfo{author}{Moradi, S.} \emph{et~al.}
\newblock \bibinfo{title}{High-throughput exploration of halide perovskite
  compositionally-graded films and degradation mechanisms}.
\newblock \emph{\bibinfo{journal}{Commun. Mater.}}
  \textbf{\bibinfo{volume}{3}}, \bibinfo{pages}{1--5} (\bibinfo{year}{2022}).

\bibitem{Sun2019}
\bibinfo{author}{Sun, S.} \emph{et~al.}
\newblock \bibinfo{title}{Accelerated development of perovskite-inspired
  materials via high-throughput synthesis and machine-learning diagnosis}.
\newblock \emph{\bibinfo{journal}{Joule}} \textbf{\bibinfo{volume}{3}},
  \bibinfo{pages}{1437--1451} (\bibinfo{year}{2019}).

\bibitem{Yao2020}
\bibinfo{author}{Yao, Y.} \emph{et~al.}
\newblock \bibinfo{title}{High-throughput, combinatorial synthesis of
  multimetallic nanoclusters}.
\newblock \emph{\bibinfo{journal}{Proc. National Acad. Sci. United States
  America}} \textbf{\bibinfo{volume}{117}}, \bibinfo{pages}{6316--6322}
  (\bibinfo{year}{2020}).

\bibitem{Clayson2020}
\bibinfo{author}{Clayson, I.~G.} \emph{et~al.}
\newblock \bibinfo{title}{High throughput methods in the synthesis,
  characterization, and optimization of porous materials}.
\newblock \emph{\bibinfo{journal}{Adv. Mater.}} \textbf{\bibinfo{volume}{32}},
  \bibinfo{pages}{2002780} (\bibinfo{year}{2020}).

\bibitem{Zeng2023}
\bibinfo{author}{Zeng, M.} \emph{et~al.}
\newblock \bibinfo{title}{High-throughput printing of combinatorial materials
  from aerosols}.
\newblock \emph{\bibinfo{journal}{Nature}} \textbf{\bibinfo{volume}{617}},
  \bibinfo{pages}{292--298} (\bibinfo{year}{2023}).

\bibitem{Liu2017}
\bibinfo{author}{Liu, P.} \emph{et~al.}
\newblock \bibinfo{title}{High throughput materials research and development
  for lithium ion batteries}.
\newblock \emph{\bibinfo{journal}{J. Materiomics}}
  \textbf{\bibinfo{volume}{3}}, \bibinfo{pages}{202--208}
  (\bibinfo{year}{2017}).

\bibitem{Makula2018}
\bibinfo{author}{Makuła, P.}, \bibinfo{author}{Pacia, M.} \&
  \bibinfo{author}{Macyk, W.}
\newblock \bibinfo{title}{How to correctly determine the band gap energy of
  modified semiconductor photocatalysts based on uv-vis spectra}.
\newblock \emph{\bibinfo{journal}{J. Phys. Chem. Lett.}}
  \textbf{\bibinfo{volume}{9}}, \bibinfo{pages}{6814--6817}
  (\bibinfo{year}{2018}).

\bibitem{Du2021}
\bibinfo{author}{Du, X.} \emph{et~al.}
\newblock \bibinfo{title}{Elucidating the full potential of opv materials
  utilizing a high-throughput robot-based platform and machine learning}.
\newblock \emph{\bibinfo{journal}{Joule}} \textbf{\bibinfo{volume}{5}},
  \bibinfo{pages}{495--506} (\bibinfo{year}{2021}).

\bibitem{Surmiak2020}
\bibinfo{author}{Surmiak, M.~A.} \emph{et~al.}
\newblock \bibinfo{title}{High-throughput characterization of perovskite solar
  cells for rapid combinatorial screening}.
\newblock \emph{\bibinfo{journal}{Sol. RRL}} \textbf{\bibinfo{volume}{4}},
  \bibinfo{pages}{2000097} (\bibinfo{year}{2020}).

\bibitem{Reinhardt2020}
\bibinfo{author}{Reinhardt, E.}, \bibinfo{author}{Salaheldin, A.~M.},
  \bibinfo{author}{Distaso, M.}, \bibinfo{author}{Segets, D.} \&
  \bibinfo{author}{Peukert, W.}
\newblock \bibinfo{title}{Rapid characterization and parameter space
  exploration of perovskites using an automated routine}.
\newblock \emph{\bibinfo{journal}{ACS Comb. Sci.}}
  \textbf{\bibinfo{volume}{22}}, \bibinfo{pages}{6--17} (\bibinfo{year}{2020}).

\bibitem{Ahmadi2021}
\bibinfo{author}{Ahmadi, M.}, \bibinfo{author}{Ziatdinov, M.},
  \bibinfo{author}{Zhou, Y.}, \bibinfo{author}{Lass, E.~A.} \&
  \bibinfo{author}{Kalinin, S.~V.}
\newblock \bibinfo{title}{Machine learning for high-throughput experimental
  exploration of metal halide perovskites}.
\newblock \emph{\bibinfo{journal}{Joule}} \textbf{\bibinfo{volume}{5}},
  \bibinfo{pages}{2797--2822} (\bibinfo{year}{2021}).

\bibitem{Wang2017}
\bibinfo{author}{Wang, Z.} \emph{et~al.}
\newblock \bibinfo{title}{Efficient ambient-air-stable solar cells with 2d–3d
  heterostructured butylammonium-caesium- formamidinium lead halide
  perovskites}.
\newblock \emph{\bibinfo{journal}{Nat. Energy}} \textbf{\bibinfo{volume}{2}},
  \bibinfo{pages}{1--10} (\bibinfo{year}{2017}).

\bibitem{Liu2023}
\bibinfo{author}{Liu, A.} \emph{et~al.}
\newblock \bibinfo{title}{High-performance metal halide perovskite
  transistors}.
\newblock \emph{\bibinfo{journal}{Nat. Electron.}} \bibinfo{pages}{1--13}
  (\bibinfo{year}{2023}).

\bibitem{escobedo2019}
\bibinfo{author}{Escobedo-Morales, A.} \emph{et~al.}
\newblock \bibinfo{title}{Automated method for the determination of the band
  gap energy of pure and mixed powder samples using diffuse reflectance
  spectroscopy}.
\newblock \emph{\bibinfo{journal}{Heliyon}} \textbf{\bibinfo{volume}{5}},
  \bibinfo{pages}{e01505} (\bibinfo{year}{2019}).

\bibitem{Wu2023_advmat}
\bibinfo{author}{Wu, T.~C.} \emph{et~al.}
\newblock \bibinfo{title}{A materials acceleration platform for organic laser
  discovery}.
\newblock \emph{\bibinfo{journal}{Adv. Mater.}} \textbf{\bibinfo{volume}{35}},
  \bibinfo{pages}{2207070} (\bibinfo{year}{2023}).

\bibitem{Siemenn2022}
\bibinfo{author}{Siemenn, A.~E.} \emph{et~al.}
\newblock \bibinfo{title}{A machine learning and computer vision approach to
  rapidly optimize multiscale droplet generation}.
\newblock \emph{\bibinfo{journal}{ACS Appl. Mater. Interfaces}}
  \textbf{\bibinfo{volume}{14}}, \bibinfo{pages}{4668--4679}
  (\bibinfo{year}{2022}).

\bibitem{Zhu2020}
\bibinfo{author}{Zhu, Y.} \emph{et~al.}
\newblock \bibinfo{title}{Unraveling pore evolution in post-processing of
  binder jetting materials: X-ray computed tomography, computer vision, and
  machine learning}.
\newblock \emph{\bibinfo{journal}{Additive Manuf.}}
  \textbf{\bibinfo{volume}{34}}, \bibinfo{pages}{101183}
  (\bibinfo{year}{2020}).

\bibitem{Tung2010}
\bibinfo{author}{Tung, F.}, \bibinfo{author}{Wong, A.} \&
  \bibinfo{author}{Clausi, D.~A.}
\newblock \bibinfo{title}{Enabling scalable spectral clustering for image
  segmentation}.
\newblock \emph{\bibinfo{journal}{Pattern Recogn.}}
  \textbf{\bibinfo{volume}{43}}, \bibinfo{pages}{4069--4076}
  (\bibinfo{year}{2010}).

\bibitem{Jain_2021_ICCV}
\bibinfo{author}{Jain, S.}, \bibinfo{author}{Paudel, D.~P.},
  \bibinfo{author}{Danelljan, M.} \& \bibinfo{author}{Van~Gool, L.}
\newblock \bibinfo{title}{Scaling semantic segmentation beyond 1k classes on a
  single gpu} (\bibinfo{year}{2021}).

\bibitem{garnot2021panoptic}
\bibinfo{author}{Garnot, V. S.~F.} \& \bibinfo{author}{Landrieu, L.}
\newblock \bibinfo{title}{Panoptic segmentation of satellite image time series
  with convolutional temporal attention networks}.
\newblock \emph{\bibinfo{journal}{Proceedings of the IEEE/CVF International
  Conference on Computer Vision (CVPR)}} \bibinfo{pages}{4872--4881}
  (\bibinfo{year}{2021}).

\bibitem{Li_2022_CVPR}
\bibinfo{author}{Li, H.}, \bibinfo{author}{Pan, X.}, \bibinfo{author}{Yan, K.},
  \bibinfo{author}{Tang, F.} \& \bibinfo{author}{Zheng, W.-S.}
\newblock \bibinfo{title}{Siod: Single instance annotated per category per
  image for object detection}.
\newblock \emph{\bibinfo{journal}{Proceedings of the IEEE/CVF Conference on
  Computer Vision and Pattern Recognition (CVPR)}}
  \bibinfo{pages}{14197--14206} (\bibinfo{year}{2022}).

\bibitem{park_ding_2019}
\bibinfo{author}{Park, C.} \& \bibinfo{author}{Ding, Y.}
\newblock \bibinfo{title}{Automating material image analysis for material
  discovery}.
\newblock \emph{\bibinfo{journal}{MRS Commun.}} \textbf{\bibinfo{volume}{9}},
  \bibinfo{pages}{545–555} (\bibinfo{year}{2019}).

\bibitem{doi:10.1080/0740817X.2011.587867}
\bibinfo{author}{Park, C.} \emph{et~al.}
\newblock \bibinfo{title}{A multistage, semi-automated procedure for analyzing
  the morphology of nanoparticles}.
\newblock \emph{\bibinfo{journal}{IIE Trans.}} \textbf{\bibinfo{volume}{44}},
  \bibinfo{pages}{507--522} (\bibinfo{year}{2012}).

\bibitem{CHOWDHURY2016176}
\bibinfo{author}{Chowdhury, A.}, \bibinfo{author}{Kautz, E.},
  \bibinfo{author}{Yener, B.} \& \bibinfo{author}{Lewis, D.}
\newblock \bibinfo{title}{Image driven machine learning methods for
  microstructure recognition}.
\newblock \emph{\bibinfo{journal}{Comput. Mater. Sci.}}
  \textbf{\bibinfo{volume}{123}}, \bibinfo{pages}{176--187}
  (\bibinfo{year}{2016}).

\bibitem{chen_wolff_1998}
\bibinfo{author}{Chen, H.} \& \bibinfo{author}{Wolff, L.~B.}
\newblock \bibinfo{title}{Polarization phase-based method for material
  classification in computer vision}.
\newblock \emph{\bibinfo{journal}{Int. J. Comput. Vis.}}
  \textbf{\bibinfo{volume}{28}}, \bibinfo{pages}{73–83}
  (\bibinfo{year}{1998}).

\bibitem{leung_malik_2001}
\bibinfo{author}{Leung, T.} \& \bibinfo{author}{Malik, J.}
\newblock \bibinfo{title}{Representing and recognizing the visual appearance of
  materials using three-dimensional textons}.
\newblock \emph{\bibinfo{journal}{Int. J. Comput. Vis.}}
  \textbf{\bibinfo{volume}{43}}, \bibinfo{pages}{29–44}
  (\bibinfo{year}{2001}).

\bibitem{Ma2023}
\bibinfo{author}{Ma, Y.} \emph{et~al.}
\newblock \bibinfo{title}{Improving 3d food printing performance using computer
  vision and feedforward nozzle motion control}.
\newblock \emph{\bibinfo{journal}{J. Food Eng.}}
  \textbf{\bibinfo{volume}{339}}, \bibinfo{pages}{111277}
  (\bibinfo{year}{2023}).

\bibitem{Wang_2022_CVPR}
\bibinfo{author}{Wang, Z.}, \bibinfo{author}{Liu, J.}, \bibinfo{author}{Li, G.}
  \& \bibinfo{author}{Han, H.}
\newblock \bibinfo{title}{Blind2unblind: Self-supervised image denoising with
  visible blind spots}.
\newblock \emph{\bibinfo{journal}{Proceedings of the IEEE/CVF Conference on
  Computer Vision and Pattern Recognition (CVPR)}} \bibinfo{pages}{2027--2036}
  (\bibinfo{year}{2022}).

\bibitem{Neshatavar_2022_CVPR}
\bibinfo{author}{Neshatavar, R.}, \bibinfo{author}{Yavartanoo, M.},
  \bibinfo{author}{Son, S.} \& \bibinfo{author}{Lee, K.~M.}
\newblock \bibinfo{title}{Cvf-sid: Cyclic multi-variate function for
  self-supervised image denoising by disentangling noise from image}.
\newblock \emph{\bibinfo{journal}{Proceedings of the IEEE/CVF Conference on
  Computer Vision and Pattern Recognition (CVPR)}}
  \bibinfo{pages}{17583--17591} (\bibinfo{year}{2022}).

\bibitem{Tian2020}
\bibinfo{author}{Tian, J.} \emph{et~al.}
\newblock \bibinfo{title}{Composition engineering of all-inorganic perovskite
  film for efficient and operationally stable solar cells}.
\newblock \emph{\bibinfo{journal}{Adv. Funct. Mater.}}
  \textbf{\bibinfo{volume}{30}}, \bibinfo{pages}{2001764}
  (\bibinfo{year}{2020}).

\bibitem{Jeon2015}
\bibinfo{author}{Jeon, N.~J.} \emph{et~al.}
\newblock \bibinfo{title}{Compositional engineering of perovskite materials for
  high-performance solar cells}.
\newblock \emph{\bibinfo{journal}{Nature}} \textbf{\bibinfo{volume}{517}},
  \bibinfo{pages}{476--480} (\bibinfo{year}{2015}).

\bibitem{Massuyeau2022}
\bibinfo{author}{Massuyeau, F.} \emph{et~al.}
\newblock \bibinfo{title}{Perovskite or not perovskite? a deep-learning
  approach to automatically identify new hybrid perovskites from x-ray
  diffraction patterns}.
\newblock \emph{\bibinfo{journal}{Adv. Mater.}} \textbf{\bibinfo{volume}{34}},
  \bibinfo{pages}{2203879} (\bibinfo{year}{2022}).

\bibitem{Mundt2020}
\bibinfo{author}{Mundt, L.~E.} \& \bibinfo{author}{Schelhas, L.~T.}
\newblock \bibinfo{title}{Structural evolution during perovskite crystal
  formation and degradation: In situ and operando x-ray diffraction studies}.
\newblock \emph{\bibinfo{journal}{Adv. Energy Mater.}}
  \textbf{\bibinfo{volume}{10}}, \bibinfo{pages}{1903074}
  (\bibinfo{year}{2020}).

\bibitem{Zhidkov2021}
\bibinfo{author}{Zhidkov, I.~S.} \emph{et~al.}
\newblock \bibinfo{title}{Xps spectra as a tool for studying photochemical and
  thermal degradation in apbx3 hybrid halide perovskites}.
\newblock \emph{\bibinfo{journal}{Nano Energy}} \textbf{\bibinfo{volume}{79}},
  \bibinfo{pages}{105421} (\bibinfo{year}{2021}).

\bibitem{Ahmad2017}
\bibinfo{author}{Ahmad, Z.} \emph{et~al.}
\newblock \bibinfo{title}{Instability in ch3nh3pbi3 perovskite solar cells due
  to elemental migration and chemical composition changes}.
\newblock \emph{\bibinfo{journal}{Sci. Reports}} \textbf{\bibinfo{volume}{7}},
  \bibinfo{pages}{1--8} (\bibinfo{year}{2017}).

\bibitem{opencv_library}
\bibinfo{author}{Bradski, G.}
\newblock \bibinfo{title}{{The OpenCV Library}}.
\newblock \emph{\bibinfo{journal}{Dr. Dobb’s J. Softw. Tools}}
  (\bibinfo{year}{2000}).

\bibitem{batagelj2003efficient}
\bibinfo{author}{Batagelj, V.}
\newblock \bibinfo{title}{Efficient algorithms for citation network analysis}.
\newblock \emph{\bibinfo{journal}{arXiv preprint cs/0309023}}
  (\bibinfo{year}{2003}).

\bibitem{Meyer1994}
\bibinfo{author}{Meyer, F.}
\newblock \bibinfo{title}{Topographic distance and watershed lines}.
\newblock \emph{\bibinfo{journal}{Signal Process.}}
  \textbf{\bibinfo{volume}{38}}, \bibinfo{pages}{113--125}
  (\bibinfo{year}{1994}).

\bibitem{Elsayed2023}
\bibinfo{author}{Elsayed, M.~R.}, \bibinfo{author}{Elseman, A.~M.},
  \bibinfo{author}{Abdelmageed, A.~A.}, \bibinfo{author}{Hashem, H.~M.} \&
  \bibinfo{author}{Hassen, A.}
\newblock \bibinfo{title}{Synthesis and numerical simulation of
  formamidinium-based perovskite solar cells: a predictable device performance
  at nis-egypt}.
\newblock \emph{\bibinfo{journal}{Sci. Reports}} \textbf{\bibinfo{volume}{13}},
  \bibinfo{pages}{1--16} (\bibinfo{year}{2023}).

\bibitem{Murugadoss2021}
\bibinfo{author}{Murugadoss, G.} \emph{et~al.}
\newblock \bibinfo{title}{Crystal stabilization of $\alpha$-fapbi3 perovskite
  by rapid annealing method in industrial scale}.
\newblock \emph{\bibinfo{journal}{J. Mater. Research Technol.}}
  \textbf{\bibinfo{volume}{12}}, \bibinfo{pages}{1924--1930}
  (\bibinfo{year}{2021}).

\bibitem{Zhang2019}
\bibinfo{author}{Zhang, J.} \emph{et~al.}
\newblock \bibinfo{title}{Boosting photovoltaic performance for lead halide
  perovskites solar cells with bf4- anion substitutions}.
\newblock \emph{\bibinfo{journal}{Adv. Funct. Mater.}}
  \textbf{\bibinfo{volume}{29}}, \bibinfo{pages}{1808833}
  (\bibinfo{year}{2019}).

\bibitem{Maqsood2020}
\bibinfo{author}{Maqsood, A.} \emph{et~al.}
\newblock \bibinfo{title}{Perovskite solar cells based on compact, smooth
  fa0.1ma0.9pbi3 film with efficiency exceeding 22\%}.
\newblock \emph{\bibinfo{journal}{Nanoscale Research Lett.}}
  \textbf{\bibinfo{volume}{15}}, \bibinfo{pages}{1--9} (\bibinfo{year}{2020}).

\bibitem{Tauc1966}
\bibinfo{author}{Tauc, J.}, \bibinfo{author}{Grigorovici, R.} \&
  \bibinfo{author}{Vancu, A.}
\newblock \bibinfo{title}{Optical properties and electronic structure of
  amorphous germanium}.
\newblock \emph{\bibinfo{journal}{Phys. Status Solidi B}}
  \textbf{\bibinfo{volume}{15}}, \bibinfo{pages}{627--637}
  (\bibinfo{year}{1966}).

\bibitem{Kubelka1931}
\bibinfo{author}{Kubelka, P.} \& \bibinfo{author}{Munk, F.}
\newblock \bibinfo{title}{A contribution to the optics of pigments}.
\newblock \emph{\bibinfo{journal}{Z. Technol. Phys.}}
  \textbf{\bibinfo{volume}{12}}, \bibinfo{pages}{593--599}
  (\bibinfo{year}{1931}).

\bibitem{Weber2016}
\bibinfo{author}{Weber, O.~J.}, \bibinfo{author}{Charles, B.} \&
  \bibinfo{author}{Weller, M.~T.}
\newblock \bibinfo{title}{Phase behaviour and composition in the
  formamidinium–methylammonium hybrid lead iodide perovskite solid solution}.
\newblock \emph{\bibinfo{journal}{J. Mater. Chem. A}}
  \textbf{\bibinfo{volume}{4}}, \bibinfo{pages}{15375--15382}
  (\bibinfo{year}{2016}).

\bibitem{gil2020efficient}
\bibinfo{author}{Gil-Escrig, L.} \emph{et~al.}
\newblock \bibinfo{title}{Efficient vacuum-deposited perovskite solar cells
  with stable cubic fa1--x ma x pbi3}.
\newblock \emph{\bibinfo{journal}{ACS Energy Lett.}}
  \textbf{\bibinfo{volume}{5}}, \bibinfo{pages}{3053--3061}
  (\bibinfo{year}{2020}).

\bibitem{lu2020compositional}
\bibinfo{author}{Lu, H.}, \bibinfo{author}{Krishna, A.},
  \bibinfo{author}{Zakeeruddin, S.~M.}, \bibinfo{author}{Gr{\"a}tzel, M.} \&
  \bibinfo{author}{Hagfeldt, A.}
\newblock \bibinfo{title}{Compositional and interface engineering of
  organic-inorganic lead halide perovskite solar cells}.
\newblock \emph{\bibinfo{journal}{Iscience}} \textbf{\bibinfo{volume}{23}}
  (\bibinfo{year}{2020}).

\bibitem{slimi2016perovskite}
\bibinfo{author}{Slimi, B.} \emph{et~al.}
\newblock \bibinfo{title}{Perovskite fa1-xmaxpbi3 for solar cells: films
  formation and properties}.
\newblock \emph{\bibinfo{journal}{Energy Proc.}}
  \textbf{\bibinfo{volume}{102}}, \bibinfo{pages}{87--95}
  (\bibinfo{year}{2016}).

\bibitem{Wang2016}
\bibinfo{author}{Wang, D.}, \bibinfo{author}{Wright, M.},
  \bibinfo{author}{Elumalai, N.~K.} \& \bibinfo{author}{Uddin, A.}
\newblock \bibinfo{title}{Stability of perovskite solar cells}.
\newblock \emph{\bibinfo{journal}{Sol. Energy Mater.}}
  \textbf{\bibinfo{volume}{147}}, \bibinfo{pages}{255--275}
  (\bibinfo{year}{2016}).

\bibitem{Nan2021}
\bibinfo{author}{Nan, Z.~A.} \emph{et~al.}
\newblock \bibinfo{title}{Revealing phase evolution mechanism for stabilizing
  formamidinium-based lead halide perovskites by a key intermediate phase}.
\newblock \emph{\bibinfo{journal}{Chem}} \textbf{\bibinfo{volume}{7}},
  \bibinfo{pages}{2513--2526} (\bibinfo{year}{2021}).

\bibitem{Wu2023}
\bibinfo{author}{Wu, J.}, \bibinfo{author}{Chen, J.} \& \bibinfo{author}{Wang,
  H.}
\newblock \bibinfo{title}{Phase transition kinetics of mapbi3 for
  tetragonal-to-orthorhombic evolution}.
\newblock \emph{\bibinfo{journal}{JACS Au}} \textbf{\bibinfo{volume}{3}},
  \bibinfo{pages}{1205--1212} (\bibinfo{year}{2023}).

\bibitem{Stoumpos2017}
\bibinfo{author}{Stoumpos, C.~C.}, \bibinfo{author}{Mao, L.},
  \bibinfo{author}{Malliakas, C.~D.} \& \bibinfo{author}{Kanatzidis, M.~G.}
\newblock \bibinfo{title}{Structure-band gap relationships in hexagonal
  polytypes and low-dimensional structures of hybrid tin iodide perovskites}.
\newblock \emph{\bibinfo{journal}{Inorg. Chem.}} \textbf{\bibinfo{volume}{56}},
  \bibinfo{pages}{56--73} (\bibinfo{year}{2017}).

\bibitem{Charles2017}
\bibinfo{author}{Charles, B.}, \bibinfo{author}{Dillon, J.},
  \bibinfo{author}{Weber, O.~J.}, \bibinfo{author}{Islam, M.~S.} \&
  \bibinfo{author}{Weller, M.~T.}
\newblock \bibinfo{title}{Understanding the stability of mixed a-cation lead
  iodide perovskites}.
\newblock \emph{\bibinfo{journal}{J. Mater. Chem. A}}
  \textbf{\bibinfo{volume}{5}}, \bibinfo{pages}{22495--22499}
  (\bibinfo{year}{2017}).

\bibitem{Pisanu2017}
\bibinfo{author}{Pisanu, A.} \emph{et~al.}
\newblock \bibinfo{title}{The fa1–xmaxpbi3 system: Correlations among
  stoichiometry control, crystal structure, optical properties, and phase
  stability}.
\newblock \emph{\bibinfo{journal}{J. Phys. Chem. C}}
  \textbf{\bibinfo{volume}{121}}, \bibinfo{pages}{8746--8751}
  (\bibinfo{year}{2017}).

\bibitem{Binek2015}
\bibinfo{author}{Binek, A.}, \bibinfo{author}{Hanusch, F.~C.},
  \bibinfo{author}{Docampo, P.} \& \bibinfo{author}{Bein, T.}
\newblock \bibinfo{title}{Stabilization of the trigonal high-temperature phase
  of formamidinium lead iodide}.
\newblock \emph{\bibinfo{journal}{J. Phys. Chem. Lett.}}
  \textbf{\bibinfo{volume}{6}}, \bibinfo{pages}{1249--1253}
  (\bibinfo{year}{2015}).

\bibitem{Wang2017b}
\bibinfo{author}{Wang, T.} \emph{et~al.}
\newblock \bibinfo{title}{Indirect to direct bandgap transition in
  methylammonium lead halide perovskite}.
\newblock \emph{\bibinfo{journal}{Energy Environ. Sci.}}
  \textbf{\bibinfo{volume}{10}}, \bibinfo{pages}{509--515}
  (\bibinfo{year}{2017}).

\bibitem{Targhi2018}
\bibinfo{author}{Targhi, F.~F.}, \bibinfo{author}{Jalili, Y.~S.} \&
  \bibinfo{author}{Kanjouri, F.}
\newblock \bibinfo{title}{Mapbi3 and fapbi3 perovskites as solar cells: Case
  study on structural, electrical and optical properties}.
\newblock \emph{\bibinfo{journal}{Results Phys.}}
  \textbf{\bibinfo{volume}{10}}, \bibinfo{pages}{616--627}
  (\bibinfo{year}{2018}).

\bibitem{s120607063}
\bibinfo{author}{Menesatti, P.} \emph{et~al.}
\newblock \bibinfo{title}{Rgb color calibration for quantitative image
  analysis: The “3d thin-plate spline” warping approach}.
\newblock \emph{\bibinfo{journal}{Sensors}} \textbf{\bibinfo{volume}{12}},
  \bibinfo{pages}{7063--7079} (\bibinfo{year}{2012}).

\end{thebibliography}
}

\newpage
\beginsupplement
\onecolumn
\begin{centering}
\large{\textbf{Supplementary Information:}\\Using Scalable Computer Vision to Automate High-throughput Semiconductor Characterization} \par
\end{centering}

\begin{description}
\begin{centering}

\item \textbf{Alexander E. Siemenn}$^{1 * \dagger}$, \textbf{Eunice Aissi}$^{1 * \dagger}$, Fang Sheng$^{1}$, Armi Tiihonen$^{1,2}$, Hamide Kavak$^{1,3}$, Basita Das$^{1}$, Tonio Buonassisi$^{1}$
\item $^{1}$Department of Mechanical Engineering, Massachusetts Institute of Technology, Cambridge, MA 02139, USA
\item $^{2}$Department of Applied Physics, Aalto University, Espoo, 02150, Finland
\item $^{3}$Department of Physics, Cukurova University, Adana, 01330, Turkiye
\item 
\item $^*$Corresponding authors: \{asiemenn@mit.edu, eunicea@mit.edu\}
\item $^\dagger$Contributed equally.

\end{centering}
\end{description}

\section*{Experiments}

\noindent \textbf{Processing Times}

\begin{figure}[h!]
\begin{center}
\includegraphics[width=1\columnwidth]{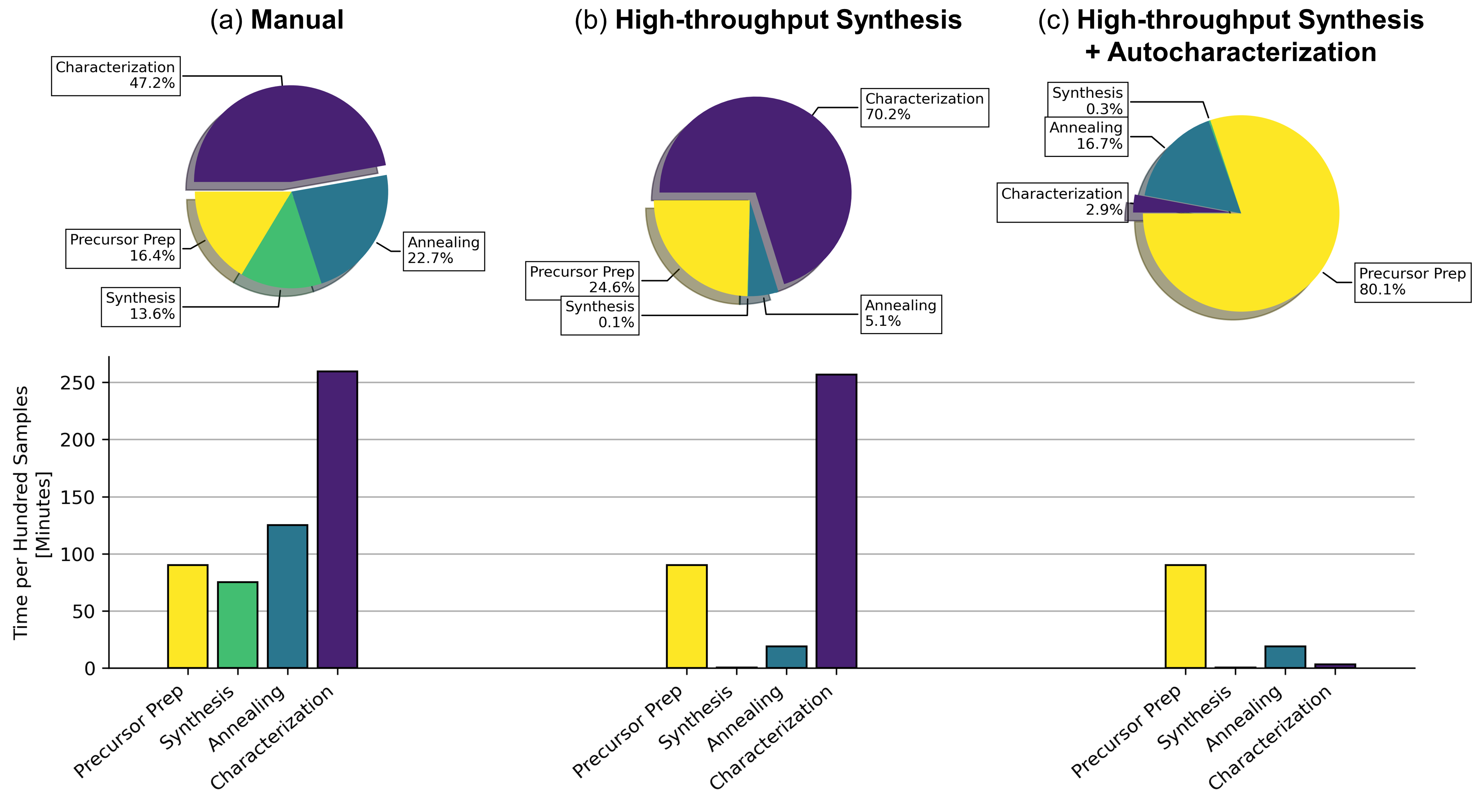}
\end{center}
   \caption{Minimum time to process 100 perovskite samples and compute band gap. The processing times are shown for three different scenarios: (a) manual synthesis manual and characterization, (b) high-throughput synthesis and manual characterization, and (c) high-throughput synthesis and high-throughput autocharacterization.}
\label{sfig:times}
\end{figure}

In this paper, we aim to achieve higher rate-matching between the synthesis and characterization of materials for high-throughput screening. In Figure \ref{sfig:times}, the minimum processing time required to go from precursor to data is illustrated for perovskite samples. The minimum processing times to collect band gap data for 100 perovskite samples are shown and represent the times experimentally recorded during this study. Three different scenarios are represented that use combinations of both manual and high-throughput methods. The processing time to collect these band gap data is further broken down into four steps: (1) precursor preparation, (2) synthesis, (3) annealing, and (4) characterization. These are the \textit{minimum} processing times as they do not account for sample transfer times or reloading, \textit{e.g.}, moving a sample to the hotplate. In Figure \ref{sfig:times}b, moving from manual to high-throughput synthesis results in a higher discrepancy of throughputs between synthesis and characterization, which bottlenecks the materials screening loop. However, in Figure \ref{sfig:times}c, by using high-throughput autocharacterization, the rate of high-throughput synthesis is closely matched to that of high-throughput synthesis, in turn, enabling more efficient materials screening. The bottleneck then becomes precursor preparation, which is out of the scope of this study.

To detail the individual processing time contributions in this study, precursor preparation is always conducted manually and takes 90 minutes to make 100 samples worth of solution for all three scenarios. Synthesis takes 45 seconds per sample for manual spin coating and takes 20 seconds per 100 samples for high-throughput manufacturing. Annealing takes 10 minutes for manual thin films where 8 samples can fit on a single hot plate at a time. It takes 15 minutes for the thicker, high-throughput samples but all 100 samples can fit on a single hot plate. Characterization for computing band gap takes 255 minutes per 100 samples manually by a domain expert. It takes 3 minutes per 100 samples using the band gap extractor autocharacterization algorithm developed in this paper.

\bigbreak

\noindent \textbf{Experimental Reproducibility}

\begin{figure}[h!]
\centering
\begin{subfigure}[b]{0.49\textwidth} 
\includegraphics[width=\textwidth]{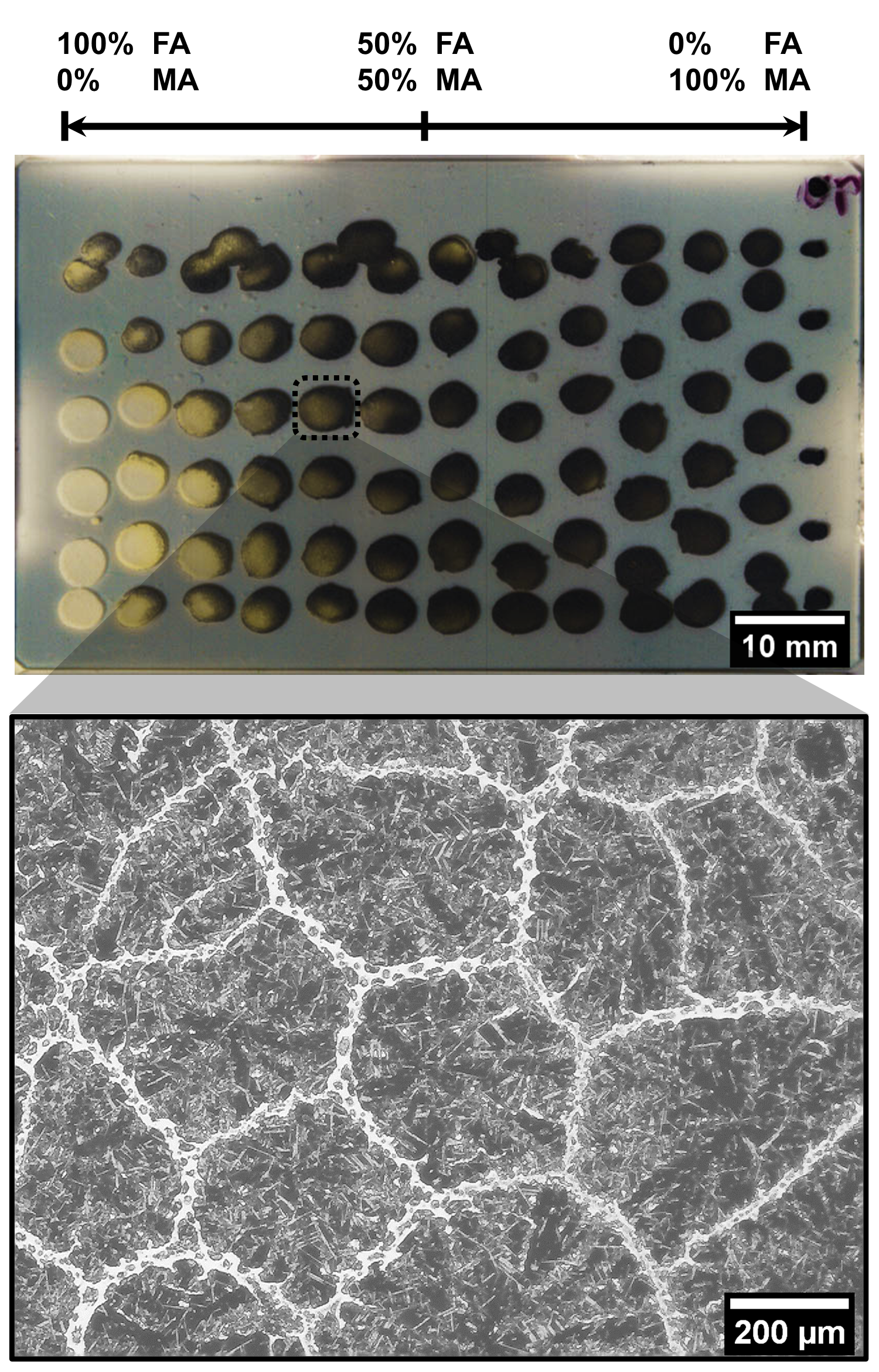}
\caption{Batch A}
\end{subfigure}
\begin{subfigure}[b]{0.49\textwidth} 
\includegraphics[width=\textwidth]{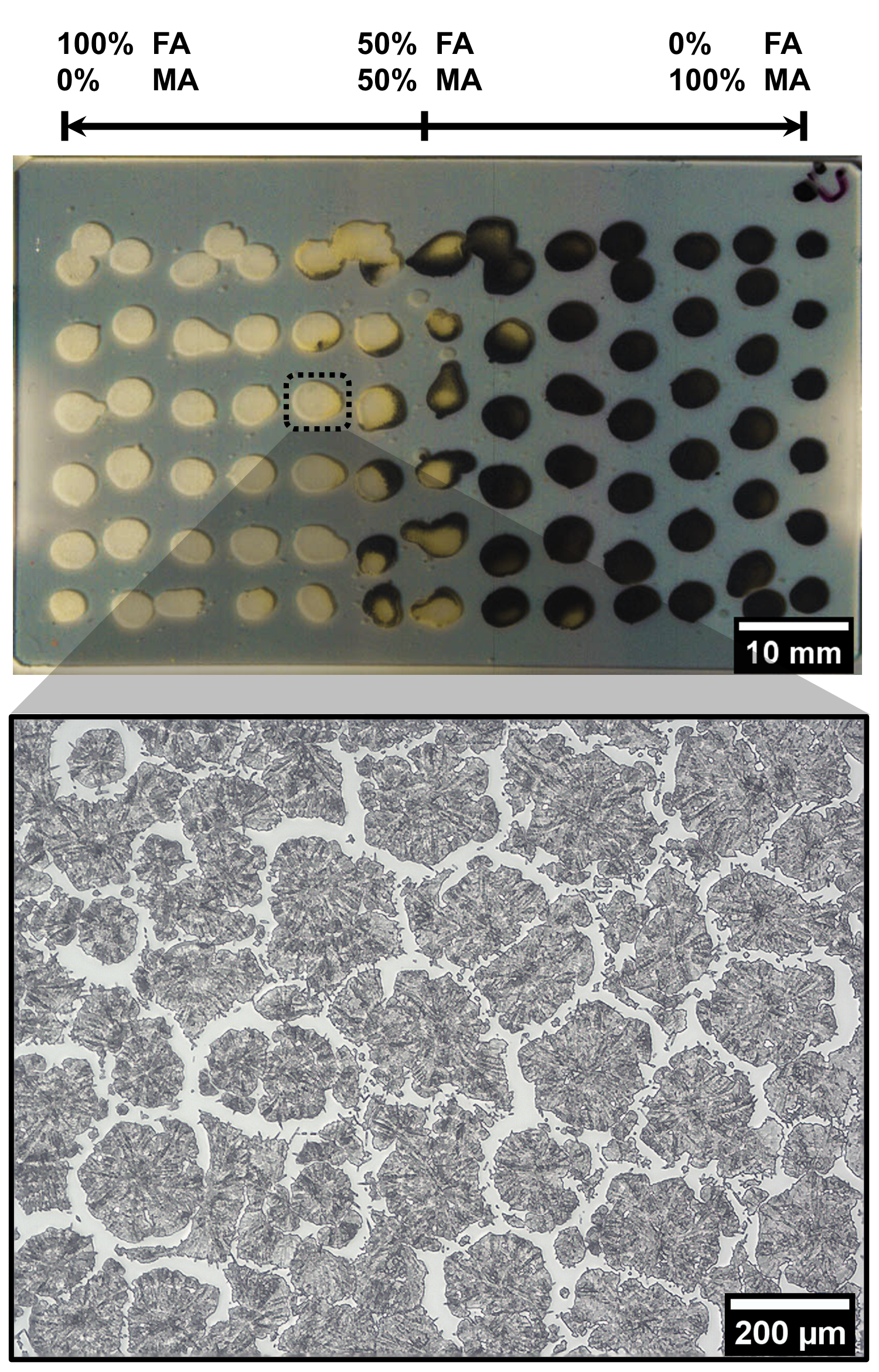}
\caption{Batch B}
\end{subfigure}
\caption{Optical microscopy of deposit morphology between two different batches of FA$_{0.67}$MA$_{0.33}$PbI$_3$ after controlled degradation.}
\label{sfig:morphology}
\end{figure}

Variability exists across samples manufactured using our high-throughput combinatorial printer, as the setup is designed for low-cost, scalable, and high-throughput screening rather than high-fidelity experiments. Although the purpose of the paper is not to highlight the perovskite manufacturing method but to focus on the rate matching of characterization, it is important to understand the sources of variability from experiment to experiment, as these sources of variability will arise for those who replicate the proposed approaches. Figure \ref{sfig:morphology} shows the post-degradation structural morphology differences between two of the same sample compositions across two separate batches, manufactured using the same printing conditions. Both batches were degraded under the same environmental conditions for 2 hours at 35$^\circ$C, 40\% relative humidity, and 0.5 suns of AM1.5 illumination (without UV). During crystallization, batch A achieved more uniform and compact grain boundaries, whereas the crystallization in sample B produced jagged boundaries, inducing more pathways for degradation. This crystallization mismatch explains the accelerated degradation noted in batch B relative to batch A. Samples that are prone to phase changes (\textit{e.g.}, compositions near phase boundaries such as $0.0\leq x \leq 0.15$, for FA$_{1-x}$MA$_{x}$PbI$_3$), may experience high sample-to-sample variances using any sample preparation approach. Therefore, the effects of these variations must be carefully considered in high-throughput manufacturing scenarios, where not every sample can be fully characterized at high fidelity. These morphological variations in the high-throughput manufacturing process must be further studied to ascertain better control over experimental reproducibility. In this study, three batches of samples were selected with similar morphology in an attempt to minimize this variation and maintain the emphasis of the paper on the assessment of the proposed automatic characterization methods, rather than on the method of synthesis.

\section*{Material Characterization}

\noindent \textbf{Computer Vision Segmentation and Composition Mapping}

\begin{figure}[h!]
\begin{center}
\includegraphics[width=0.6\columnwidth]{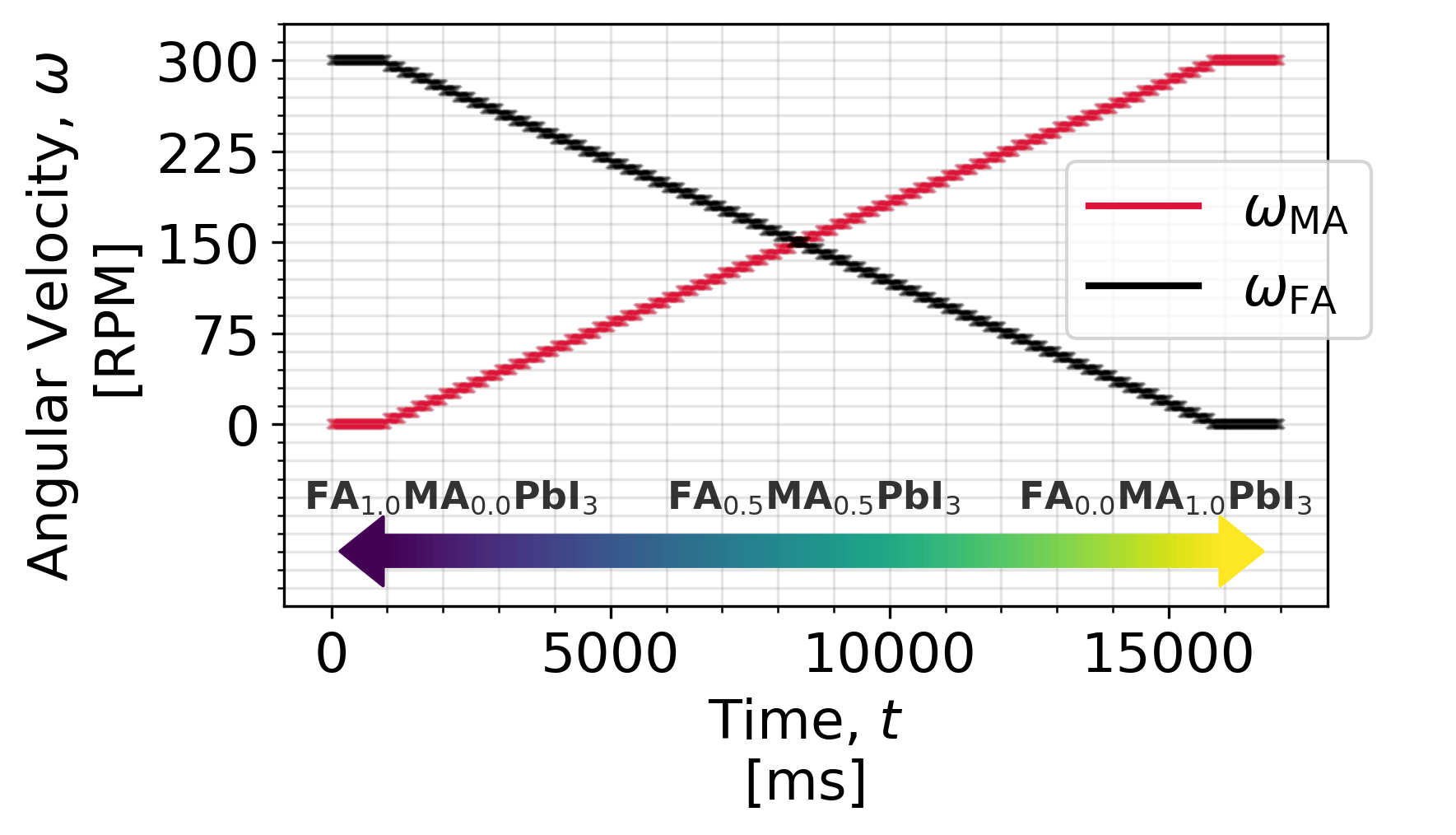}
\end{center}
   \caption{Traces of motor velocity over time for each perovskite precursor, MAPbI$_3$ and FAPbI$_3$, for high-throughput combinatorial deposition as droplets.}
\label{sfig:motor-traces}
\end{figure}

\begin{figure}[h!]
\centering
\begin{subfigure}[b]{0.6\textwidth} 
\includegraphics[width=\textwidth]{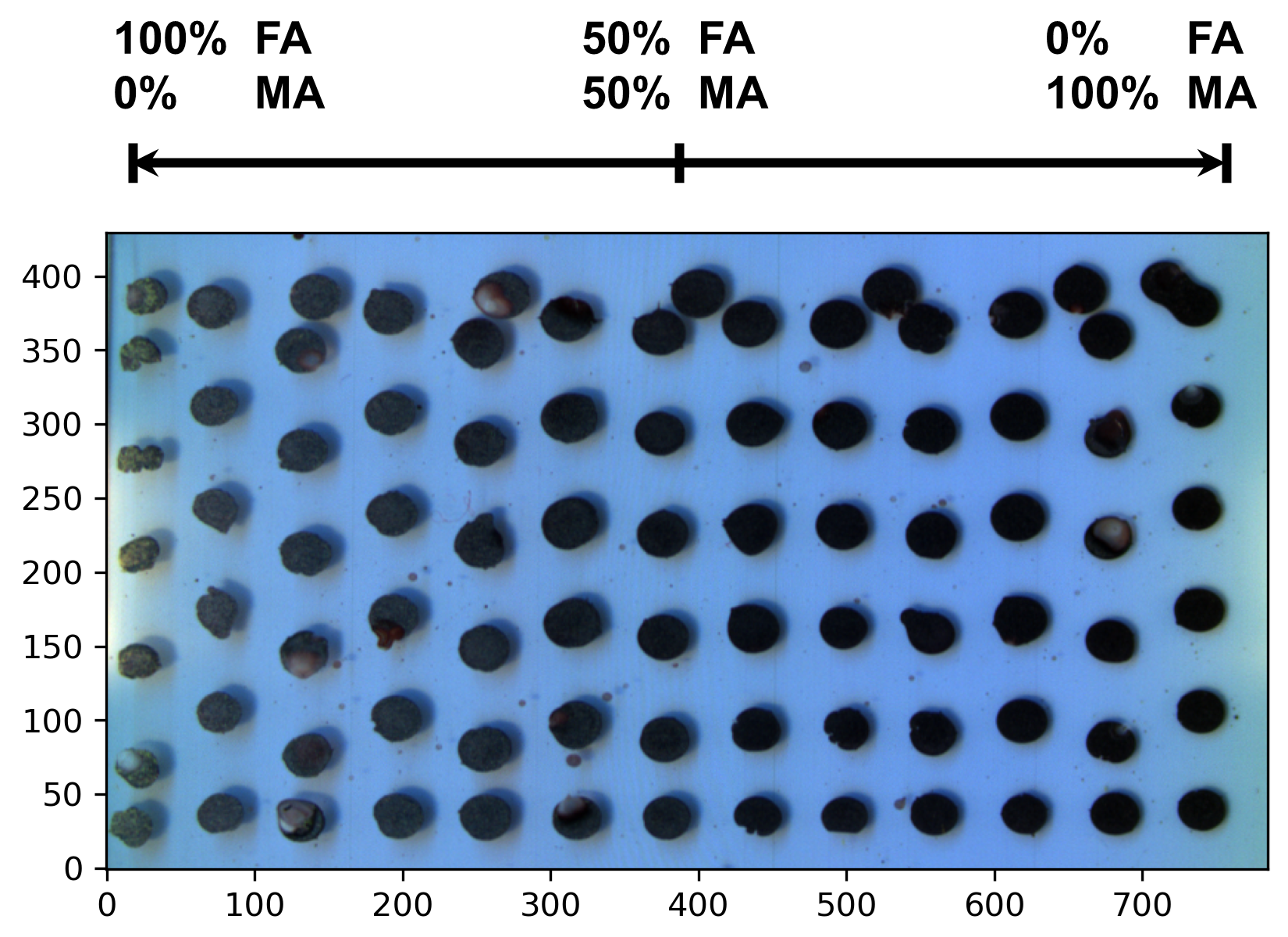}
\caption{Raw Image}
\end{subfigure}
\begin{subfigure}[b]{0.6\textwidth} 
\includegraphics[width=\textwidth]{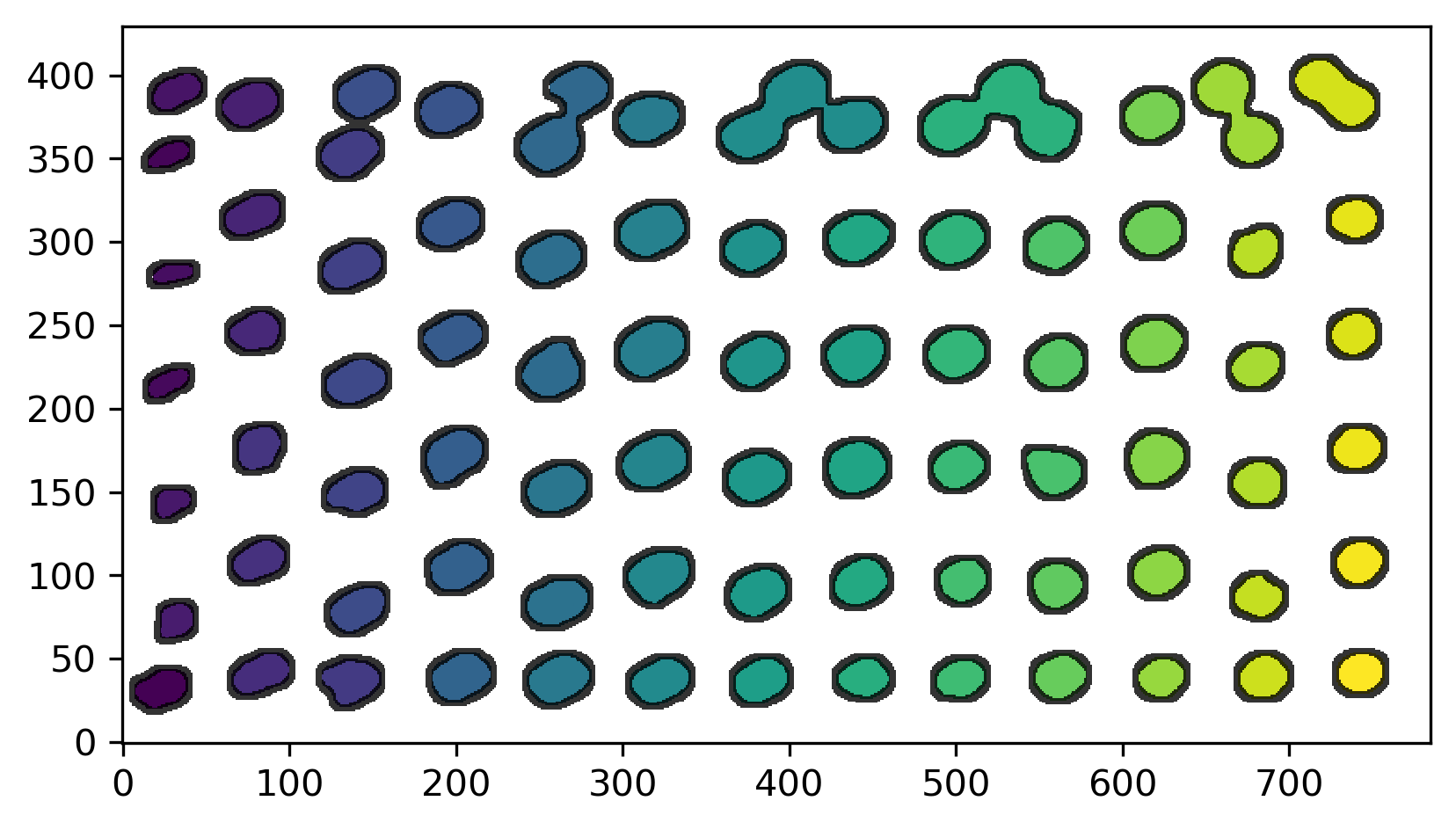}
\caption{Computer Vision-segmented Image}
\end{subfigure}
\begin{subfigure}[b]{0.75\textwidth} 
\includegraphics[width=\textwidth]{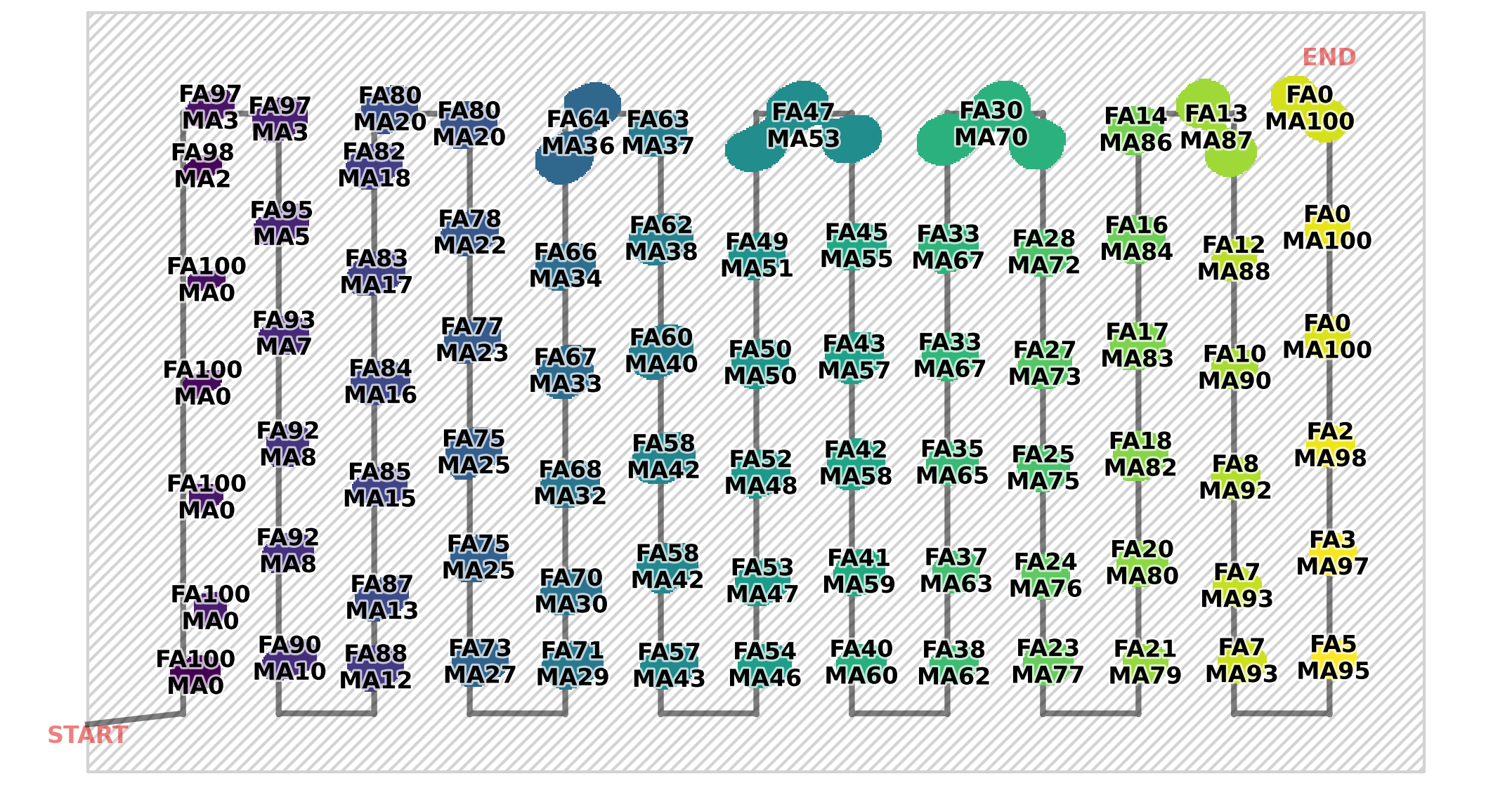}
\caption{Automated Composition Extraction Overlay}
\end{subfigure}
\caption{Computer vision segmentation and composition mapping.}
\label{sfig:comp-extract}
\end{figure}

Figures \ref{sfig:comp-extract}a-b illustrate the process of going from a raw hyperspectral datacube ($\Omega$) to computer vision-segmented data ($\Phi$), which is used as input to the autocharacterization methods developed in this paper. An image, $\Omega$, which can be in either Hyperspectral or RGB format, is segmented using Algorithm \ref{alg:segmentation}, producing the segmented pixels, $(\widehat{X}, \widehat{Y})$, and their corresponding set of reflectance values, $R(\lambda)$. The matched sets of $(\widehat{X}, \widehat{Y})$ and $R(\lambda)$ are denoted as $\Phi$. The compositions of each deposited sample are then able to be mapped onto the segmented $\Phi$ using the G-code raster path of the printer head, the pump speed traces from Figure \ref{sfig:motor-traces}, and Equation \ref{eq:composition}. Figure \ref{sfig:comp-extract}c illustrates this complete mapping of all material deposits with their derived compositions within the FA$_{1-x}$MA$_{x}$PbI$_3$ series.

\bigbreak
\noindent \textbf{Band Gap}

\begin{figure}[h!]
\begin{center}
\includegraphics[width=0.7\columnwidth]{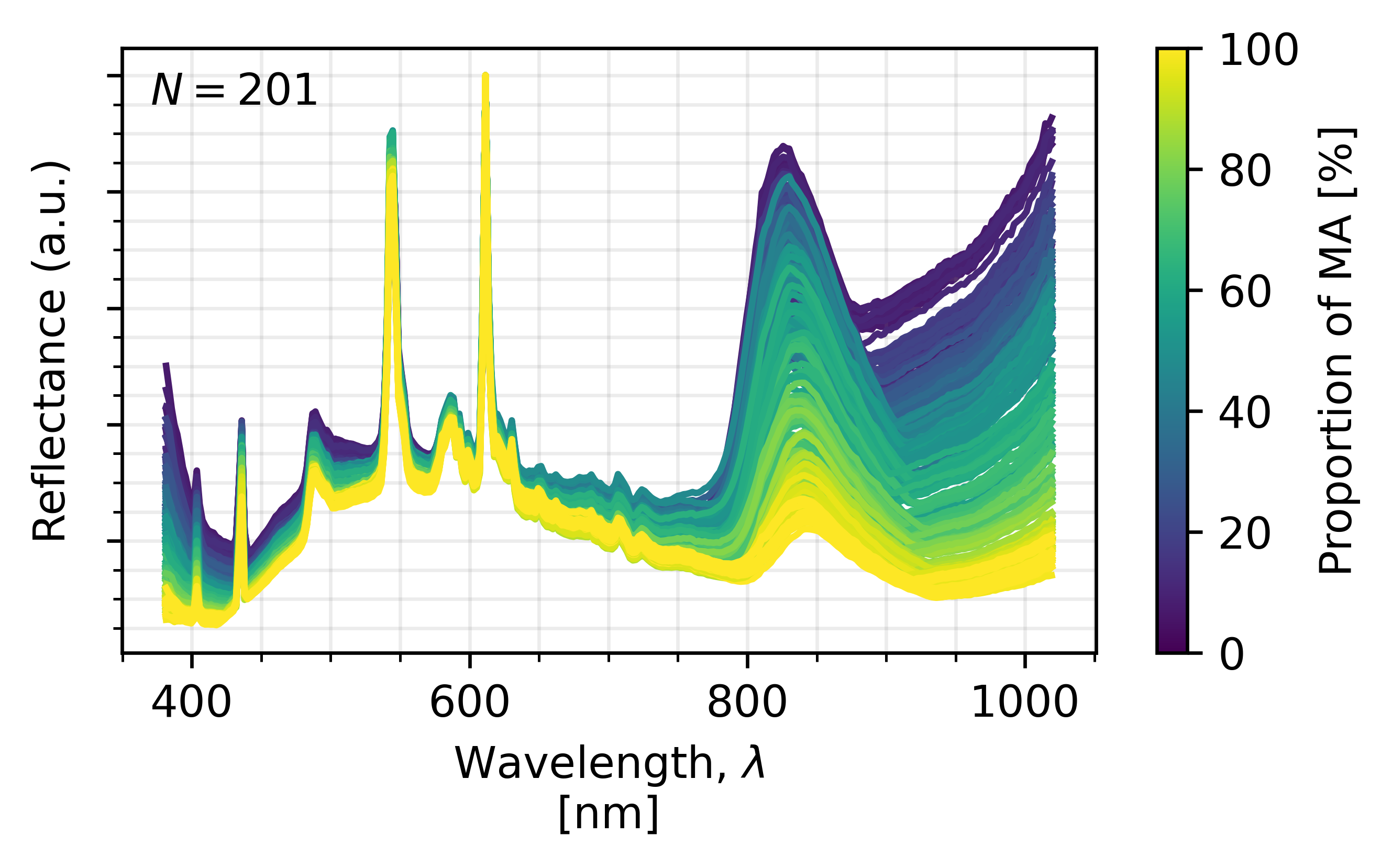}
\end{center}
   \caption{Hyperspectral reflectance of all $N=201$ samples synthesized in this study, color mapped as a function of composition.}
   \label{sfig:hs}
\end{figure}

The materials within the FA$_{1-x}$MA$_{x}$PbI$_3$ compositional series are direct band gap semiconductors \citeSM{Wang2017b, Targhi2018}. To compute the band gap of these materials, first, the reflectance spectra of all samples are measured using a hyperspectral camera (Resonon Pika L) that measures reflectance within the wavelength range $\lambda \in [380\mathrm{nm},1020\mathrm{nm}]$ at 2nm resolution. Figure \ref{sfig:hs} illustrates the measured reflectance spectra for all $N=201$ samples in the paper, gathered using the computer vision-segmentation of the raw hyperspectral datacube, as shown in Figure \ref{sfig:comp-extract}. Then, the reflectance spectra are converted to their corresponding absorption spectra using the Kubelka-Munk equation \citeSM{Kubelka1931,Makula2018} for sufficiently thick samples (the thickness of our samples is approximately 300$\mu$m -- which is considered sufficiently thick for reflectance measurement):

\begin{equation}
    F(R) = \frac{(1-R)^2}{2R},
\end{equation}
where $R$ is the reflectance for the entire range of $\lambda$ for a given segmented pixel in the reflectance hypercube, $(\widehat{X},\widehat{Y})$. Next, the Tauc curves are computed from $F(R)$ \citeSM{Tauc1966}:

\begin{equation}
\label{seq:eg}
    (F(R)\cdot h\nu)^{1/\gamma} = B(h\nu -E_g),
\end{equation}
where $h\nu$ is energy ($h\nu=\frac{1240}{\lambda}$), $\gamma=\frac{1}{2}$ for direct band gap and $\gamma=2$ for indirect band gap, $B$ is a constant that allows the band gap, $E_g$, to be the x-intercept of a regression fit line to the slope of Tauc curve. Hence, the following equation arises that enables computation of the direct band gap from the initial reflectance spectra by equating Equation \ref{seq:final} to Equation \ref{seq:eg}:

\begin{equation}
\label{seq:final}
    (F(R)\cdot h\nu)^{2} = \left(\frac{(1-R)^2}{2R}\cdot\frac{1240}{\lambda}\right)^2=\left(\frac{620(1-R)^2}{\lambda R}\right)^2.
\end{equation}

\begin{figure}[h!]
\centering
\begin{subfigure}[b]{0.49\textwidth} 
\includegraphics[width=\textwidth]{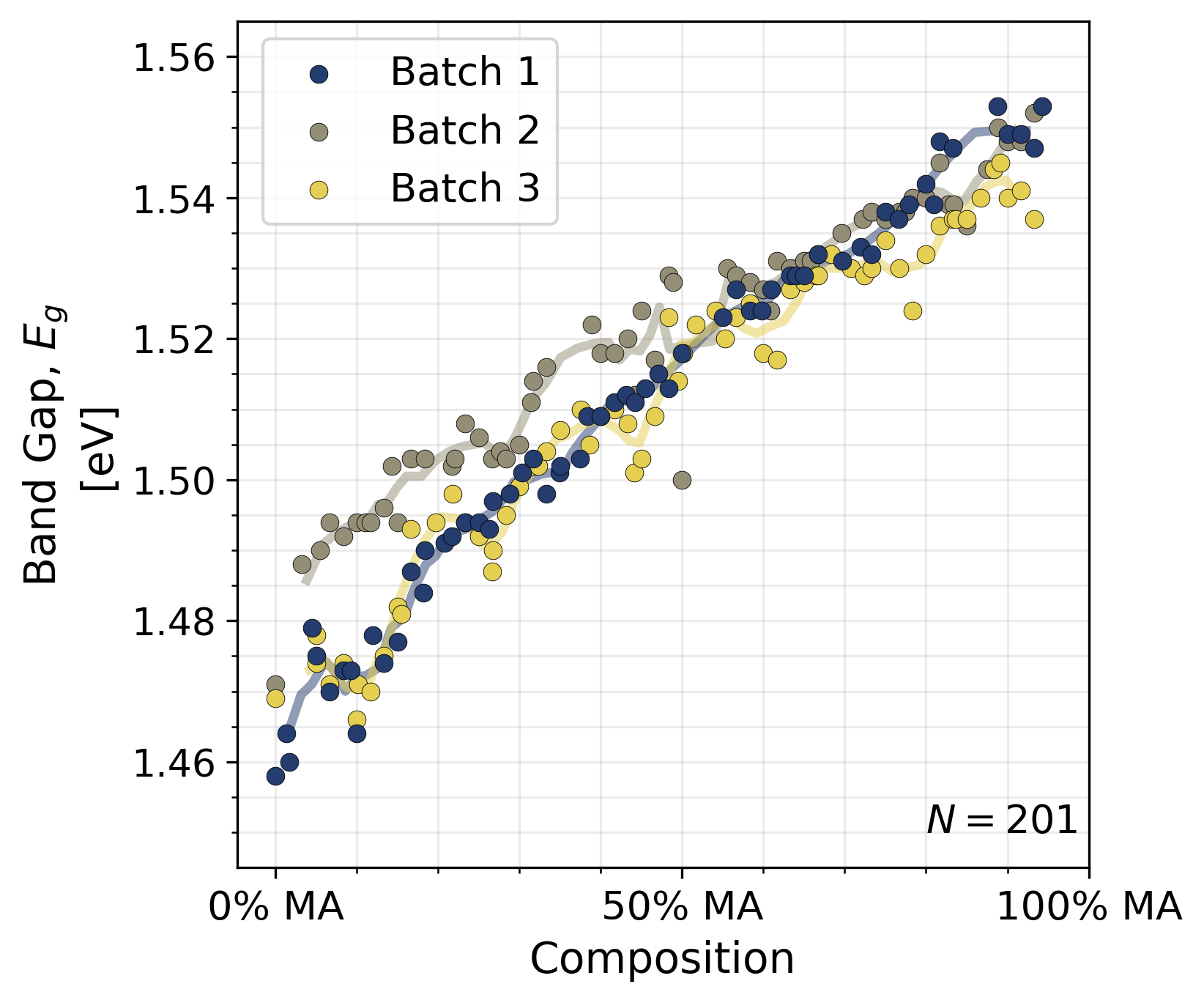}
\caption{Algorithm Batched Output}
\end{subfigure}
\begin{subfigure}[b]{0.49\textwidth} 
\includegraphics[width=\textwidth]{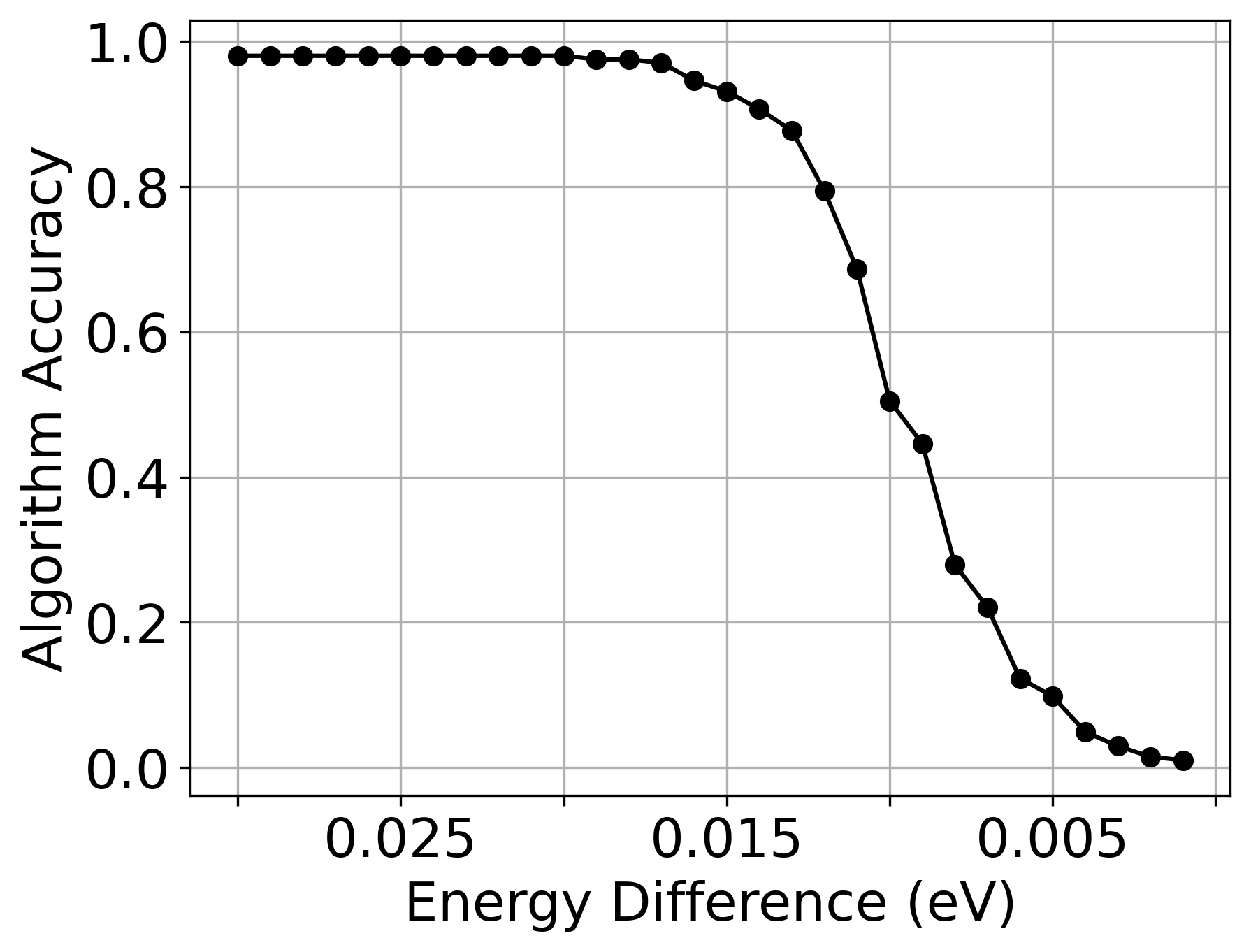}
\caption{Algorithm Accuracy}
\end{subfigure}
   \caption{(a) Direct band gap values output by the automatic band gap extraction algorithm, split by batch, as a function of FA$_{1-x}$MA$_{x}$PbI$_3$ composition. (b) Automatic band gap extraction algorithm accuracy as a function of the maximum allowable difference in energy between the domain expert-calculated and automatically calculated band gaps.}
\label{sfig:bandgap-acc}
\end{figure}

In this paper, we use the theory formulated above from optics to automatically compute the median band gap across all vision-segmented pixels $(\widehat{X},\widehat{Y})$ for a given perovskite sample, as described by the autocharacterization algorithm illustrated in Figure \ref{fig:bandgap}. Currently, the autocharacterization algorithm is configured to operate only on materials with a single direct band gap. The band gaps computed using the automatic Tauc segmentation and regression RMSE minimization processes employed by the autocharacterization algorithms are benchmarked against the band gaps calculated manually by a domain expert. This band gap comparison between algorithm and expert is used to determine an accuracy metric for the algorithm, assuming the expert-calculated output as ground truth. The accuracy is calculated by taking the average of a binary $0/1$ for all $N=201$ samples, determined based on whether or not the differences between the automatic and the expert band gap values fall within a specified energy difference threshold (shown along the $x$-axis of Figure \ref{sfig:bandgap-acc}b). 

% \begin{figure}[h!]
% \begin{center}
% \includegraphics[width=0.4\columnwidth]{figs/bandgap_extractor_acc_curve_R1.png}
% \end{center}
%    \caption{Automatic band gap extraction algorithm accuracy as a function of the maximum allowable difference in energy between the domain expert-calculated and automatically calculated band gaps.}
% \label{sfig:bandgap-acc}
% \end{figure}

Figure \ref{sfig:bandgap-acc}a illustrates the autocharacterization output band gaps as a function of composition for the three independent batches. Figure \ref{sfig:bandgap-acc}b illustrates the accuracy of the automatic algorithm as a function of the energy difference threshold. The algorithm achieves 98.5\% accuracy within 0.02eV and as the threshold becomes tighter, the algorithm accuracy is expected to decrease.

\bigbreak

\noindent \textbf{Stability}

\begin{figure}[h!]
\centering
\begin{subfigure}[b]{0.42\textwidth} 
\includegraphics[width=\textwidth]{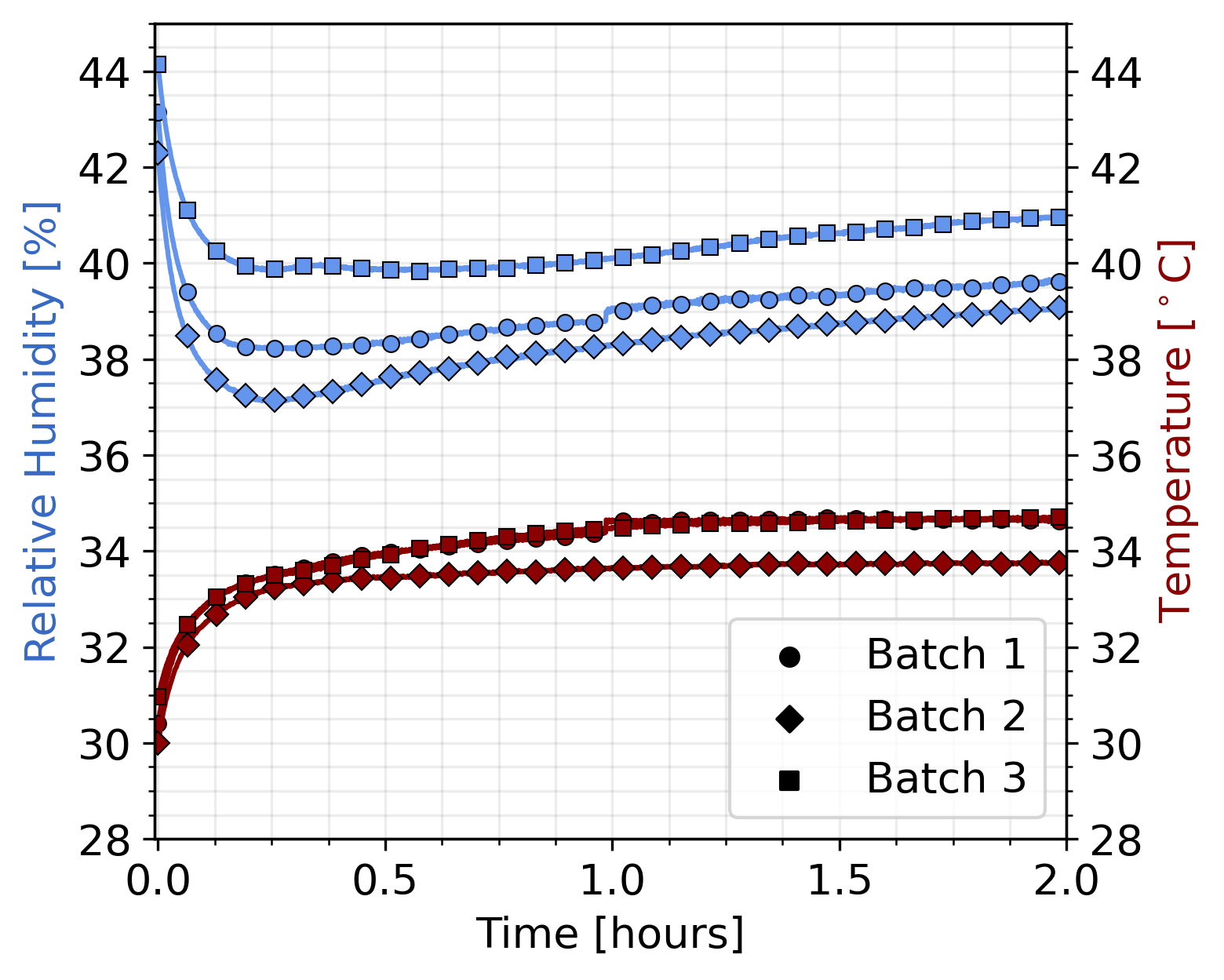}
\caption{Temporal Degradation Conditions}
\end{subfigure}
\begin{subfigure}[b]{0.53\textwidth} 
\includegraphics[width=\textwidth]{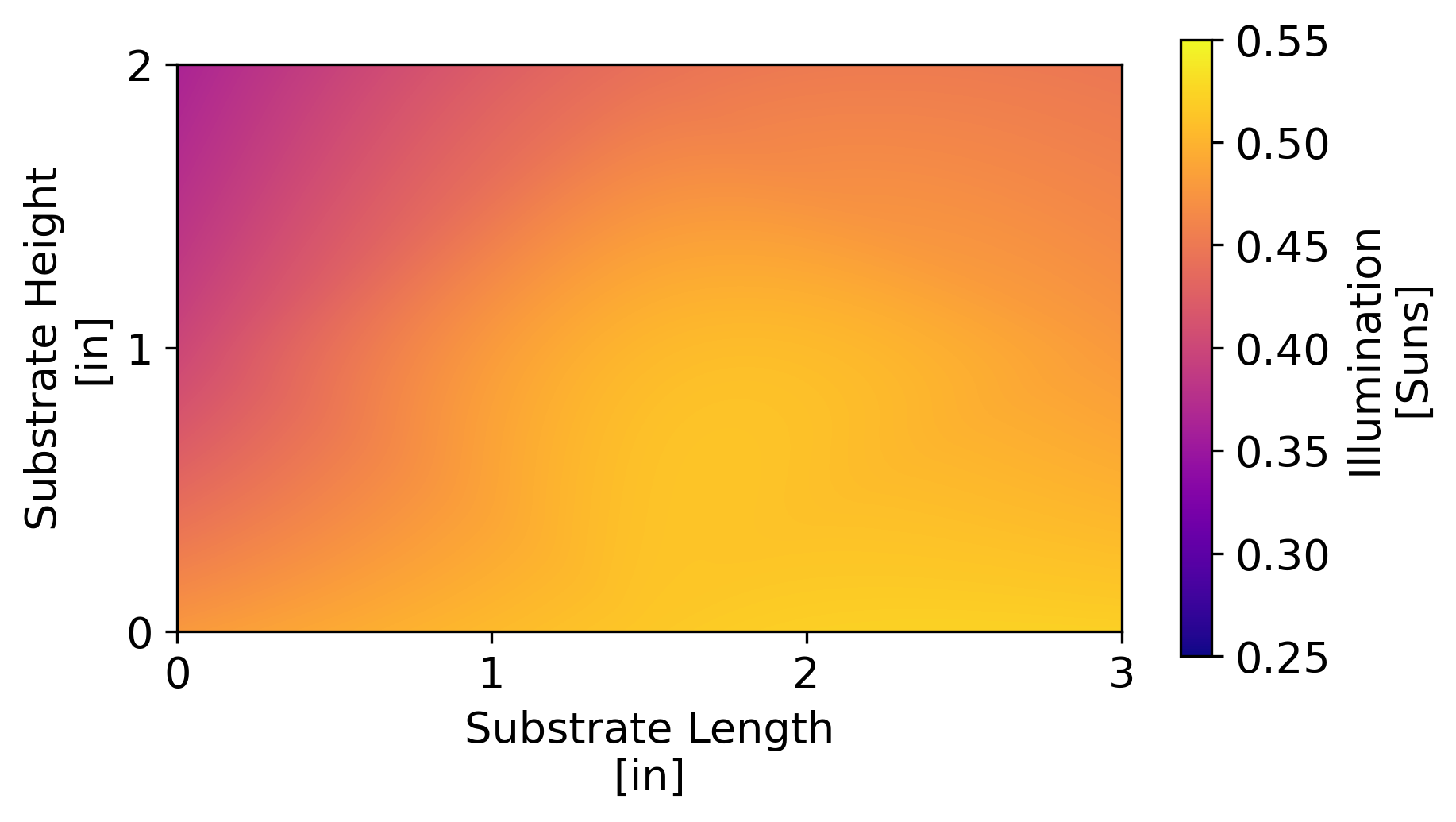}
\caption{Fixed Degradation Illumination}
\end{subfigure}
\caption{(a) Temporal degradation conditions for each of the 3 batches of samples over the course of 2 hours, measured using a temperature and humidity sensor (Adafruit, AHT10). (b) Spatial uniformity of illumination measured across the substrate within the degradation chamber, measured using a lux sensor (Adafruit, VEML7700).}
\label{sfig:deg}
\end{figure}

\begin{figure}[h!]
\begin{center}
\includegraphics[width=1\columnwidth]{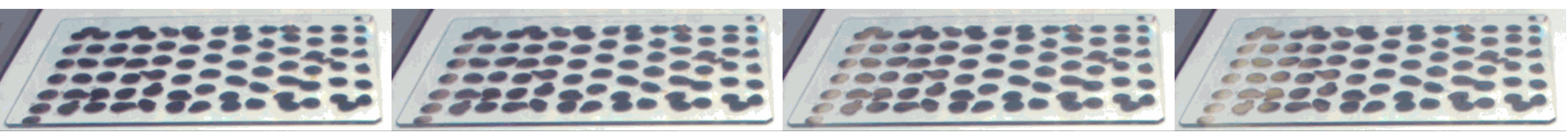}
\end{center}
   \caption{Degradation of the high-throughput manufactured FA$_{1-x}$MA$_{x}$PbI$_3$ compositional series over 2 hours.}
\label{sfig:sequence}
\end{figure}

To conduct the degradation experiments in this paper, we put the samples in a degradation chamber and monitor the conditions for 2 hours, capturing RGB images (Thorlabs DCC1645C camera with the infrared filter removed to increase sensitivity towards dark samples) every 30 seconds. The construction and operation of the degradation chamber setup are detailed in Keesey \& Tiihonen \textit{et al.} \citeSM{keesey_tiihonen_siemenn_colburn_sun_hartono_serdy_zeile_he_gurtner_et}. Figure \ref{sfig:deg}(a) shows the time series of temperature and humidity conditions over the course of the experiment. Over the 2-hour degradation experiment, the temperature conditions were maintained at $34.5^\circ \mathrm{C} \pm 0.5^\circ \mathrm{C}$ with a relative humidity of $40\% \pm 1\%$. An initial jump in temperature with a respective dip in humidity is noted as the samples are placed into the degradation chamber before the internal environment equilibrates. A class AAA solar simulator (G2V Optics Sunbrick Base, with visible-only part of the AM1.5G spectrum) is used to illuminate the 3"$\times$2" substrate of samples during the 2-hour degradation experiment. Figure \ref{sfig:deg}(b) shows the spatial uniformity of the illumination across the substrate internal to the degradation chamber, measured using a lux sensor (Adafruit, VEML7700). These illumination conditions are held fixed over the course of the experiment. Most of the substrate experiences 0.50-0.55 suns of illumination, however, in the top-left corner of the substrate, a dip in illumination occurs as a result of minor occlusion due to sensor placement. Figure \ref{sfig:sequence} illustrates that this dip in illumination at the top-left corner of the substrate does not have a major effect on the degradation pattern of the FA$_{1-x}$MA$_{x}$PbI$_3$ compositional series. Degradation, as indicated by the yellowing, is shown to begin in the formamidinium (FA)-rich end of the samples and migrates towards the methylammonium (MA)-rich end over time.

\begin{figure}[h!]

\centering
\begin{subfigure}[b]{1.1\textwidth} 
\includegraphics[width=\textwidth]{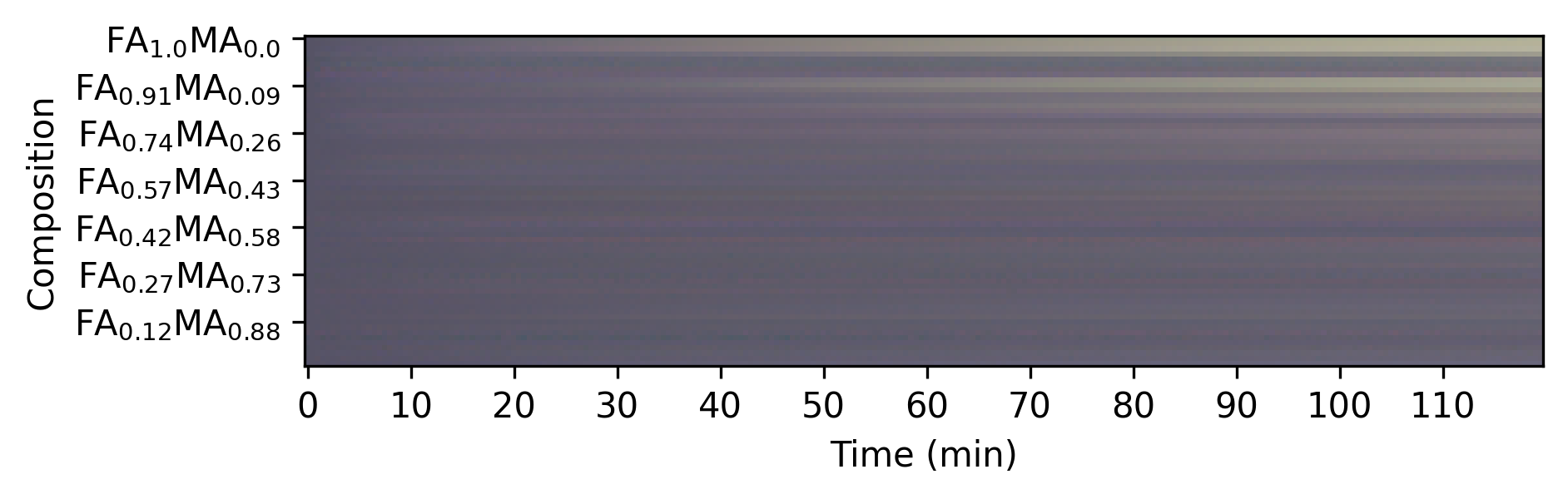}
\caption{Stability Time Series for Batch 1}
\end{subfigure}
\begin{subfigure}[b]{1.1\textwidth} 
\includegraphics[width=\textwidth]{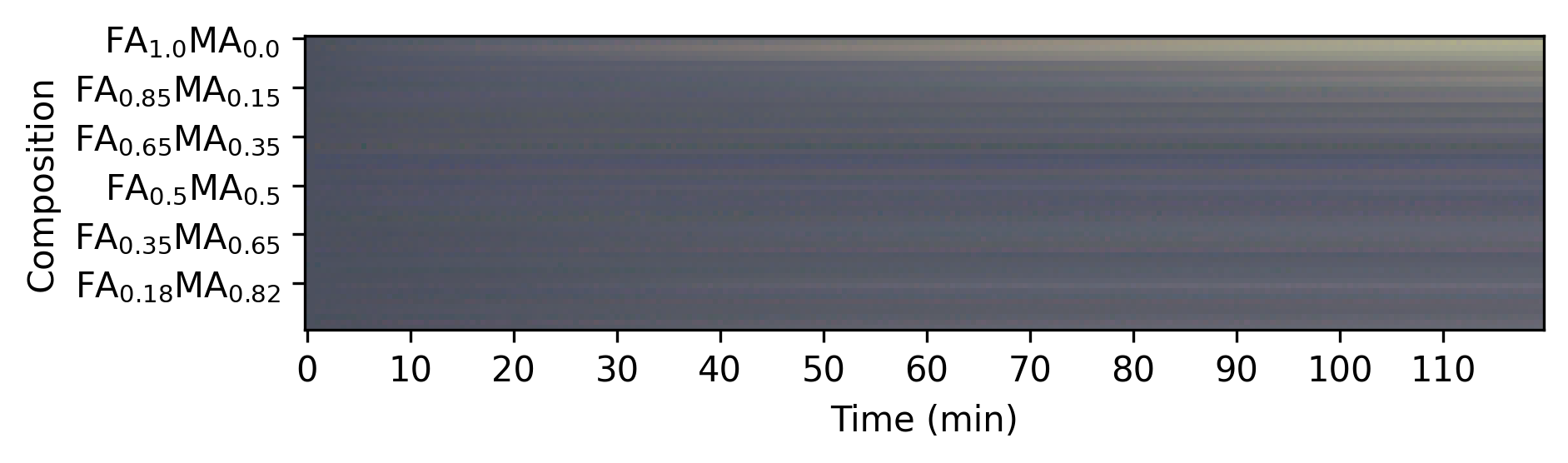}
\caption{Stability Time Series for Batch 2}
\end{subfigure}
\begin{subfigure}[b]{1.1\textwidth} 
\includegraphics[width=\textwidth]{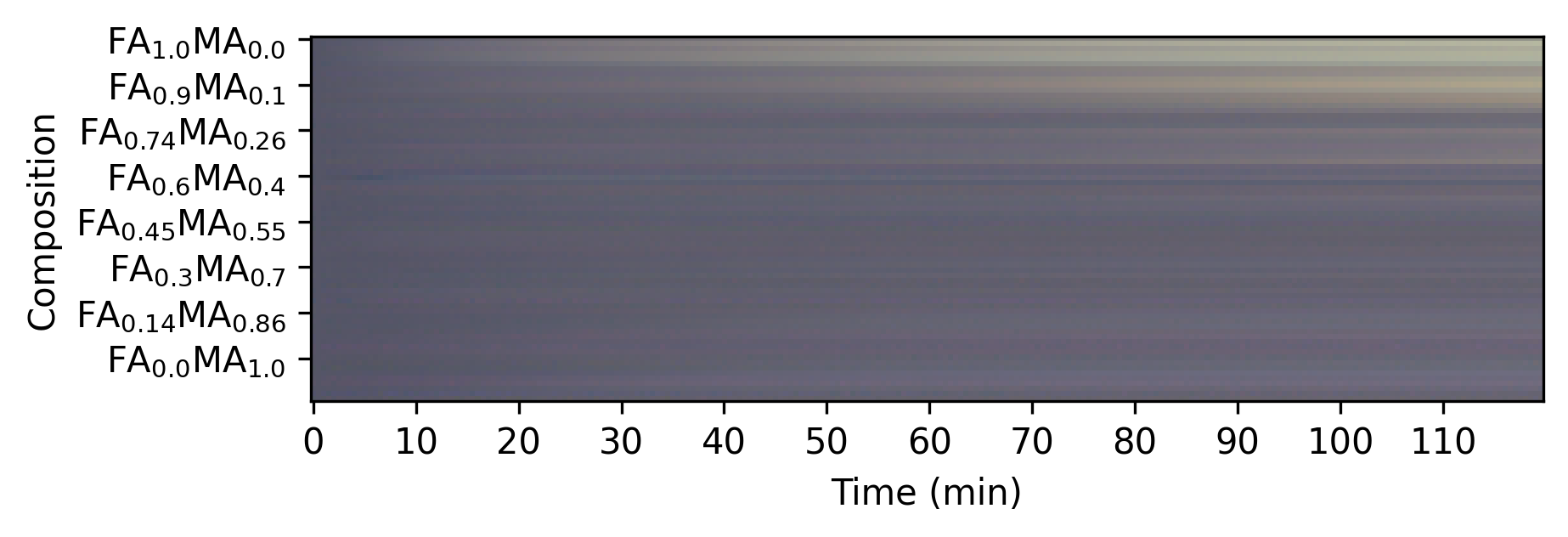}
\caption{Stability Time Series for Batch 3}
\end{subfigure}
   \caption{Final stability time series matrix of all batches of segmented droplets over the course of a 2-hour degradation experiment. All samples begin the experiment colored as dark gray and over the course of degradation, samples with MA proportions between 0\% to 20\% exhibit yellowing around the 20min to 40min mark, thus indicating degradation.}
   \label{sfig:degmatrix}
\end{figure}

After segmenting each sample across the entire time domain of the experiment, the full matrix of degradation time series is populated and color calibrated automatically, as shown in Figure \ref{sfig:degmatrix}. In this specific experiment, spatial non-uniformity of reflected surfaces was detected in the post-analysis, which does not noticeably affect the instability index calculation but does give rise to artificial color differences in the samples, depending on their location on the substrate. These spatially-dependent color differences arise due to the physical configuration of the environmental chamber and the RGB camera. To account for these color differences in the final output matrix time series, an additional color correction step was applied that initializes all deposited samples to the same color. This color correction is only cosmetic and was not used in the calculations for determining the degradation intensity, $I_c$. Color correction aids in the interpretation of the visualized time series data by making the color changes due to degradation easier to see while diminishing the unwanted effects of spatially-dependent reflectivity aberrations. After the color correction, a fully calibrated matrix of degradation time-series is acquired, showing the color change over time for each perovskite sample within its batch, shown in Figure \ref{sfig:degmatrix}.

\begin{figure}[h!]
\centering
\begin{subfigure}[b]{0.55\textwidth} 
\includegraphics[width=\textwidth]{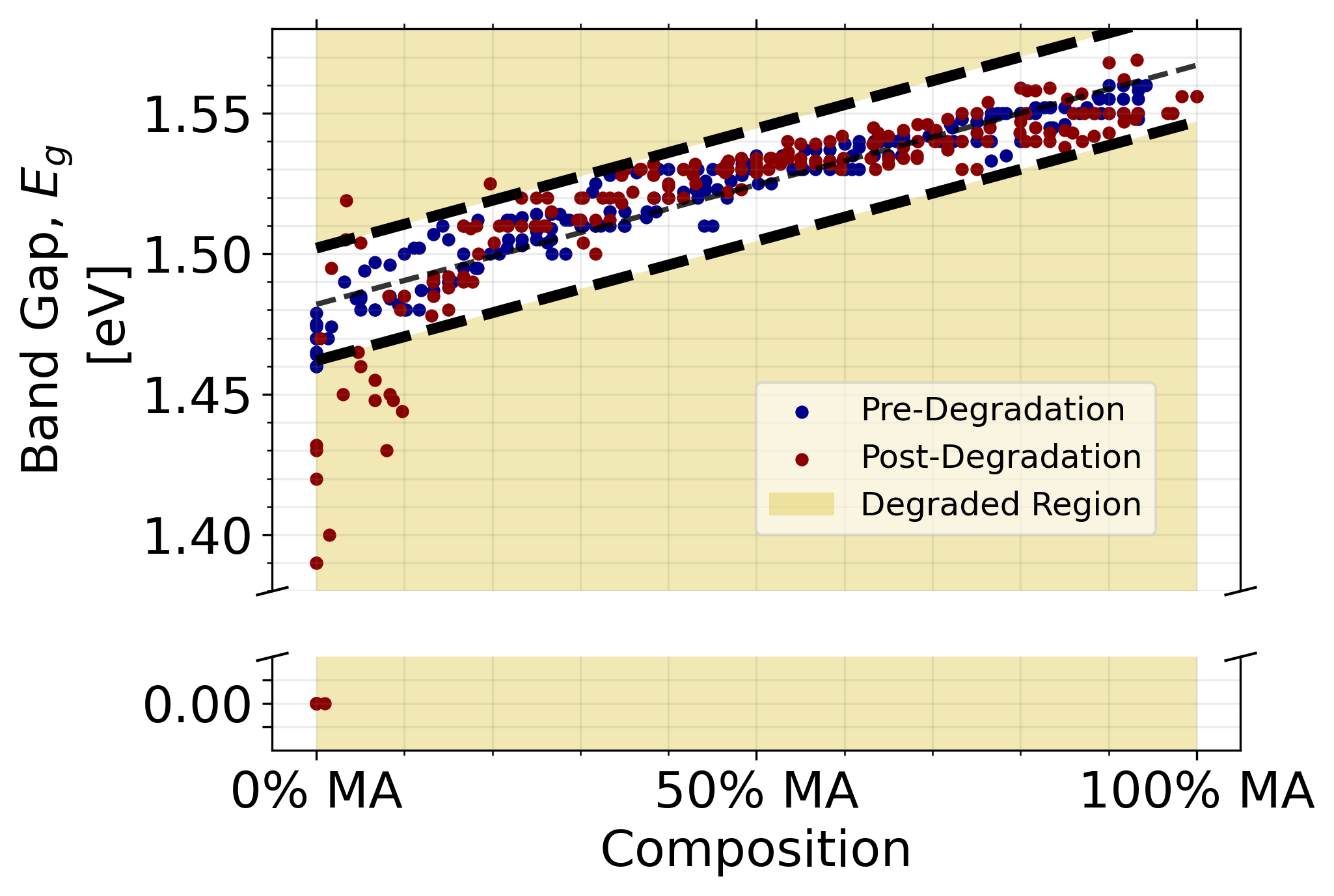}
\caption{Ground Truth Determination}
\end{subfigure}
\begin{subfigure}[b]{0.44\textwidth} 
\includegraphics[width=\textwidth]{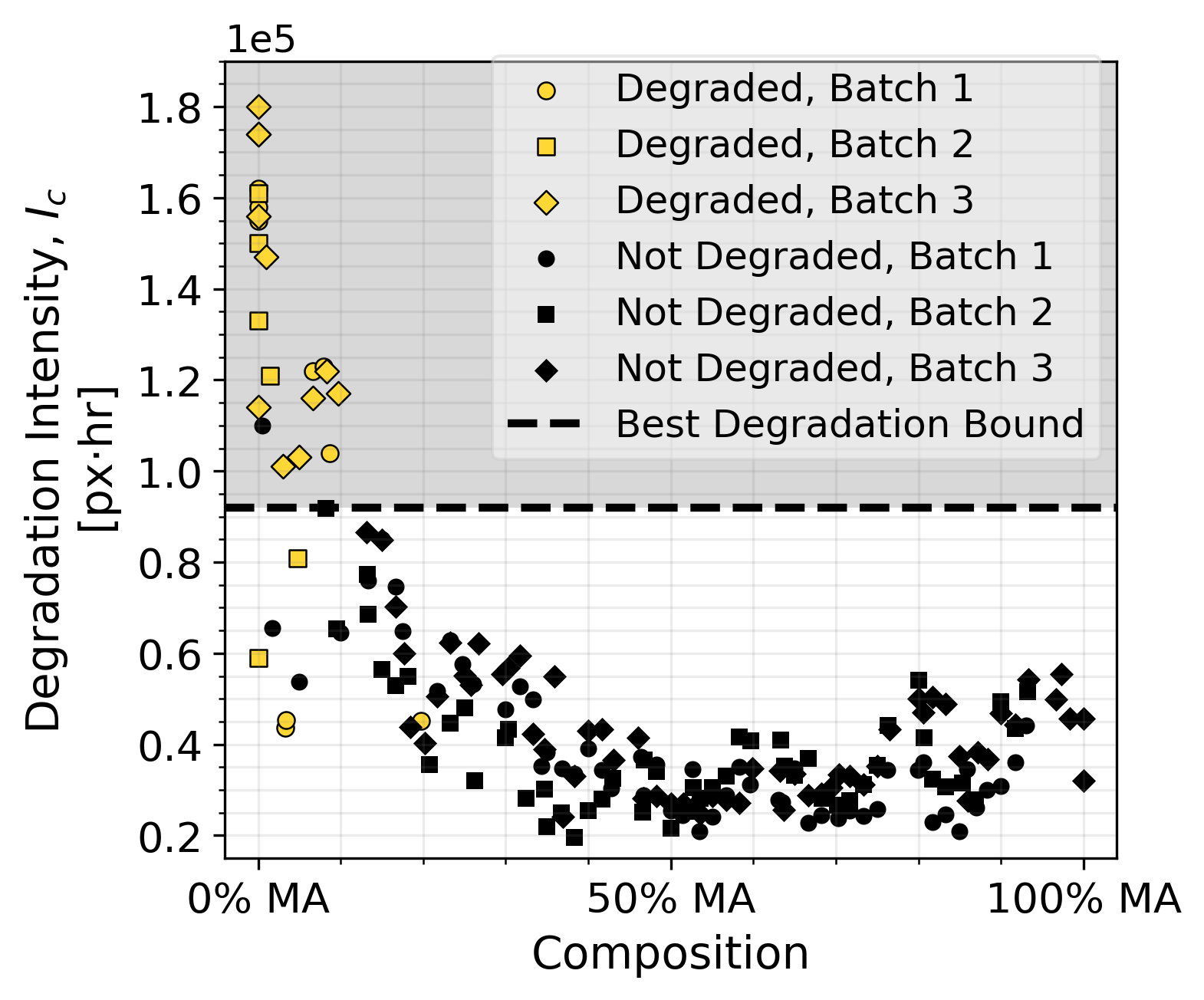}
\caption{Algorithm Extracted Degradation}
\end{subfigure}
\\

\begin{subfigure}[b]{0.49\textwidth} 
\includegraphics[width=\textwidth]{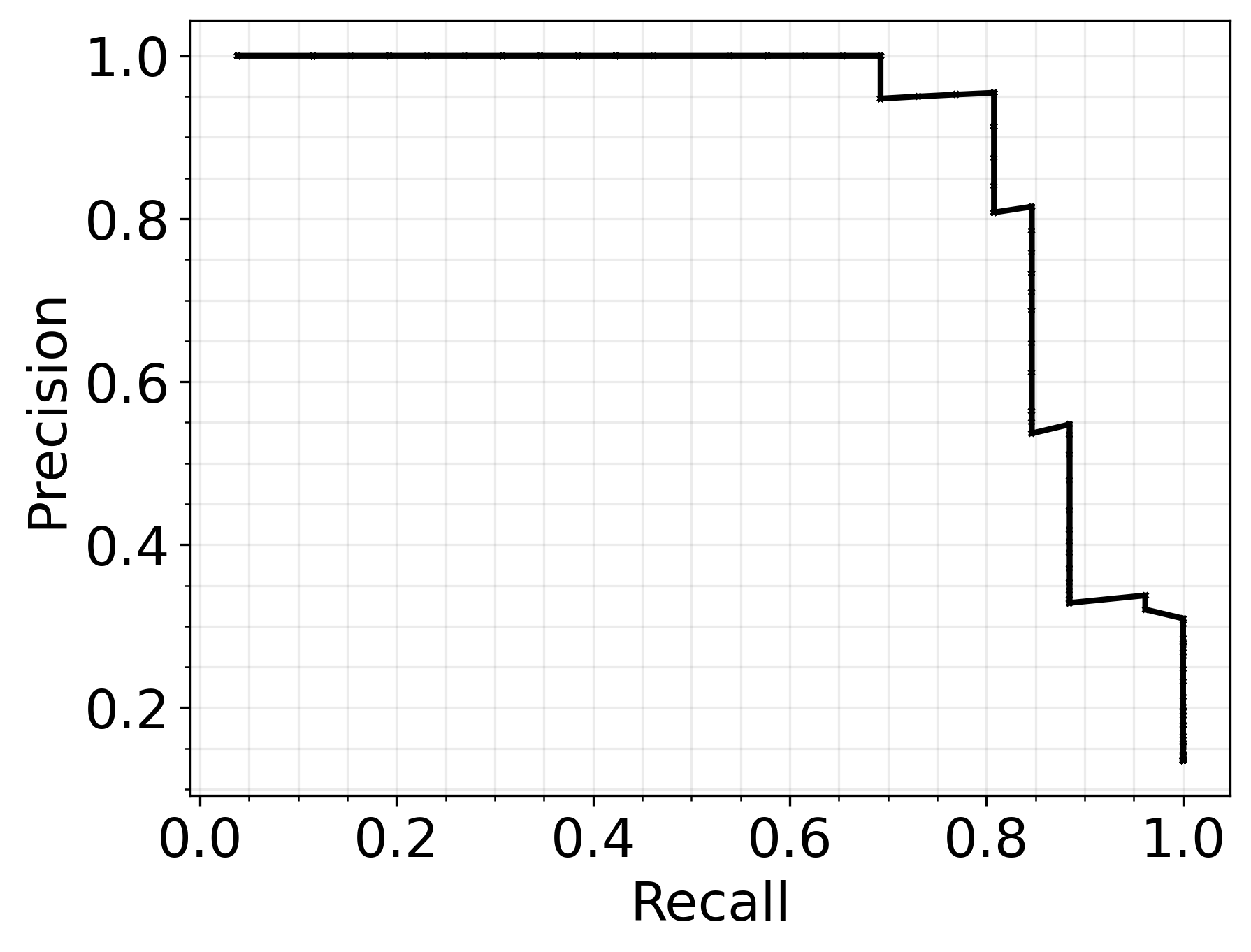}
\caption{Algorithm Precision-Recall}
\end{subfigure}
\begin{subfigure}[b]{0.49\textwidth} 
\includegraphics[width=\textwidth]{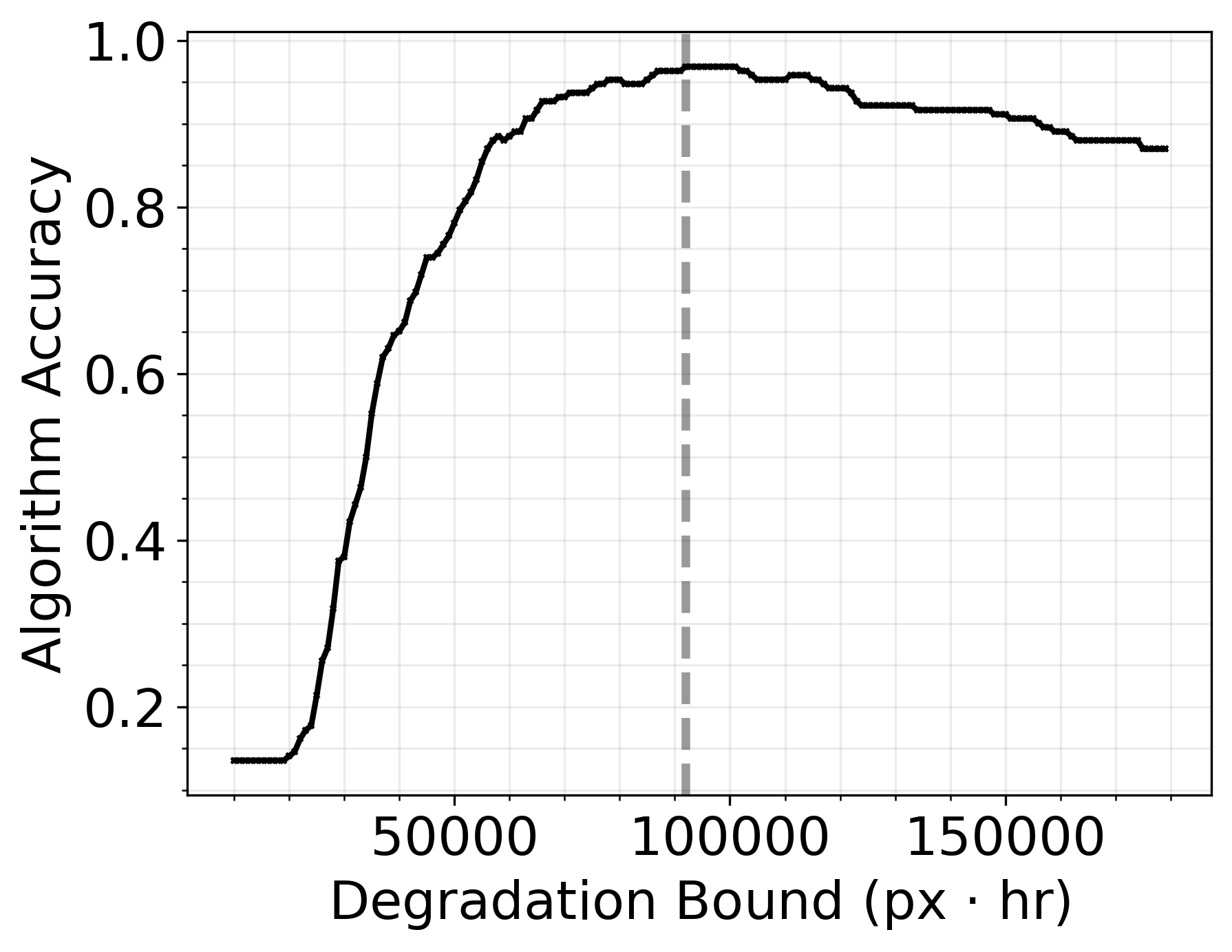}
\caption{Algorithm Accuracy}
\end{subfigure}
   \caption{Automatic degradation detection algorithm performance benchmarking. (a) Shows how the ground truth degradation is determined using post-degradation band gap measurement. The break in the graph is used to visualize which compositions had no band gap after degradation, reported as 0.0eV. (b) Values of $I_c$ as a function of composition, split by batch with a total of $N=201$ samples. $I_c$ is used as a classifier for degradation, where the dashed line indicates the separation of high $I_c$ versus low $I_c$. (c) Precision-recall performance of $I_c$ as a classifier for degradation based on the classification rate of false negatives and false positives. (d) Accuracy of $I_c$ as a classifier for determining high and low degradation. The $x$-axis indicates the decision boundary value of $I_c$, where values above are considered degraded and values below are not. The optimal value of $I_c$ as a decision boundary for degradation is $I_c = 0.92\times10^5$px$\cdot$hr, and is shown as a dashed vertical line. This value is where the accuracy of the algorithm is maximum at 96.9\%.}
   \label{sfig:deg-acc}
\end{figure}
% \vspace{50px}

The degradation intensity, $I_c$, is computed from the color-changing time series per sample using Equation \ref{eq:5} \citeSM{Sun2021, keesey_tiihonen_siemenn_colburn_sun_hartono_serdy_zeile_he_gurtner_et}. The degradation detection algorithm automatically computes $I_c$ for every sample. High values of $I_c$ correspond to high degradation. To benchmark the algorithm, the difference in domain expert-computed band gaps before and after degradation is used as a ground truth to quantify whether the material has truly degraded or not. This is possible due to the change in band gap that occurs in perovskites during either a phase transition or chemical decomposition \citeSM{Stoumpos2017, Sun2021}. Hence, in Figure \ref{sfig:deg-acc}(a), any expert-calculated post-degraded band gaps (red points) samples that fall outside of the $\pm0.02$eV bounds, with respect to the regression fit line to the expert-calculated pre-degradation band gaps (blue points), are considered to exhibit ``Ground Truth Degradation". These ground truth-degraded samples are denoted by yellow scatter points in Figure \ref{sfig:deg-acc}(b). The bounds of $\pm0.02$eV are used because they empirically fit the expert-calculated pre-degradation band gaps (blue points) with little to no tolerance. Thus, the region where ``Ground Truth Degradation" occurs in the post-degraded samples is indicated by the yellow shaded region in Figure \ref{sfig:deg-acc}(a). 

\clearpage
\newpage

Figure \ref{sfig:deg-acc}(b) illustrates that the magnitude of $I_c$ strongly corresponds with the ground truth determination of degradation using band gap difference as a metric. This correspondence can be quantified using the precision-recall (PR) of the autocharacterization algorithm. A PR curve quantifies the performance of using a classifier, in this case, $I_c$, to predict a ground truth, in this case, degradation:
\begin{equation}
\begin{split}
    \mathrm{Recall} &= \frac{TP}{TP+FN}\\
    \mathrm{Precision} &= \frac{TP}{TP+FP},
\end{split}
\end{equation}
where $TP$ are the true positives, $FN$ are the false negatives, $FP$ are the false positives. We use the PR curve instead of the ROC (receiver operating characteristic) curve here due to the large class imbalance between the number of degraded samples versus non-degraded samples (there are significantly more non-degraded samples than there are degraded samples). 

Figure \ref{sfig:deg-acc}c illustrates the PR curve of the automatic degradation detection algorithm based on the degradation decision boundary (horizontal black dashed line in Figure \ref{sfig:deg-acc}b). The goal is to have both high precision and high recall simultaneously. The PR-AUC (precision-recall area under the curve) figure of merit boils the PR curve down to a single number that determines the performance of $I_c$ as a good predictor for degradation. The value of PR-AUC falls between $0.0$ and $1.0$, where a value of $1.0$ represents perfect performance. The $I_c$ values computed by the autocharacterization algorithm achieve a PR-AUC of $0.853 \in [0,1]$, implying that high values of $I_c$ do strongly correspond to ground truth degradation. Figure \ref{sfig:deg-acc}d shows the effect of moving the decision boundary on the accuracy in detecting the degradation. Considering recall, precision, and accuracy, $I_c$ performs optimally, with an accuracy of 96.9\%, in detecting degraded samples when the decision boundary is set to  $0.92\times10^5$px$\cdot$hr. However, accuracies of over 90\% are achieved for a wide range of $I_c$ decision boundaries: $0.7\times10^5$px$\cdot$hr $\leq I_c \leq 1.6\times10^5$px$\cdot$hr. Hence, indicating that $I_c$ is a general yet strong predictor of degradation.

\bigbreak

\noindent \textbf{Phase and Elemental Analysis}

\begin{figure}[h!]
\centering
\begin{subfigure}[b]{0.48\textwidth} 
\includegraphics[width=\textwidth]{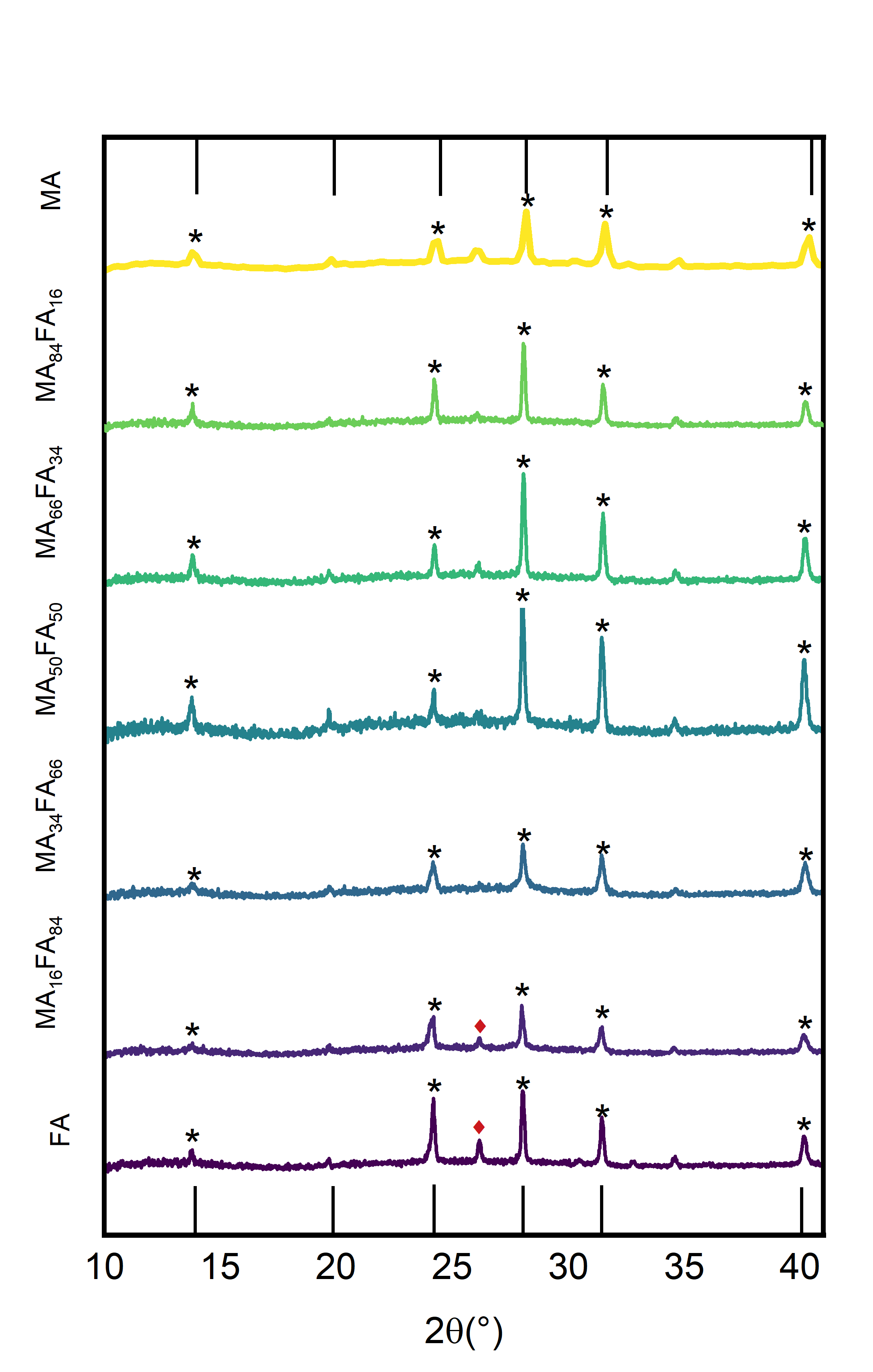}
\caption{Pre-degradation XRD}
\end{subfigure}
\begin{subfigure}[b]{0.492\textwidth} 
\includegraphics[width=\textwidth]{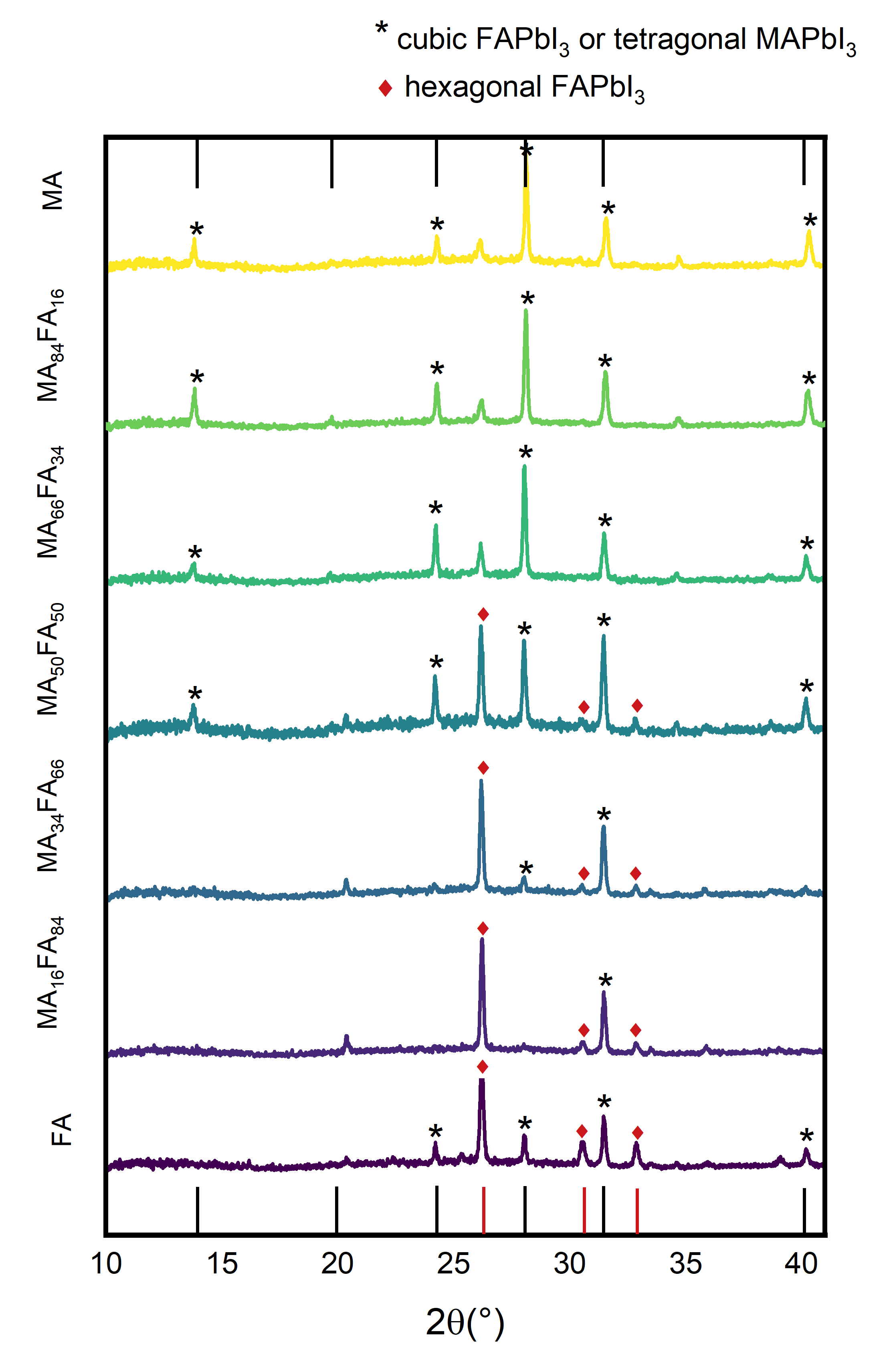}
\caption{Post-degradation XRD}
\end{subfigure}
\caption{XRD peak intensities for uniformly-spaced compositions along the FA$_{1-x}$MA$_{x}$PbI$_3$ series before and after degradation.}
\label{sfig:xrd}
\end{figure}

Phase analysis, such as X-ray diffraction (XRD), of perovskites is used to determine the structure and quality of the manufactured samples. In this study, we measure our samples using the Bruker X-ray Diffractometer with a Cobalt Source D8 and General Area Detector Diffraction System. Figure \ref{sfig:xrd} illustrates the pre- and post-degradation XRD traces for equally spaced compositions along the FA$_{1-x}$MA$_{x}$PbI$_3$ series. The reference peak locations for both the favorable cubic $\alpha$-FAPbI$_3$ \citeSM{Nan2021} and favorable tetragonal MAPbI$_3$ phases \citeSM{Wu2023} are shown together as black vertical lines and ``$*$" symbols since a shift of only $\Delta 2 \theta \approx 0.16^\circ$ is seen from FAPbI$_3$ to MAPbI$_3$ at around the $2\theta=31.5^\circ$ peak. High-resolution scans of this peak shift along FA$_{1-x}$MA$_{x}$PbI$_3$ are shown in Figure \ref{fig:integral}b and can be used as an additional validation tool for composition shift. During degradation, FAPbI$_3$ phase transitions from a favorable cubic $\alpha$-phase to a non-perovskite hexagonal $\delta$-phase \citeSM{Nan2021}. Hence, XRD is a useful validation tool for pre- and post-degradation determination. Prior to degradation, Figure \ref{sfig:xrd}a illustrates that the samples exhibit high adherence to their favorable phases with only minor contribution of $\delta$-FAPbI$_3$ phase. However, after the degradation test, Figure \ref{sfig:xrd}b illustrates high degradation (indicated by the high peak intensities of the $\delta$-FAPbI$_3$ phase at $2\theta=26.3 ^\circ$, denoted by the red diamond) in the FA-rich compositions. This XRD result matches the detected yellowing degradation pattern of the FA-rich samples, as shown in Figure \ref{sfig:sequence} and Figure \ref{sfig:degmatrix}.

\begin{figure}[h!]
\begin{center}
\includegraphics[width=0.5\columnwidth]{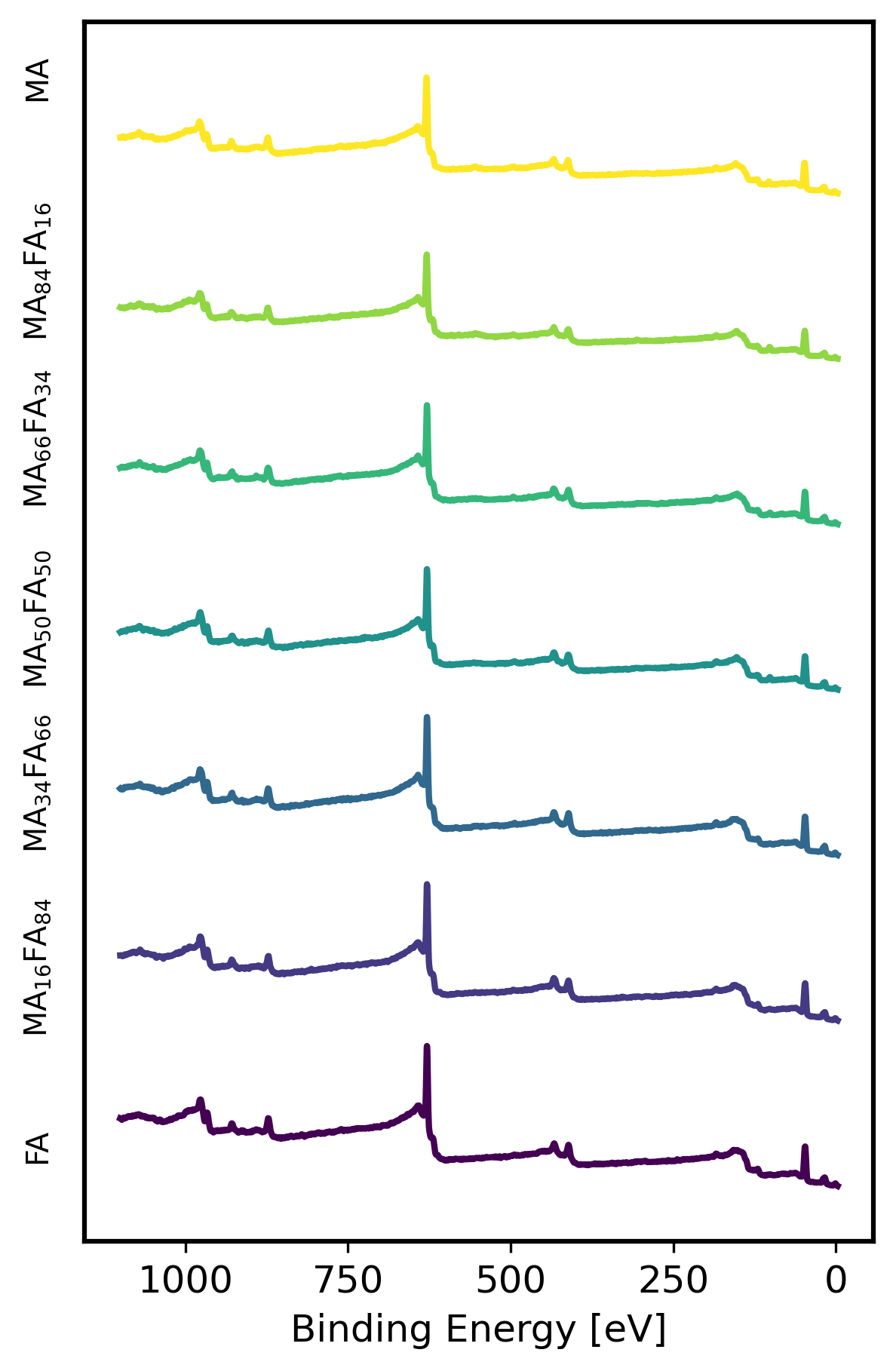}
\end{center}
   \caption{Full XPS spectra for uniformly-spaced compositions along the FA$_{1-x}$MA$_{x}$PbI$_3$ series.}
\label{sfig:xps}
\end{figure}

Elemental analysis, such as X-ray photoelectron spectroscopy (XPS), is used to determine the shift in the binding energy of bonds present within the different A-site cations (FA and MA) along the FA$_{1-x}$MA$_{x}$PbI$_3$ series, in turn corresponding to a ``composition" shift. XPS is a surface-sensitive quantitative spectroscopic technique that can identify crystalline phases. In this study, we measure our samples using the PHI 5000 Versa Probe II Focus X-ray Photoelectron Spectrometer, equipped with a monochromated AlK$\alpha$ X-ray source for excitation at 1486.6eV with an X-ray beam size of 200$\mu$m. Survey spectra of FAPbI$_3$ to MAPbI$_3$ series are depicted in Figure \ref{sfig:xps}. Using the survey spectra alone, there are variations in A-site composition that can not be distinguished clearly in survey spectra due to the many similarities between the FA and MA molecules \citeSM{Elsayed2023}. However, the primary distinguishing feature of these is the presence of the carbon-nitrogen double bond (C=N), clearly detectable in the high-resolution XPS scans, as shown in Figure \ref{fig:integral}c. In the high-resolution scans of C1s1 and N1s2, the calibration is performed on the lowest C1s energy peak of 284.8eV. Hence, the shift in C=N peak intensity quantifies the presence of FA relative to MA, in turn, determining composition along the FA$_{1-x}$MA$_{x}$PbI$_3$ series.

\begin{figure}[h!]
\centering
\begin{subfigure}[b]{0.465\textwidth} 
\includegraphics[width=\textwidth]{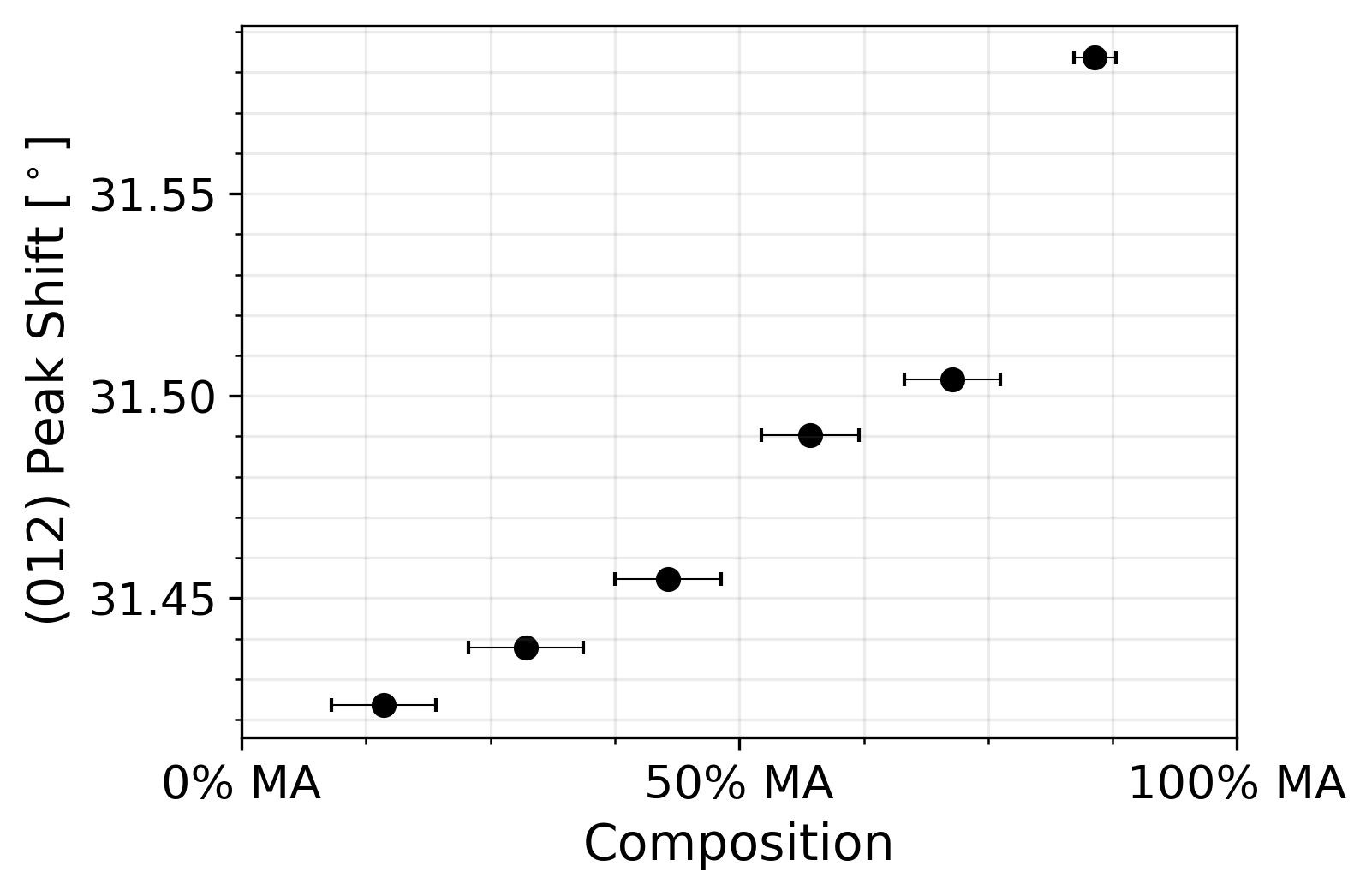}
\caption{XRD (012) Peak Angle Shift}
\end{subfigure}
\begin{subfigure}[b]{0.4\textwidth} 
\includegraphics[width=\textwidth]{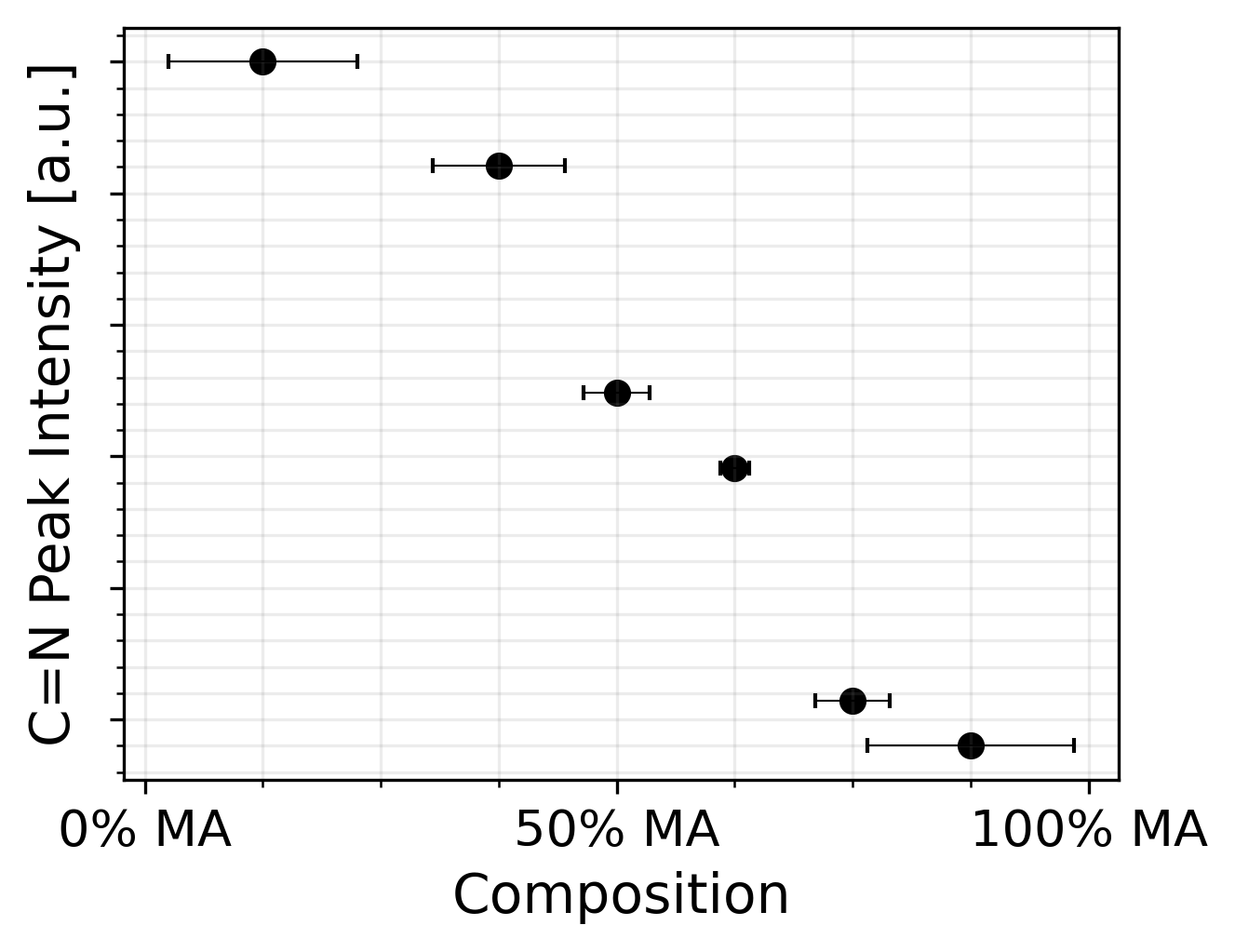}
\caption{XPS C=N Peak Intensity Shift}
\end{subfigure}
\caption{(a) XRD $2\theta$ values for the (012) crystallographic plane peak shift from FA-rich to MA-rich FA$_{1-x}$MA$_{x}$PbI$_3$ compositions. (b) XPS intensity shift for the C=N peak from FA-rich to MA-rich FA$_{1-x}$MA$_{x}$PbI$_3$ compositions. The horizontal error bars illustrate the relative widths of the XRD and XPS peaks.}
\vspace{-10px}
\label{sfig:quant-xrd-xps}
\end{figure}

Figure \ref{sfig:quant-xrd-xps} shows the quantitative XRD peak shifts and XPS peak intensities from the high-resolution scans for the phase and elemental shifts that occur along the FA$_{1-x}$MA$_{x}$PbI$_3$ series. The XRD peak of the (012) crystallographic plane shifts from lower to higher $2\theta$ angles as more MA is added to the composition. Conversely, the XPS peak for the presence of C=N bonds shifts from higher to lower intensity as more MA is added to the composition. Thus, both of these measurements validate the presence of a compositional gradient occurring across the synthesized batches of samples.\\ \\ \\

\bigbreak
\noindent \textbf{Size of the Perovskite Search Space}

\begin{figure}[h!]
\begin{center}
\includegraphics[width=0.95\columnwidth]{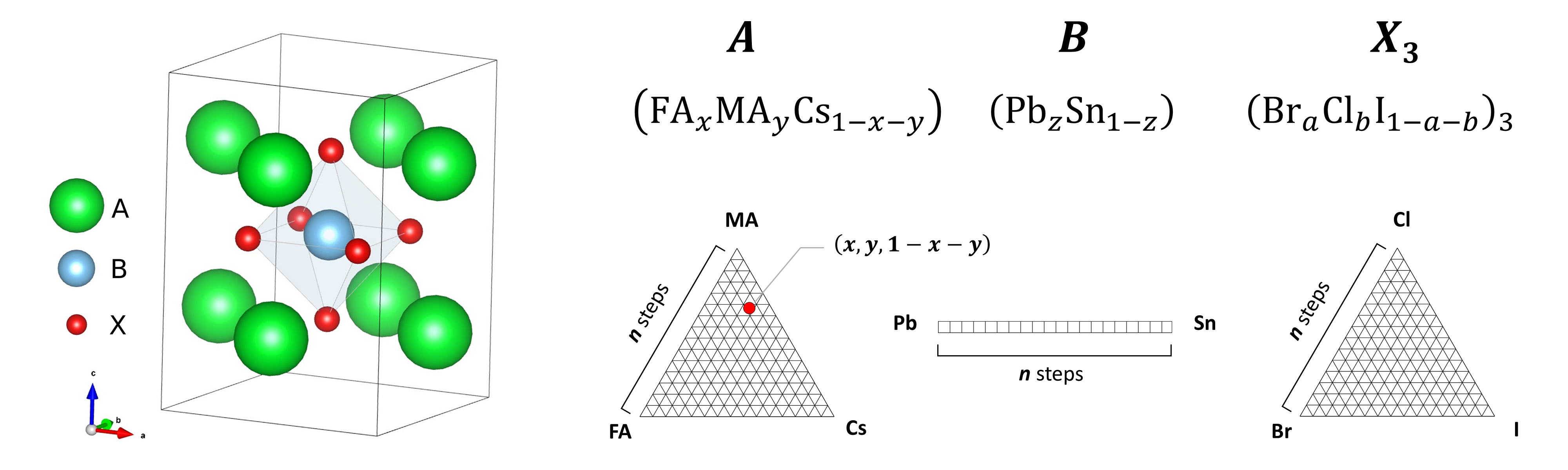}
\end{center}
   \caption{Archetypal metal halide perovskite compositional search space.}
\label{sfig:search}
\end{figure}

A commonly explored metal halide perovskite search space for photovoltaic applications from literature consists of the following eight-component material system: $(\textrm{FA}_x \textrm{MA}_y \textrm{Cs}_{1-x-y})(\textrm{Pb}_z \textrm{Sn}_{1-z}) (\textrm{Br}_a \textrm{Cl}_b \textrm{I}_{1-a-b})_3$ \citeSM{Wang2023, Ahmadi2021, Sun2021, Wang2017, Sun2019, Liu2023}. Figure \ref{sfig:search} shows the discretization of these eight components within the archetypal ABX$_3$ perovskite structure. The number of steps per edge, $n$, determines the compositional resolution for each subspace. As the number of steps increases, the number of potential compositions increases, and, in turn, the search space becomes more vast. For this eight-component search space, the number of possible compositions is proportional to the product of each subspace's (A, B, and X) step size to the power of the number of components within each subspace (here, 3-components for A (FA, MA, and Cs), 2-components for B (Pb and Sn), and 3-components for X (Br, Cl, and I)) $\implies n^3\times n\times n^3$ for the A$\times$B$\times$X subspaces. A caveat in this equation is that for binary subspaces, the power of the step size is 1 instead of 2 since the space is linear, as shown by the B-site subspace in Figure \ref{sfig:search}. Hence, for a low-resolution search space of $n=10$ steps, $1\times10^6$ total compositions are considered; and for a high-resolution search space of $n=100$ steps, $7\times10^{12}$ compositions are considered.

% \section{Characterization Methods}
% \subsection{Hyperspectral Imaging}
% Immediately after the samples are synthesized, the samples are removed from the glovebox and their hyperspectral images are taken using Resonon, Inc. Pika L camera. Figure \ref{sfig:hyperspectral} illustrates the reflectance trace of all $80$ samples. One of the sample glass slides is measured using XRD, and the other undergoes a degradation experiment and thereafter is measured using XRD post-degradation.

% \begin{figure}[h]
% \begin{center}
% \includegraphics[width=0.55\columnwidth]{figs/degrmatrix_no_clean-up.jpeg}
% \end{center}
%    \caption{Stability timeseries matrix of all segmented droplets over the course of a 6-hour degradation experiment. After color calibration and before spatial stratification. The two lighter points at approx. 110 minutes are droplets of water in the optical path of the camera and are excluded form the analysis as such.}
%    \label{sfig:degmatrix_no_clean-up}
% \end{figure}

% \begin{center}
% \includegraphics[width=0.55\columnwidth]{figs/stability_timeseries_matrix_R1.png}
% \end{center}

\section*{Full Experimental Results}

Table \ref{table4} contains the full readout of the characterization results extracted by the autocharacterization algorithms for all $N=201$ samples. We report the numerical values of calculated composition, autocharacterization-calculated band gap (Auto $E_g$), human domain expert-calculated band gap (Expert $E_g$), autocharacterization-calculated degree of degradation ($I_c$), and the ground truth degradation determined by the human domain expert. Dashes indicate values unable to be determined for that composition due to missing data.

\begin{longtable}{cccccc}
\caption{Results readout for all perovskite samples produced in this study.} \\
\toprule
Sample & Computed Composition & Auto $E_g$ & Expert $E_g$ & $I_c$ & Degradation \\
 &  & (eV) & (eV) & (px$\cdot$hr)$\times10^5$ & (Ground Truth) \\
\midrule
\endfirsthead
\caption{Continued from previous page.} \\
\toprule
Sample & Computed Composition & Auto $E_g$ & Expert $E_g$ & $I_c$ & Degradation \\
 &  & (eV) & (eV) & (px$\cdot$hr)$\times10^5$ & (Ground Truth) \\
\midrule
\endhead
\bottomrule
\endfoot
1      &  FA$_{1.000}$MA$_{0.000}$PbI$_3$ &         1.471 &           1.460 &  1.581 &      Yes \\
2      &  FA$_{1.000}$MA$_{0.000}$PbI$_3$ &         1.463 &           1.474 &  1.737 &      Yes \\
3      &  FA$_{1.000}$MA$_{0.000}$PbI$_3$ &         1.463 &           1.475 &  1.562 &      Yes \\
4      &  FA$_{1.000}$MA$_{0.000}$PbI$_3$ &         1.467 &           1.479 &  1.799 &      Yes \\
5      &  FA$_{1.000}$MA$_{0.000}$PbI$_3$ &         1.458 &           1.464 &  1.335 &      Yes \\
6      &  FA$_{1.000}$MA$_{0.000}$PbI$_3$ &         1.469 &           1.470 &  1.739 &      Yes \\
7      &  FA$_{1.000}$MA$_{0.000}$PbI$_3$ &         1.454 &           1.460 &  1.465 &      Yes \\
8      &  FA$_{1.000}$MA$_{0.000}$PbI$_3$ &         1.455 &           1.470 &      - &      Yes \\
9      &  FA$_{1.000}$MA$_{0.000}$PbI$_3$ &         1.450 &           1.460 &      - &      Yes \\
10     &  FA$_{1.000}$MA$_{0.000}$PbI$_3$ &         1.457 &           1.470 &      - &      Yes \\
11     &  FA$_{1.000}$MA$_{0.000}$PbI$_3$ &         1.457 &           1.465 &  1.607 &      Yes \\
12     &  FA$_{1.000}$MA$_{0.000}$PbI$_3$ &         1.475 &           1.475 &  1.501 &       No \\
13     &  FA$_{1.000}$MA$_{0.000}$PbI$_3$ &         1.473 &           1.470 &  1.098 &      Yes \\
14     &  FA$_{1.000}$MA$_{0.000}$PbI$_3$ &         1.471 &           1.470 &  1.623 &      Yes \\
15     &  FA$_{0.987}$MA$_{0.013}$PbI$_3$ &         1.464 &           1.470 &  1.209 &      Yes \\
16     &  FA$_{0.983}$MA$_{0.017}$PbI$_3$ &         1.460 &           1.474 &  0.655 &       No \\
17     &  FA$_{0.968}$MA$_{0.032}$PbI$_3$ &         1.488 &           1.490 &  0.436 &      Yes \\
18     &  FA$_{0.955}$MA$_{0.045}$PbI$_3$ &         1.479 &           1.484 &  0.453 &      Yes \\
19     &  FA$_{0.950}$MA$_{0.050}$PbI$_3$ &         1.474 &           1.480 &  0.809 &      Yes \\
20     &  FA$_{0.950}$MA$_{0.050}$PbI$_3$ &         1.475 &           1.484 &  0.538 &       No \\
21     &  FA$_{0.950}$MA$_{0.050}$PbI$_3$ &         1.478 &           1.485 &  1.012 &      Yes \\
22     &  FA$_{0.945}$MA$_{0.055}$PbI$_3$ &         1.490 &           1.494 &  1.034 &      Yes \\
23     &  FA$_{0.933}$MA$_{0.067}$PbI$_3$ &         1.471 &           1.480 &  1.215 &      Yes \\
24     &  FA$_{0.933}$MA$_{0.067}$PbI$_3$ &         1.470 &           1.480 &  1.165 &      Yes \\
25     &  FA$_{0.933}$MA$_{0.067}$PbI$_3$ &         1.494 &           1.497 &  1.223 &      Yes \\
26     &  FA$_{0.917}$MA$_{0.083}$PbI$_3$ &         1.473 &           1.484 &  1.040 &      Yes \\
27     &  FA$_{0.917}$MA$_{0.083}$PbI$_3$ &         1.474 &           1.485 &  0.918 &       No \\
28     &  FA$_{0.917}$MA$_{0.083}$PbI$_3$ &         1.492 &           1.496 &  1.231 &      Yes \\
29     &  FA$_{0.907}$MA$_{0.093}$PbI$_3$ &         1.473 &           1.482 &  0.652 &       No \\
30     &  FA$_{0.900}$MA$_{0.100}$PbI$_3$ &         1.464 &           1.480 &  1.170 &      Yes \\
31     &  FA$_{0.900}$MA$_{0.100}$PbI$_3$ &         1.466 &           1.485 &      - &        - \\
32     &  FA$_{0.900}$MA$_{0.100}$PbI$_3$ &         1.494 &           1.500 &  0.645 &       No \\
33     &  FA$_{0.898}$MA$_{0.102}$PbI$_3$ &         1.471 &           1.480 &      - &        - \\
34     &  FA$_{0.889}$MA$_{0.111}$PbI$_3$ &         1.494 &           1.502 &      - &        - \\
35     &  FA$_{0.883}$MA$_{0.117}$PbI$_3$ &         1.470 &           1.480 &      - &        - \\
36     &  FA$_{0.883}$MA$_{0.117}$PbI$_3$ &         1.494 &           1.502 &      - &        - \\
37     &  FA$_{0.881}$MA$_{0.119}$PbI$_3$ &         1.478 &           1.487 &      - &       No \\
38     &  FA$_{0.867}$MA$_{0.133}$PbI$_3$ &         1.496 &           1.507 &  0.759 &       No \\
39     &  FA$_{0.867}$MA$_{0.133}$PbI$_3$ &         1.474 &           1.490 &  0.773 &       No \\
40     &  FA$_{0.867}$MA$_{0.133}$PbI$_3$ &         1.475 &           1.487 &  0.684 &       No \\
41     &  FA$_{0.857}$MA$_{0.143}$PbI$_3$ &         1.502 &           1.510 &  0.866 &        - \\
42     &  FA$_{0.850}$MA$_{0.150}$PbI$_3$ &         1.494 &           1.505 &  0.851 &       No \\
43     &  FA$_{0.850}$MA$_{0.150}$PbI$_3$ &         1.477 &           1.490 &  0.565 &       No \\
44     &  FA$_{0.850}$MA$_{0.150}$PbI$_3$ &         1.482 &           1.490 &  0.849 &       No \\
45     &  FA$_{0.846}$MA$_{0.154}$PbI$_3$ &         1.481 &           1.490 &  0.438 &       No \\
46     &  FA$_{0.833}$MA$_{0.167}$PbI$_3$ &         1.487 &           1.495 &  0.529 &       No \\
47     &  FA$_{0.833}$MA$_{0.167}$PbI$_3$ &         1.493 &           1.500 &  0.703 &       No \\
48     &  FA$_{0.833}$MA$_{0.167}$PbI$_3$ &         1.503 &           1.510 &  0.747 &       No \\
49     &  FA$_{0.819}$MA$_{0.181}$PbI$_3$ &         1.484 &           1.495 &  0.649 &       No \\
50     &  FA$_{0.817}$MA$_{0.183}$PbI$_3$ &         1.503 &           1.512 &  0.549 &       No \\
51     &  FA$_{0.817}$MA$_{0.183}$PbI$_3$ &         1.490 &           1.495 &  0.600 &       No \\
52     &  FA$_{0.803}$MA$_{0.197}$PbI$_3$ &         1.494 &           1.500 &  0.452 &      Yes \\
53     &  FA$_{0.792}$MA$_{0.208}$PbI$_3$ &         1.491 &           1.500 &  0.354 &       No \\
54     &  FA$_{0.783}$MA$_{0.217}$PbI$_3$ &         1.492 &           1.502 &  0.403 &       No \\
55     &  FA$_{0.783}$MA$_{0.217}$PbI$_3$ &         1.502 &           1.512 &  0.518 &       No \\
56     &  FA$_{0.782}$MA$_{0.218}$PbI$_3$ &         1.498 &           1.505 &  0.504 &       No \\
57     &  FA$_{0.779}$MA$_{0.221}$PbI$_3$ &         1.503 &           1.512 &      - &        - \\
58     &  FA$_{0.767}$MA$_{0.233}$PbI$_3$ &         1.494 &           1.503 &  0.446 &       No \\
59     &  FA$_{0.767}$MA$_{0.233}$PbI$_3$ &         1.508 &           1.513 &  0.628 &       No \\
60     &  FA$_{0.767}$MA$_{0.233}$PbI$_3$ &         1.494 &           1.505 &  0.623 &       No \\
61     &  FA$_{0.750}$MA$_{0.250}$PbI$_3$ &         1.494 &           1.508 &  0.480 &       No \\
62     &  FA$_{0.750}$MA$_{0.250}$PbI$_3$ &         1.506 &           1.514 &  0.576 &       No \\
63     &  FA$_{0.750}$MA$_{0.250}$PbI$_3$ &         1.492 &           1.505 &  0.551 &       No \\
64     &  FA$_{0.738}$MA$_{0.262}$PbI$_3$ &         1.493 &           1.504 &  0.532 &       No \\
65     &  FA$_{0.733}$MA$_{0.267}$PbI$_3$ &         1.487 &           1.505 &  0.531 &       No \\
66     &  FA$_{0.733}$MA$_{0.267}$PbI$_3$ &         1.503 &           1.514 &  0.320 &       No \\
67     &  FA$_{0.733}$MA$_{0.267}$PbI$_3$ &         1.497 &           1.509 &  0.621 &       No \\
68     &  FA$_{0.733}$MA$_{0.267}$PbI$_3$ &         1.490 &           1.500 &      - &        - \\
69     &  FA$_{0.724}$MA$_{0.276}$PbI$_3$ &         1.504 &           1.514 &      - &        - \\
70     &  FA$_{0.717}$MA$_{0.283}$PbI$_3$ &         1.495 &           1.500 &      - &        - \\
71     &  FA$_{0.717}$MA$_{0.283}$PbI$_3$ &         1.503 &           1.512 &      - &        - \\
72     &  FA$_{0.712}$MA$_{0.288}$PbI$_3$ &         1.498 &           1.512 &  0.555 &       No \\
73     &  FA$_{0.700}$MA$_{0.300}$PbI$_3$ &         1.505 &           1.512 &  0.477 &       No \\
74     &  FA$_{0.700}$MA$_{0.300}$PbI$_3$ &         1.499 &           1.510 &  0.414 &       No \\
75     &  FA$_{0.697}$MA$_{0.303}$PbI$_3$ &         1.501 &           1.510 &  0.433 &       No \\
76     &  FA$_{0.686}$MA$_{0.314}$PbI$_3$ &         1.511 &           1.522 &  0.567 &       No \\
77     &  FA$_{0.683}$MA$_{0.317}$PbI$_3$ &         1.502 &           1.510 &  0.595 &        - \\
78     &  FA$_{0.683}$MA$_{0.317}$PbI$_3$ &         1.503 &           1.510 &  0.528 &       No \\
79     &  FA$_{0.683}$MA$_{0.317}$PbI$_3$ &         1.514 &           1.525 &      - &       No \\
80     &  FA$_{0.677}$MA$_{0.323}$PbI$_3$ &         1.502 &           1.510 &  0.282 &       No \\
81     &  FA$_{0.667}$MA$_{0.333}$PbI$_3$ &         1.516 &           1.528 &  0.498 &       No \\
82     &  FA$_{0.667}$MA$_{0.333}$PbI$_3$ &         1.504 &           1.515 &  0.423 &       No \\
83     &  FA$_{0.667}$MA$_{0.333}$PbI$_3$ &         1.498 &           1.510 &  0.219 &       No \\
84     &  FA$_{0.650}$MA$_{0.350}$PbI$_3$ &         1.501 &           1.510 &  0.353 &       No \\
85     &  FA$_{0.650}$MA$_{0.350}$PbI$_3$ &         1.502 &           1.510 &  0.383 &       No \\
86     &  FA$_{0.650}$MA$_{0.350}$PbI$_3$ &         1.507 &           1.515 &  0.302 &       No \\
87     &  FA$_{0.625}$MA$_{0.375}$PbI$_3$ &         1.503 &           1.513 &  0.250 &       No \\
88     &  FA$_{0.624}$MA$_{0.376}$PbI$_3$ &         1.510 &           1.515 &  0.389 &       No \\
89     &  FA$_{0.617}$MA$_{0.383}$PbI$_3$ &         1.509 &           1.520 &  0.335 &       No \\
90     &  FA$_{0.614}$MA$_{0.386}$PbI$_3$ &         1.505 &           1.515 &  0.196 &       No \\
91     &  FA$_{0.611}$MA$_{0.389}$PbI$_3$ &         1.522 &           1.530 &  0.549 &       No \\
92     &  FA$_{0.600}$MA$_{0.400}$PbI$_3$ &         1.518 &           1.530 &  0.391 &       No \\
93     &  FA$_{0.600}$MA$_{0.400}$PbI$_3$ &         1.509 &           1.520 &  0.255 &       No \\
94     &  FA$_{0.600}$MA$_{0.400}$PbI$_3$ &         1.509 &           1.520 &  0.241 &       No \\
95     &  FA$_{0.583}$MA$_{0.417}$PbI$_3$ &         1.510 &           1.520 &  0.331 &       No \\
96     &  FA$_{0.583}$MA$_{0.417}$PbI$_3$ &         1.511 &           1.522 &  0.280 &       No \\
97     &  FA$_{0.583}$MA$_{0.417}$PbI$_3$ &         1.518 &           1.530 &  0.344 &       No \\
98     &  FA$_{0.569}$MA$_{0.431}$PbI$_3$ &         1.512 &           1.523 &  0.303 &       No \\
99     &  FA$_{0.567}$MA$_{0.433}$PbI$_3$ &         1.520 &           1.530 &  0.325 &       No \\
100    &  FA$_{0.567}$MA$_{0.433}$PbI$_3$ &         1.508 &           1.520 &  0.429 &       No \\
101    &  FA$_{0.559}$MA$_{0.441}$PbI$_3$ &         1.501 &           1.510 &  0.433 &       No \\
102    &  FA$_{0.558}$MA$_{0.442}$PbI$_3$ &         1.512 &           1.526 &  0.366 &       No \\
103    &  FA$_{0.558}$MA$_{0.442}$PbI$_3$ &         1.511 &           1.523 &  0.320 &       No \\
104    &  FA$_{0.550}$MA$_{0.450}$PbI$_3$ &         1.524 &           1.530 &  0.414 &       No \\
105    &  FA$_{0.550}$MA$_{0.450}$PbI$_3$ &         1.503 &           1.510 &  0.270 &       No \\
106    &  FA$_{0.545}$MA$_{0.455}$PbI$_3$ &         1.513 &           1.523 &  0.365 &       No \\
107    &  FA$_{0.533}$MA$_{0.467}$PbI$_3$ &         1.517 &           1.532 &  0.372 &       No \\
108    &  FA$_{0.533}$MA$_{0.467}$PbI$_3$ &         1.509 &           1.520 &  0.288 &       No \\
109    &  FA$_{0.529}$MA$_{0.471}$PbI$_3$ &         1.515 &           1.526 &  0.252 &       No \\
110    &  FA$_{0.517}$MA$_{0.483}$PbI$_3$ &         1.529 &           1.532 &  0.282 &       No \\
111    &  FA$_{0.517}$MA$_{0.483}$PbI$_3$ &         1.523 &           1.530 &  0.340 &       No \\
112    &  FA$_{0.517}$MA$_{0.483}$PbI$_3$ &         1.513 &           1.528 &  0.356 &       No \\
113    &  FA$_{0.511}$MA$_{0.489}$PbI$_3$ &         1.528 &           1.532 &  0.286 &       No \\
114    &  FA$_{0.504}$MA$_{0.496}$PbI$_3$ &         1.514 &           1.530 &  0.270 &       No \\
115    &  FA$_{0.500}$MA$_{0.500}$PbI$_3$ &         1.500 &           1.535 &  0.254 &       No \\
116    &  FA$_{0.500}$MA$_{0.500}$PbI$_3$ &         1.518 &           1.530 &  0.216 &       No \\
117    &  FA$_{0.498}$MA$_{0.502}$PbI$_3$ &         1.518 &           1.525 &  0.253 &       No \\
118    &  FA$_{0.483}$MA$_{0.517}$PbI$_3$ &         1.522 &           1.525 &  0.244 &       No \\
119    &  FA$_{0.459}$MA$_{0.541}$PbI$_3$ &         1.524 &           1.530 &  0.210 &       No \\
120    &  FA$_{0.450}$MA$_{0.550}$PbI$_3$ &         1.523 &           1.530 &  0.241 &       No \\
121    &  FA$_{0.447}$MA$_{0.553}$PbI$_3$ &         1.520 &           1.530 &  0.306 &       No \\
122    &  FA$_{0.445}$MA$_{0.555}$PbI$_3$ &         1.530 &           1.537 &  0.282 &       No \\
123    &  FA$_{0.433}$MA$_{0.567}$PbI$_3$ &         1.523 &           1.530 &  0.331 &       No \\
124    &  FA$_{0.433}$MA$_{0.567}$PbI$_3$ &         1.527 &           1.533 &  0.305 &       No \\
125    &  FA$_{0.433}$MA$_{0.567}$PbI$_3$ &         1.529 &           1.537 &  0.288 &       No \\
126    &  FA$_{0.417}$MA$_{0.583}$PbI$_3$ &         1.524 &           1.533 &  0.416 &       No \\
127    &  FA$_{0.417}$MA$_{0.583}$PbI$_3$ &         1.525 &           1.530 &  0.262 &       No \\
128    &  FA$_{0.417}$MA$_{0.583}$PbI$_3$ &         1.528 &           1.537 &  0.351 &       No \\
129    &  FA$_{0.402}$MA$_{0.598}$PbI$_3$ &         1.524 &           1.533 &  0.312 &       No \\
130    &  FA$_{0.400}$MA$_{0.600}$PbI$_3$ &         1.518 &           1.530 &  0.248 &       No \\
131    &  FA$_{0.400}$MA$_{0.600}$PbI$_3$ &         1.521 &           1.530 &  0.287 &       No \\
132    &  FA$_{0.400}$MA$_{0.600}$PbI$_3$ &         1.527 &           1.539 &  0.408 &       No \\
133    &  FA$_{0.392}$MA$_{0.608}$PbI$_3$ &         1.524 &           1.530 &  0.279 &       No \\
134    &  FA$_{0.391}$MA$_{0.609}$PbI$_3$ &         1.527 &           1.535 &  0.271 &       No \\
135    &  FA$_{0.383}$MA$_{0.617}$PbI$_3$ &         1.531 &           1.538 &  0.347 &       No \\
136    &  FA$_{0.383}$MA$_{0.617}$PbI$_3$ &         1.517 &           1.530 &  0.457 &       No \\
137    &  FA$_{0.367}$MA$_{0.633}$PbI$_3$ &         1.529 &           1.539 &  0.274 &       No \\
138    &  FA$_{0.367}$MA$_{0.633}$PbI$_3$ &         1.527 &           1.535 &  0.410 &       No \\
139    &  FA$_{0.367}$MA$_{0.633}$PbI$_3$ &         1.530 &           1.540 &  0.278 &       No \\
140    &  FA$_{0.360}$MA$_{0.640}$PbI$_3$ &         1.529 &           1.540 &  0.352 &       No \\
141    &  FA$_{0.350}$MA$_{0.650}$PbI$_3$ &         1.529 &           1.540 &  0.332 &       No \\
142    &  FA$_{0.350}$MA$_{0.650}$PbI$_3$ &         1.528 &           1.535 &  0.256 &       No \\
143    &  FA$_{0.350}$MA$_{0.650}$PbI$_3$ &         1.531 &           1.540 &  0.347 &       No \\
144    &  FA$_{0.342}$MA$_{0.658}$PbI$_3$ &         1.531 &           1.540 &  0.335 &       No \\
145    &  FA$_{0.336}$MA$_{0.664}$PbI$_3$ &         1.529 &           1.535 &  0.291 &       No \\
146    &  FA$_{0.333}$MA$_{0.667}$PbI$_3$ &         1.532 &           1.542 &  0.227 &       No \\
147    &  FA$_{0.333}$MA$_{0.667}$PbI$_3$ &         1.532 &           1.542 &  0.369 &       No \\
148    &  FA$_{0.332}$MA$_{0.668}$PbI$_3$ &         1.529 &           1.540 &  0.289 &       No \\
149    &  FA$_{0.317}$MA$_{0.683}$PbI$_3$ &         1.532 &           1.540 &  0.244 &       No \\
150    &  FA$_{0.304}$MA$_{0.696}$PbI$_3$ &         1.535 &           1.542 &  0.283 &       No \\
151    &  FA$_{0.303}$MA$_{0.697}$PbI$_3$ &         1.531 &           1.542 &  0.281 &       No \\
152    &  FA$_{0.292}$MA$_{0.708}$PbI$_3$ &         1.530 &           1.540 &  0.236 &       No \\
153    &  FA$_{0.280}$MA$_{0.720}$PbI$_3$ &         1.533 &           1.546 &  0.254 &       No \\
154    &  FA$_{0.278}$MA$_{0.722}$PbI$_3$ &         1.537 &           1.545 &  0.266 &       No \\
155    &  FA$_{0.276}$MA$_{0.724}$PbI$_3$ &         1.529 &           1.540 &  0.276 &       No \\
156    &  FA$_{0.267}$MA$_{0.733}$PbI$_3$ &         1.538 &           1.548 &  0.243 &       No \\
157    &  FA$_{0.267}$MA$_{0.733}$PbI$_3$ &         1.532 &           1.548 &  0.312 &       No \\
158    &  FA$_{0.267}$MA$_{0.733}$PbI$_3$ &         1.530 &           1.540 &  0.305 &       No \\
159    &  FA$_{0.250}$MA$_{0.750}$PbI$_3$ &         1.537 &           1.547 &  0.258 &       No \\
160    &  FA$_{0.250}$MA$_{0.750}$PbI$_3$ &         1.534 &           1.540 &  0.333 &       No \\
161    &  FA$_{0.250}$MA$_{0.750}$PbI$_3$ &         1.538 &           1.547 &  0.354 &       No \\
162    &  FA$_{0.234}$MA$_{0.766}$PbI$_3$ &         1.537 &           1.548 &  0.344 &       No \\
163    &  FA$_{0.233}$MA$_{0.767}$PbI$_3$ &         1.530 &           1.540 &  0.331 &       No \\
164    &  FA$_{0.233}$MA$_{0.767}$PbI$_3$ &         1.538 &           1.550 &  0.441 &       No \\
165    &  FA$_{0.233}$MA$_{0.767}$PbI$_3$ &         1.525 &           1.533 &  0.312 &       No \\
166    &  FA$_{0.226}$MA$_{0.774}$PbI$_3$ &         1.538 &           1.550 &  0.353 &       No \\
167    &  FA$_{0.221}$MA$_{0.779}$PbI$_3$ &         1.539 &           1.550 &  0.433 &       No \\
168    &  FA$_{0.217}$MA$_{0.783}$PbI$_3$ &         1.540 &           1.550 &  0.501 &       No \\
169    &  FA$_{0.217}$MA$_{0.783}$PbI$_3$ &         1.524 &           1.535 &  0.457 &       No \\
170    &  FA$_{0.200}$MA$_{0.800}$PbI$_3$ &         1.540 &           1.550 &  0.344 &       No \\
171    &  FA$_{0.200}$MA$_{0.800}$PbI$_3$ &         1.532 &           1.540 &  0.415 &       No \\
172    &  FA$_{0.200}$MA$_{0.800}$PbI$_3$ &         1.542 &           1.550 &  0.541 &       No \\
173    &  FA$_{0.191}$MA$_{0.809}$PbI$_3$ &         1.539 &           1.550 &  0.361 &       No \\
174    &  FA$_{0.183}$MA$_{0.817}$PbI$_3$ &         1.536 &           1.540 &  0.470 &       No \\
175    &  FA$_{0.183}$MA$_{0.817}$PbI$_3$ &         1.545 &           1.550 &  0.229 &       No \\
176    &  FA$_{0.183}$MA$_{0.817}$PbI$_3$ &         1.548 &           1.552 &  0.324 &       No \\
177    &  FA$_{0.172}$MA$_{0.828}$PbI$_3$ &         1.539 &           1.552 &  0.503 &       No \\
178    &  FA$_{0.167}$MA$_{0.833}$PbI$_3$ &         1.537 &           1.545 &  0.307 &       No \\
179    &  FA$_{0.167}$MA$_{0.833}$PbI$_3$ &         1.547 &           1.552 &  0.247 &       No \\
180    &  FA$_{0.166}$MA$_{0.834}$PbI$_3$ &         1.539 &           1.552 &  0.489 &       No \\
181    &  FA$_{0.163}$MA$_{0.837}$PbI$_3$ &         1.537 &           1.545 &  0.374 &       No \\
182    &  FA$_{0.150}$MA$_{0.850}$PbI$_3$ &         1.537 &           1.546 &  0.316 &       No \\
183    &  FA$_{0.150}$MA$_{0.850}$PbI$_3$ &         1.536 &           1.552 &  0.209 &       No \\
184    &  FA$_{0.133}$MA$_{0.867}$PbI$_3$ &         1.540 &           1.550 &  0.345 &       No \\
185    &  FA$_{0.125}$MA$_{0.875}$PbI$_3$ &         1.544 &           1.552 &  0.261 &       No \\
186    &  FA$_{0.117}$MA$_{0.883}$PbI$_3$ &         1.544 &           1.550 &  0.300 &       No \\
187    &  FA$_{0.112}$MA$_{0.888}$PbI$_3$ &         1.553 &           1.556 &  0.279 &       No \\
188    &  FA$_{0.111}$MA$_{0.889}$PbI$_3$ &         1.550 &           1.555 &  0.276 &       No \\
189    &  FA$_{0.109}$MA$_{0.891}$PbI$_3$ &         1.545 &           1.550 &  0.383 &       No \\
190    &  FA$_{0.100}$MA$_{0.900}$PbI$_3$ &         1.548 &           1.555 &  0.309 &       No \\
191    &  FA$_{0.100}$MA$_{0.900}$PbI$_3$ &         1.549 &           1.560 &  0.494 &       No \\
192    &  FA$_{0.100}$MA$_{0.900}$PbI$_3$ &         1.540 &           1.550 &  0.368 &       No \\
193    &  FA$_{0.083}$MA$_{0.917}$PbI$_3$ &         1.541 &           1.550 &  0.468 &       No \\
194    &  FA$_{0.083}$MA$_{0.917}$PbI$_3$ &         1.549 &           1.560 &  0.434 &       No \\
195    &  FA$_{0.083}$MA$_{0.917}$PbI$_3$ &         1.548 &           1.555 &  0.361 &       No \\
196    &  FA$_{0.067}$MA$_{0.933}$PbI$_3$ &         1.552 &           1.555 &  0.442 &       No \\
197    &  FA$_{0.067}$MA$_{0.933}$PbI$_3$ &         1.547 &           1.560 &  0.444 &       No \\
198    &  FA$_{0.067}$MA$_{0.933}$PbI$_3$ &         1.537 &           1.550 &  0.543 &       No \\
199    &  FA$_{0.067}$MA$_{0.933}$PbI$_3$ &         1.551 &           1.558 &  0.516 &       No \\
200    &  FA$_{0.067}$MA$_{0.933}$PbI$_3$ &         1.533 &           1.548 &  0.554 &       No \\
201    &  FA$_{0.058}$MA$_{0.942}$PbI$_3$ &         1.553 &           1.560 &  0.498 &       No \\
\label{table4}
\end{longtable}

\newpage
{\small
\bibliographystyleSM{sn-nature}
\bibliographySM{references}
}

\end{document}